\documentclass[twocolumn,10pt]{asme2e}

\usepackage{graphicx}
\usepackage{amssymb,amsmath}
\usepackage{algorithm2e}
\usepackage{multirow}
\usepackage{epstopdf}
\usepackage{subfig}

\SetAlgoCaptionSeparator{.}
\SetAlCapNameFnt{\small}
\usepackage{bm}
\usepackage{amsfonts}
\confshortname{OMAE 2020}
\conffullname{the ASME 2020 39th International Conference on Ocean, Offshore and Arctic Engineering}

\confdate{28-3}
\confmonth{June-July}
\confyear{2020}
\confcity{Fort Lauderdale}
\confcountry{USA}

\papernum{OMAE2020-18556}

\title{Deep Convolutional Recurrent Autoencoders for Flow Field Prediction}

\author{Sandeep R. Bukka\thanks{Email address of correspondance-sandeep@u.nus.edu},$\quad$Allan Ross Magee
    \affiliation{
    Keppel-NUS Corp Lab\\
    National University of Singapore\\
    Department of Civil and Environmental Engineering\\
    Singapore
    }    
}

 \author{Rajeev K. Jaiman\thanks{Currently at University of British Columbia}
    \affiliation{
    Department of Mechanical Engineering\\
    National University of Singapore \\ 
    Singapore
    }
}

\def\buf{\textbf{u}^\mathrm{f}}
\def\bbf{\textbf{b}^\mathrm{f}}
\def\stf{\bm{\sigma}^{\mathrm{f}}}
\def\df{\Omega^\mathrm{f}(t)}
\def\vf{\mu^\mathrm{f}}
\def\bus{\textbf{u}^\mathrm{s}}
\def\ps{\bm{\phi}^\mathrm{s}}
\def\bbs{\textbf{b}^\mathrm{s}}
\def\Fts{\textbf{F}^\mathrm{s}}
\def\ds{\Omega^\mathrm{s}}

\begin{document}

\maketitle

\begin{abstract}
In this paper, an end-to-end nonlinear model reduction methodology is presented based on the convolutional recurrent autoencoder networks. The methodology is developed in the context of overall data-driven reduced order model framework proposed in the paper. The basic idea behind the methodology is
to obtain the low dimensional representations via convolutional neural networks and evolve these
low dimensional features via recurrent neural networks in time domain. The high dimensional
representations are constructed from the evolved low dimensional features via transpose
convolutional neural networks. With an unsupervised training strategy, the model serves as an end
to end tool which can evolve the flow state of the nonlinear dynamical system. The convolutional recurrent
autoencoder network model is applied on the problem of flow past bluff bodies for the first
time. To demonstrate the effectiveness of the methodology, two canonical problems namely the flow past plain cylinder and the flow past side-by-side cylinders are
explored in this paper. Pressure and velocity fields of the unsteady flow are predicted in
future via the convolutional recurrent autoencoder model. The performance of the model is satisfactory for both the problems. Specifically, the multiscale nature and  the gap flow dynamics of the side-by-side cylinders are captured by the proposed data-driven model reduction methodology. The error metrics, the normalized squared error and the normalized reconstruction error are considered for the assessment of the data-driven framework.


\end{abstract}



\section*{INTRODUCTION}

In the past decade, the hardware and techniques in experiments and computations
have become much more advanced and resulted in the generation of massive amounts
of data. It is definitely a welcoming note to have more data, but the gigantic scale
of modern datasets poses serious challenges in terms of analysis. The demand for
data driven methods
rises from these challenges and the primary motivation of such
methods is to come up with efficient models which can characterize the data in
meaningful ways. One other way to put the analysis via data-driven methods is learning from experience. In nature, various animals ranging from fish, cheetah, birds
and other aerial, aquatic and terrestrial lifeforms are able to achieve remarkable levels
of efficiency in terms of speed, manoeuvring and control by manipulating the fluid
environment around them. The critical point here is to question the learning process of these organisms.

It is very clear that they do not have any prior knowledge of calculus or the Navier-Stokes equations. Humans also achieved remarkable engineering feats of fluid manipulation before the knowledge of current equation based analysis came into existence.
The feasible explanation for such learning is that it came from intuition and experience
\cite{bruntonreview19}.
It is also an undeniable fact that the current approach of physics-based engineering is an epitome of human achievement which revolutionized the performance of the
engineering systems. However, there are serious hurdles associated with physics-based
engineering such as high-dimensionality and non-linearity, which defy closed-form
solutions and limit real-time optimization and control efforts. With the recent surge
in computational capabilities and advancements in machine learning, deep learning,
artificial intelligence, learning from experience has taken the spotlight.

The marine/offshore industry is going through a paradigm shift with regard to automation and autonomy. While automation generally implies to the activity performed
devoid of human intervention, autonomy is the ability of a system to achieve goals
while operating independently of external control (i.e., self- sufficient and self-directed
system). Major industries such as automobile, aerospace, health care, and logistics
operations are already seeing the benefits of having smart autonomous systems. The
marine/offshore industry is not an exception. Since the presence of technologies related
to smart autonomous systems (SAS) is burgeoning at a rapid pace, the mechanics
community needs to integrate their tools and techniques with these SAS technologies so
that the systems can be made more efficient, reliable and adaptable to marine/offshore
applications. The construction of an efficient ROM and its
integration with the sensory data is necessary to achieve a smart autonomous system \cite{reddy2019reduced}.

\begin{figure} 
\centering
\includegraphics[width = 0.45\textwidth]{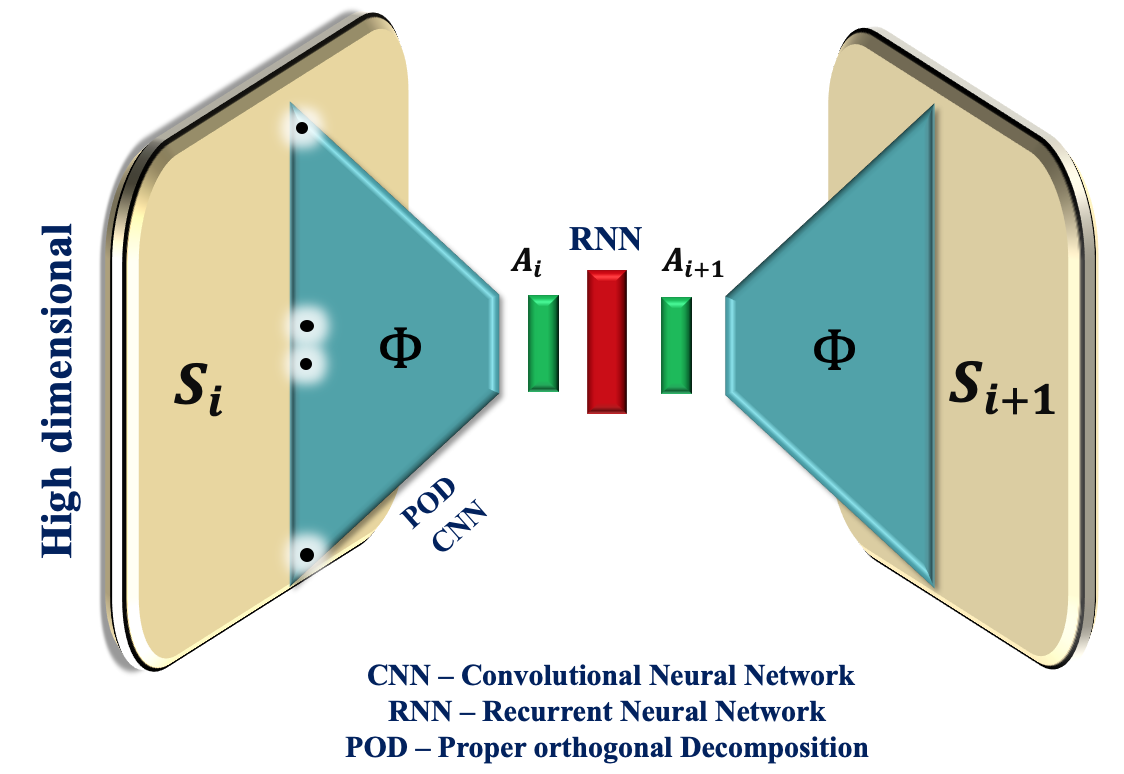}
\caption{A generic sketch of  data-driven reduced order framework using POD/CNN and RNN techniques}
\label{fig16}
\end{figure}

An overall representation of the proposed data-driven framework is presented in Figure \ref{fig16}. The framework encompasses the previously reported POD-RNN hybrid model in \cite{reddy2019reduced}. The main idea of this hybrid model comprises the following two steps: 
\begin{itemize}
    \item 1.  The identification of a low-dimensional manifold $\boldsymbol{\Phi}$ embedded in 
    $\mathbb{R}^{N}$ on which most of the data is supported. This yields, in some sense, an optimal low-dimensional representations $\textbf{A} = f(\textbf{S})$ of the data $\textbf{S}$ in terms of the intrinsic coordinates on $\boldsymbol{\Phi}$, and
    \item 2. The identification of a dynamic model which effectively evolves the low dimensional representation $\textbf{A}$ on the manifold $\boldsymbol{\Phi}$.

\end{itemize}

While the POD based model reduction is optimal and can generate physically interpretable modes it is not good enough for practical problems. In such cases, the number of dominant modes will increase and their computation can pose serious challenges. Up to 27 modes were required to capture the dominant energy of the side-by-side cylinder whereas only 4 modes were required for the flow past an isolated (single) cylinder. Similarly it was shown in \cite{miyanawala2019decomposition}, that 123 modes were required for the flow past a square cylinder at $Re = 22000$. As a natural consequence, the question of completely bypassing POD in building reduced order model arises and the alternative technique based on convolutional neural networks (CNN) will be explored in this paper for extracting the low dimensional features.

\begin{figure}
    \centering
    \includegraphics[width = 0.45\textwidth]{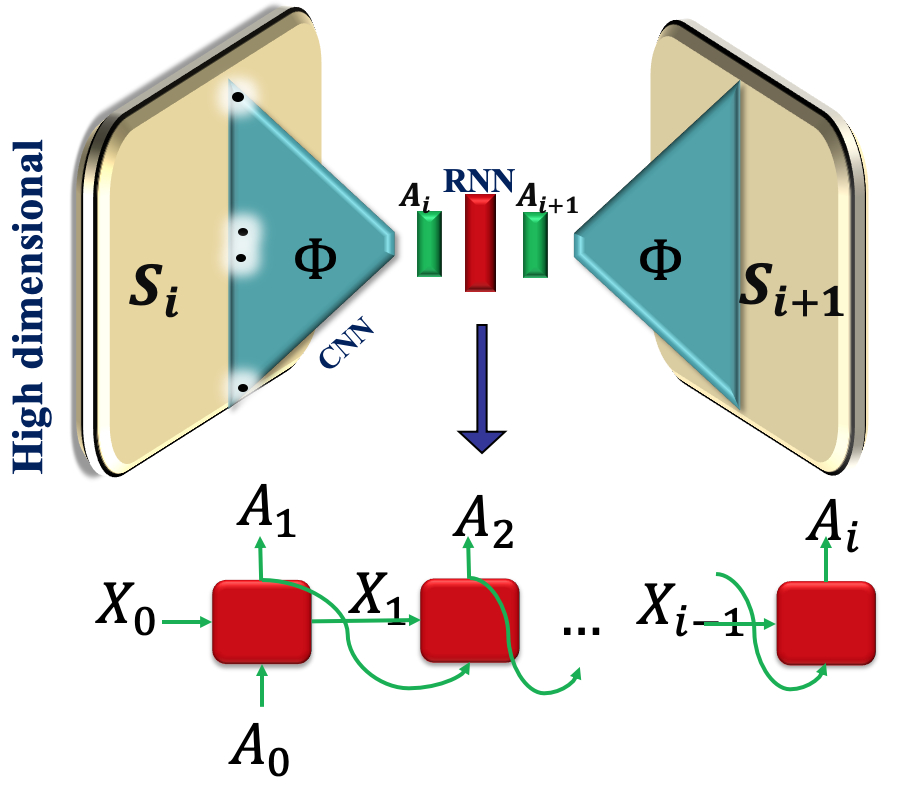}
    \caption{Schematic of Convolutional Recurrent Autoencoder network}
    \label{figcnnrnn}
\end{figure}

In addition to the POD-RNN, and also inspired from the overall framework, is a complete data-driven one, where both the low-dimensional representation of the state variable and its temporal evolution are learned via machine learning techniques. Such approach has been explored in \cite{otto2019linearly} in which an autoencoder is used to learn a low-dimensional representation of the high-dimensional state,
\begin{equation}
    \textbf{A} = \mathcal{E}(\textbf{S}),
     \label{eq62}
\end{equation}
where $\textbf{S} \in \mathbb{R}^{N}$ high-dimensional state of the system, $\textbf{A}\in \mathbb{R}^{N_{A}},N_{A}<N$, $\mathcal{E}$ is a nonlinear enocoder network and a linear recurrent model is used to evolve the low dimensional features
\begin{equation}
    \textbf{A}^{n+1} = \textbf{K}\textbf{A}^{n},
    \label{eq63}
\end{equation}
where $\textbf{K}\in \mathbb{R}^{N_{A}\times N_{A}}$. This approach was first introduced in the context of
learning a dictionary of functions used in extended dynamic mode decomposition to approximate the Koopman operator of a nonlinear system \cite{li2017extended}.
Based on the framework introduced in \cite{kani2017dr} \cite{otto2019linearly} \cite{wang2018model}, \cite{gonzalez2018deep} introduced a deep convolutional autoencoder architecture which provides certain advantages in identifying low-dimensional representations of the input data. Figure \ref{figcnnrnn} depicts the schematic of the model reduction via convolutional recurrent autoencoder network. Benchmark problems of lid-driven cavity, viscous Burgers equation were explored via the above approach in \cite{gonzalez2018deep}. 
However, such problems are far away of the practical scenarios. There is a need to address the development of efficient models for more complex problems such as flow past side-by-side cylinders, flexible offshore riser, and ocean wave predictions. Such problems contain more complex physics with multiscale features and various physical phenomena such as vortex-induced vibration, vortex shedding, gap flow, flip flopping, bifurcation and turbulence. 
In this paper, we extend the capabilities of the convolutional recurrent autoencoder model by applying it for the first time on the more complex problems of fluid-structure interaction. \cite{gonzalez2018deep} contains more details on the model reduction via convolutional recurrent networks and they are also presented here for the completeness of the present paper.


\section*{NUMERICAL METHODOLOGY}
This section deals with the formulation used for high-dimensional full-order model followed by the implementation of deep convolutional recurrent autoencoder model for a coupled fluid-structure problem. To begin with, the formulation of the coupled variational fluid-structure solver is outlined in the following section.
\subsection*{Full-order model formulation}
A numerical scheme implementing Petrov-Galerkin finite-element and semi-discrete time stepping are adopted in the present work \cite{jaiman2016partitioned,jaiman2016stable}. The incompressible Navier-Stokes equations are used in the Arbitrary Lagrangian-Eulerian (ALE) reference frame and formulated in the following form:
\begin{align}
\rho^\mathrm{f}\left(\frac{\partial \buf}{\partial t}\bigg\rvert_{\widehat{x}}+\left(\buf-\textbf{w}\right)\cdot\bm{\nabla}\buf\right)& = \bm{\nabla}\cdot\stf+\bbf \quad \text{on} \quad \Omega^{f}(t), \label{eq1}\\
\bm{\nabla}\cdot\buf& = 0 \quad \text{on}\quad \df, \label{eq2}
\end{align}
where $\buf = \buf(\textbf{x},t)$ and $\textbf{w}=\textbf{w}(\textbf{x},t)$ are the fluid and mesh velocities, respectively. In Eq.~ (\ref{eq1}) the partial time derivative with respect to the ALE referential coordinate $\widehat{x}$ is constant. 
Here $\bbf$ represents the body force per unit mass and $\stf$ is the Cauchy stress tensor for a Newtonian fluid which is defined as
\begin{equation}
\stf = -\textit{p}\textbf{I}+\vf\left(\bm{\nabla}\buf+(\bm{\nabla}\buf)^{T}\right),
\end{equation}
where \textit{p}, $\vf$ and \textbf{I} are the hydrodynamic pressure, the dynamic viscosity of the fluid, and the identity tensor, respectively. A rigid-body structure submerged in the fluid experiences unsteady fluid forces and consequently may undergo flow-induced vibrations if the body is mounted elastically. To simulate the translational motion of a rigid body about its center of mass, the equation along the Cartesian axes is given by:
\begin{equation}
\textbf{m}\cdot\frac{\partial \bus}{\partial t} + \textbf{c}\cdot\bus+\textbf{k}\cdot\left(\ps(\textbf{z}_{0},t)-\textbf{z}_{0}\right) = \Fts + \bbs \: \text{on} \: \ds,
\end{equation}
where $\textbf{m}$, $\textbf{c}$, $\textbf{k}$, $\Fts$ and $\bbs$ are the mass, damping coefficient, and stiffness coefficient vectors for
the translational motions, fluid traction, and body forces on the rigid body, respectively. Here $\ds$ represents the domain occupied by the rigid body and $\bus(t)$ represents the velocity of the immersed rigid body. The fluid and the structural equations are coupled by the continuity of velocity and traction along the fluid-structure interface. 

The new position of the rigid body is updated through a position vector $\ps$ , which maps the
initial position $\textbf{z}_{0}$ of the rigid body to its new position at time t. Let $\gamma$ be the Lagrangian point on $\Gamma$ and its corresponding mapping position vector to the new position after the motion of the rigid
body is $\phi(\gamma,t)$ at time $t$. Since the position and flow field around the moving rigid body is updated continuously, the no-slip and traction continuity conditions should be satisfied on the fluid-body
interface $\Gamma$
\begin{align}
\buf(\ps(\textbf{z}_{0},t),t)& = \bus(\textbf{z}_{0},t), \\
\int_{\phi(\gamma,t)} \stf(\textbf{x},t)\cdot\textbf{n}\text{d}\Gamma+\int_{\gamma}\Fts\text{d}\Gamma &= 0 \quad \forall \:\gamma\:\in\:\Gamma,
\end{align}
where \textbf{n} is the outer normal to the fluid-body interface. The characterization of the moving fluid-body interface is constructed by means of the ALE technique \cite{donaea1983arbitrary} The movement of the internal ALE nodes is constructed by solving a continuum hyperelastic model for the fluid mesh such that the mesh quality does not deteriorate as the displacement of the body increases. For the spatial and material mapping problem, we use classical Neo-Hookean material properties for the ALE variational formulation \cite{kuhl2004ale} which does not entail any additional user-defined remeshing parameter. The weak variational form of Eq.~(\ref{eq1}) is discretized in space using $\mathbb{P}_{2}/\mathbb{P}_{1}$ isoparametric finite elements for the fluid velocity and pressure. The second-order backward-differencing scheme is used for the time discretization of the Navier-Stokes system \cite{liu2014stable}. A partitioned staggered scheme is considered for the full-order simulations of fluid-structure interaction \cite{jaiman2011transient}. The above coupled variational formulation completes the presentation of full-order model for high-fidelity simulation. The employed in-house FSI solver has been extensively validated  in \cite{liu2016interaction,mysa2016origin}.
     
\section*{Model reduction via convolutional recurrent autoencoders}

\subsection*{Overview}
Fully-connected autoencoders are used in many applications for dimensionality reduction. The high dimensionality of DNS-level input data can act as a major bottleneck for applying fully-connected autoencoders. The input data should be flattened into an 1D array when we consider autoencoders. This might result in elimination of local spatial relations between the data. In other words, dominant flow features, often found in FSI simulations might be ignored or might not be captured efficiently. Therefore, direct application of fully-connected autoencoders on DNS-level input data is generally prohibited.
The current model seeks to
 extract dominant flow features present in many
fluid flow and fluid-structure interaction data by employing convolutional neural networks.

In particular, rather than applying a fully-connected autoencoder directly
to complex, high-dimensional simulation or experimental data, we apply it
to a vectorized feature map of a much lower-dimension obtained from a deep
convolutional network acting directly on the high-dimensional data. Wrapping the fully-connected autoencoder with a convolutional neural network has two significant advantages \cite{gonzalez2018deep}:
\begin{itemize}
    \item 1. The local approach of each convolutional layer helps to exploit local
    correlations in field values. Thus, much the same way finite-difference
    stencils can capture local gradients, each filter $\textbf{K}^{f}$ in a filter bank computes local low-level features from a small subset of the input. \\
    \item 2. The shared nature of each filter bank both allows identification of similar
    features throughout the input domain and reduction of the overall number
    of trainable parameters compared to a fully-connected layer with the
    same input size.
\end{itemize}

\begin{figure}
    \centering
    \includegraphics[width = 0.45\textwidth]{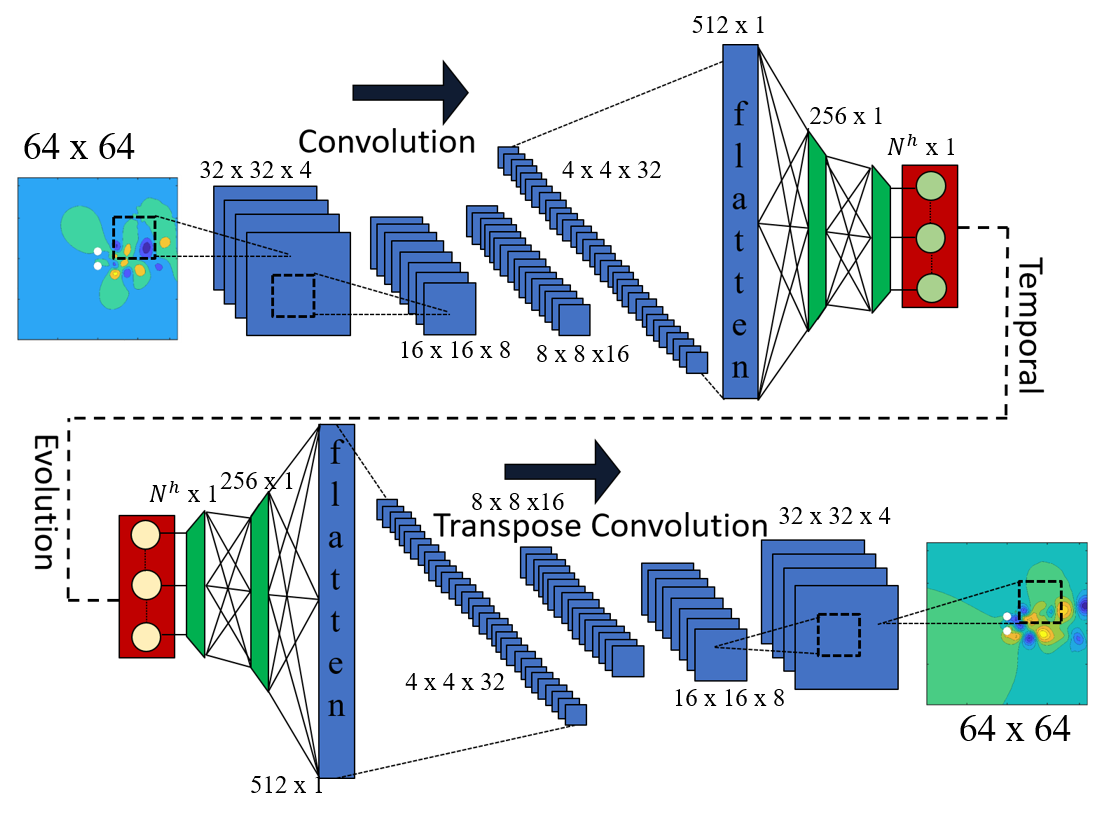}
    \caption{A sketch of 12 layer convolutional recurrent autoencoder model}
    \label{figcnnrnn1}
\end{figure}

\subsection*{Proposed architecture}
As shown in figure \ref{figcnnrnn1}, we construct a 12 layer convolutional autoencoder model. 
The first 4 layers together forms the convolution aspect of the model. The two dimensional input $S \in \mathbb{R}^{N_{x}\times N_{y}}$, with $N_{x} = N_{y} = 64$ is first passed through this convolutional block. All the layers in the convolution block used a filter bank $\textbf{K}^{f}\in \mathbb{R}^{4 \times 4}$ with the   number of filters in the first layer $f$ increasing from 4 in the first layer to 32 in the fourth layer. A stride of 2 and a padding of 1 is used across all the layers. 

The process of transpose convolution is carried out in the last four layers of the network. The tranpose convolution is not exactly the reverse of regular convolutions but it up-samples the low dimensional representations across four layers gradually to reconstruct the high dimensional state. The output size of the transpose convolutional layer is exactly the inverse of regular convolutional layer and is given by $N_o = (N_i-1)*s + f -2*P$, where $s$ is the stride, $f$ is the filter size, $P$ is the padding, $N_i$ is the input size and $N_o$ is the output size. 
Table \ref{tabcnnae} outlines the architecture of the convolutional encoder and decoder subgraphs. In this work, we consider the sigmoid activation function $\sigma(s) = \frac{1}{1+\exp^{-s}}$ for each layer of the autoencoder.

\begin{table}
    \centering
    \begin{tabular}{|c|c|c|c|c|}
    \hline
    Layer & filter size & filters & stride & padding \\ \hline
    1 & $4\times 4$ & 4 & 2 & 1 \\
    2 & $4\times 4$ & 8 & 2 & 1 \\
    3 & $4\times 4$ & 16 & 2 & 1 \\
    4 & $4\times 4$ & 32 & 2 & 1\\
    9 & $4\times 4$ & 16 & 2 & 1\\
    10 & $4\times 4$ & 8 & 2 & 1\\
    11 & $4\times 4$ & 4 & 2 & 1\\
    12 & $4\times 4$ & 1 & 2 & 1\\
    \hline
    \end{tabular}
    \caption{Filter sizes and strides of the layers in convolutional autoencoder network. Layers (1 to 4) correspond to encoder part and layers (9 to 12) correspond to decoder part}
    \label{tabcnnae}
\end{table}

A 4 layer fully connected regular autoencoder with a standard feed forward neural network is also used in the full network. 
The first two layers performs encoding by taking the vectorized form of 32 feature maps from the final layer of convolution block and returns a one dimensional vector $\textbf{A} \in \mathbb{R}^{N_{A}}$. The last two layers decode the evolved feature vector into high dimensional vector which will be reshaped as feature maps and passed into the transpose convolution block. Figure \ref{figcnnrnn1} illustrates the whole architecture with the changes in the size of the data as it flows throughout the entire model.  

The key innovation in using convolutional autoencoders in model reduction is that it allows for nonlinear autoencoders and thus nonlinear model reduction
can be applied to a large input data in a way that exploits structures inherent in many physical systems.

\subsection*{Evolution of features}
Now we proceed to the modeling the
evolution of low-dimensional features $\textbf{A}$ in a computationally efficient manner. In terms of of analysis viewpoint, it is beneficial to identify the the linear dynamics of $\textbf{A}$. However, we consider a general case of learning feature dynamics in a nonlinear setting.

Consider a set of observations $\{\textbf{s}^{n}\}_{n=0}^{m}$,
$\textbf{s}^{n}\in \mathbb{R}^{N}$ obtained from a CFD simulation or 
through experimental sampling.
An optimal low-dimensional representation $\textbf{A}^{n}\in \mathbb{R}^{N_{A}}$ where 
$N_{A}\ll N$ is obtained for each observations. These low-dimensional representations are obtained via convolutional autoencoder described in the previous section. The model for temporal evolution of this low dimensional features is constructed in a complete data-driven fashion.

The LSTM networks are employed to model the evolution of $\textbf{A}$. The details on the LSTM networks are outlined in the previous chapter. 
In a typical machine translation task, the size of the hidden states and number of layers are large when compared to the size of the feature vectors in the current work. Also, we are not interested in evolving the full high dimensional system. A single layer LSTM network is found to be sufficient for the current problem of evolving the feature vectors $\textbf{A}^{n}$. It is important to note that the the evolution
of $\textbf{A}$ does not require information from the full state $\textbf{S}$, thereby avoiding a
costly reconstruction at every step.

Initializing with a known low-dimensional representation $\textbf{A}^{0}$, one obtains a prediction for the following steps by iteratively applying
\begin{equation}
    \hat{\textbf{A}}^{n+1} = \boldsymbol{\mathcal{H}}(\hat{\textbf{A}}^{n}) \quad n = 1,2,3,\dots
    \label{eq66}
\end{equation}
where $\hat{\textbf{A}}^{1} = \boldsymbol{\mathcal{H}}(\textbf{A}^{0})$ and $\boldsymbol{\mathcal{H}}$ represents the action of a standard LSTM cell. A graphical representation of this model is depicted
in Figure \ref{figcnnrnn}.

\section*{Unsupervised training strategy}
The unsupervised training approach is the key component of the current model. The weights of both 
the convolutional autoencoder network and LSTM recurrent neural network are adjusted in a joint fashion. The main challenge here is to prevent overfitting of the CNN and RNN portion of the model. The construction of the training dataset as well as the training and evaluation algorithms are presented in this section.

Consider a dataset $\{\textbf{s}^{1},\dots,\textbf{s}^{m}\}$, where 
$\textbf{s}\in \mathbb{R}^{N_{x}\times N_{y}}$ is a 2D snapshot of
some dynamical system (e.g., a pressure field defined on a 2D grid). This dataset is broken up into a set of $\textbf{N}_{s}$
finite time training sequences
$\{\textbf{S}^{1},\dots,\textbf{S}^{N_{s}}\}$
where each training sequence
$\textbf{S}^{i}\in \mathbb{R}^{N_{x}\times N_{y}\times N_{t}}$
consists of
$N_{t}$ snapshots.

The fluctuations around the temporal mean are considered as followed generally in the data science community for improving the training. For example, we define
\begin{equation}
    \textbf{s}^{'n} = \textbf{s}^{n}-\Bar{\textbf{s}},
    \label{eq67}
\end{equation}
where $\Bar{\textbf{s}} = \frac{1}{m}\sum_{n=1}^{m}\textbf{s}^{n}$ is the temporal average over the entire dataset and $\textbf{s}^{'}$ are the fluctuations around this mean. Next the scaling of the above data is conducted as follows 
\begin{equation}
    \textbf{s}^{'n}_{s} = \frac{\textbf{s}^{'n}-\textbf{s}_{min}^{'}}{\textbf{s}_{max}^{'}-\textbf{s}_{min}^{'}}
    \label{eq68}
\end{equation}
where each $\textbf{s}^{'n}_{s} \in [0,1]^{N_{x}\times N_{y}}$.
The scaling is carried out to ensure that the data lies in the interval of $(0,1)$. This requirement stems from the fact that a sigmoid activation function is used in all the layers of convolutional autoenocoder. The training dataset with the above modifications has the following form.
\begin{equation}
    \mathcal{S} = \{\textbf{S}^{'1}_{s},\dots,\textbf{S}^{'N_{s}}_{s}\}\in [0,1]^{N_{x}\times N_{y}\times N_{t}\times N_{s}}
    \label{eq69}
\end{equation}
where each training sample $\textbf{S}^{'i}_{s} = [\textbf{S}^{'1}_{s,i},\dots,\textbf{S}_{s,i}^{N_{t}}]$ is a matrix consisting of the feature-scaled fluctuations.

\begin{figure}
    \centering
    \includegraphics[width = 0.5\textwidth]{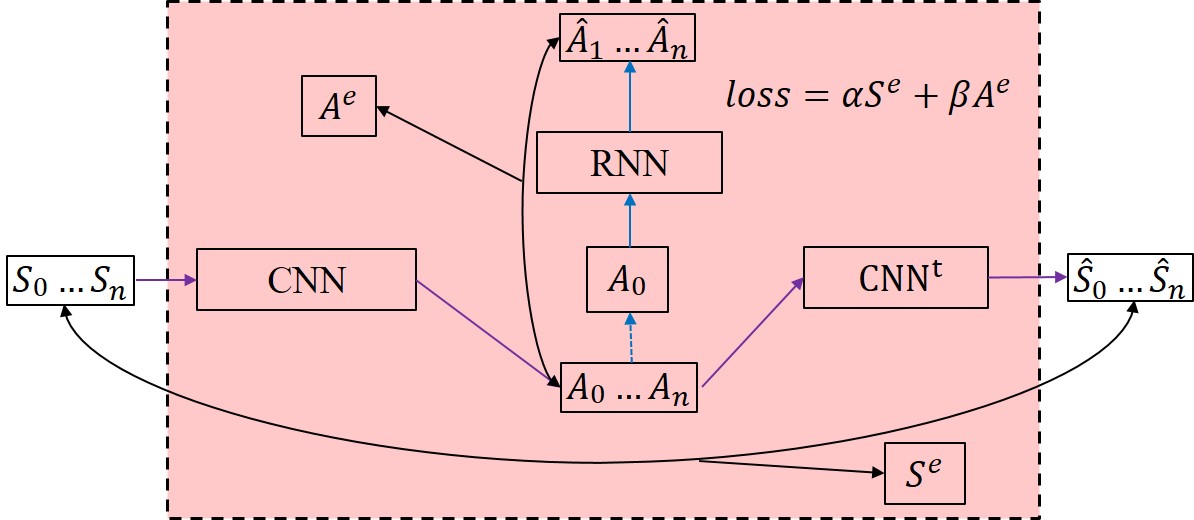}
    \caption{An illustration of training process of convolutional recurrent autoencoder model}
    \label{training}
\end{figure}

The process of training entails two stages primarily. The convolutional autoencoder takes an  $N_{b}$- sized batch of the training data $\mathcal{S}^{b}\subset \mathcal{S}$, where $\mathcal{S}^{b}\in [0,1]^{N_{x}\times N_{y}\times N_{t}\times N_{b}}$,and outputs both the current $N_{b}$-sized batch of low-dimensional representations of the training sequence
\begin{equation}
    \mathcal{A}^{b} = \{\textbf{A}^{1},\dots,\textbf{A}^{N_{b}}\}\in \mathbb{R}^{N_{A}\times N_{t}\times N_{b}},
    \label{eq610}
\end{equation}
where $\textbf{A}^{i} = [\textbf{A}^{1}_{i},\dots,\textbf{A}_{i}^{N_{t}}] \in \mathbb{R}^{N_{A}\times N_{t}}$
and a reconstruction
$\hat{\mathcal{S}}^{b}$ of the original input training batch in the first stage. This is represented as purple arrow in figure \ref{training}. 
 The second stage is indicated by the blue arrows in the same figure \ref{training}, where the first feature vector of each sequence is used to initialize and iteratively update Equation \ref{eq66} to get a reconstruction  $\hat{\mathcal{A}}^{b}$ of the low-dimensional representations of the training batch Equation \ref{eq610}.

{}

A loss function is constructed that equally weights the error in the full-state reconstruction and the evolution of the low-dimensional representations. The target of the training is to find the model parameters $\theta$  such that for any sequence 
$\textbf{S}_{s}^{'} = [\textbf{S}_{s}^{'1},\dots,\textbf{S}_{s}^{'N_{t}}]$
and its corresponding low-dimensional representation
$\textbf{A} = [\textbf{A}^{1},\dots,\textbf{A}^{N_{t}}]$, minimizes the following error between the
truth and the predictions as follows:
\begin{equation}
\begin{aligned}
    \mathcal{J}(\theta) &= [\mathcal{L}(\hat{\textbf{S}}_{s}^{'},\textbf{S}_{s}^{'},\hat{\textbf{A}},\textbf{A})] \\
    &=  \left[\frac{\alpha}{N_{t}}\operatornamewithlimits{\sum}_{n=1}^{N_{t}}\frac{\|\textbf{S}_{s}^{'n}-\hat{\textbf{S}}_{s}^{'n}\|_{2}^{2}}{\|\textbf{S}_{s}^{'n}\|_{2}^{2}}+\frac{\beta}{N_{t}}\frac{\|\textbf{A}^{n}-\hat{\textbf{A}}^{n}\|_{2}^{2}}{\|\textbf{A}^{n}\|_{2}^{2}}\right]
\end{aligned}
\label{loss}
\end{equation}
where
$\alpha = \beta = 0.5$. The values of $\alpha$ and $\beta$ are chosen in order to give equal weightage to the errors in full state reconstruction and feature prediction respectively. The predictions are denoted by the symbol $\hat{}$. In practice, the error is approximated by averaging
$\mathcal{L}(\hat{\textbf{S}}_{s}^{'},\textbf{S}_{s}^{'},\hat{\textbf{A}},\textbf{A})$ over all samples in a training batch during each backward pass. 
The convolutional autoencoder performs a regular forward pass and also constructs the low dimensional representations simultaneously at every training step. These low dimensional representations are used to train the RNN.

The ADAM optimizer \cite{kingma2014adam} is used for updating the weights during the training process. It is a modified version of the traditional stochastic gradient descent that computes adaptive learning rates for different parameters using estimates of first and second moments of the gradients. Algorithm \ref{algo:Ch6train} outlines the offline training of the convolutional recurrent autoencoder in more detail. Figure \ref{training} illustrates the overall schematic of the training process for easy comprehension. This model was built and trained using the open-source deep learning library TensorFlow.

\begin{figure}
    \centering
    \includegraphics[width = 0.5\textwidth]{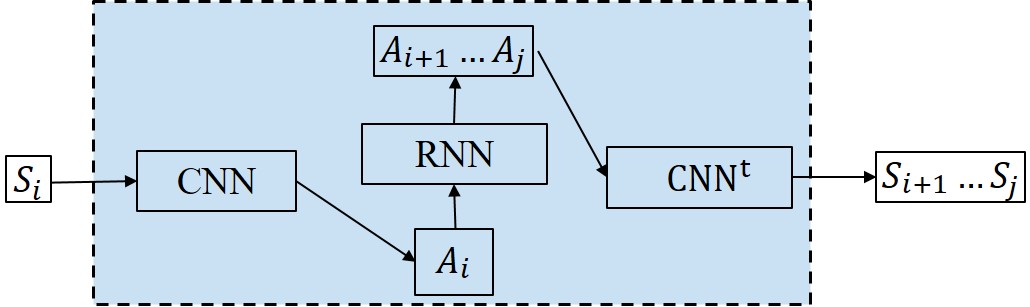}
    \caption{An illustration of prediction process of convolutional recurrent autoencoder model}
    \label{prediction}
\end{figure}{}

The online prediction is straightforward and a schematic of its process is depicted in figure \ref{prediction} for the ease of understanding. Firstly, 
 a low-dimensional representation 
$\textbf{A}^{0}\in \mathbb{R}^{N_{A}}$ is constructed using the encoder network for a given initial condition  $\textbf{S}^{0}\in [0,1]^{N_{x}\times N_{y}}$ and a set of trained parameters $\theta^{*}$.  Equation \ref{eq66} is applied iteratively for $N_{t}$ steps with  $\textbf{A}^{0}\in \mathbb{R}^{N_{A}}$ as the initial solution to get predictions of low dimensional representations. Finally, the full-dimensional state is reconstructed  $\hat{\textbf{S}}^{n}$ from the low dimensional representation $\hat{\textbf{A}}^{n}$ at every time step
or at any specific instance. The online prediction procedure is outlined in
Algorithm \ref{algo:Ch6pred}.

\begin{algorithm}[t]
    \SetKwInOut{Input}{Input}
    \SetKwInOut{Result}{Result}
    \Input{Training dataset $\mathcal{X}\in[0,1]^{N_{x}\times N_{y}\times N_{t}\times N_{s}}$, number of train-steps $N_{train}$, batch size $N_{b}$}
    \Result{Trained model parameters $\theta$}
    \BlankLine
    Randomly initialize $\theta$; \\
    \For{$i\in \{1,\dots,N_{train}\}$}
    {%
        Randomly sample batch from training data: $\mathcal{X}^{b}\subset \mathcal{X}$\;
        Flatten batch-mode: $\mathcal{X}^{b_{AE}}\leftarrow flatten(\mathcal{X}^{b})$ s.t. $\mathcal{X}^{b_{AE}}\in [0,1]^{N_{x}\times N_{y}\times (N_{t}.N_{b})}.$\;
        Encoder forward pass: $\Tilde{\mathcal{H}}^{b}\leftarrow f_{enc}(\mathcal{X}^{b_{AE}})$ where $\Tilde{\mathcal{H}}^{b}\in \mathbb{R}^{N_{h}\times (N_{t}.N_{b})}$\;
        Decoder forward pass: $\hat{\mathcal{X}}^{b_{AE}} \leftarrow f_{dec}(\Tilde{\mathcal{H}}^{b})$\;
        Reshape low-dimensional features: $\mathcal{H}^{b}\in \mathbb{R}^{N_{h}\times N_{t}\times N_{b}} \leftarrow$ reshape $(\Tilde{\mathcal{H}}^{b})$\;
        Initialize RNN subgraph loop: $\hat{\textbf{h}}_{i}^{2} \leftarrow f_{LSTM}(\textbf{h}_{i}^{1})$ for $i \in \{1,\dots,N_{b}\}$,$\textbf{h}_{i}^{1}\subset \mathcal{H}^{b}$;\;
        \For{$n \in \{2,\dots,N_{t-1}\}$}
        {%
            $\hat{\textbf{h}}_{i}^{n+1} \leftarrow f_{LSTM}(\hat{\textbf{h}}_{i}^{n})$ for $i \in \{1,\dots,N_{b}\}$,$\hat{\textbf{h}}_{i}^{n}\subset \hat{\mathcal{H}}^{b}$\;
        }
        Using $\mathcal{X}^{b},\hat{\mathcal{X}}^{b},\mathcal{H}^{b}$ and $\hat{\mathcal{H}}^{b}$ calculate approximate gradient $\hat{\textbf{g}}$ of equation\;
        Update parameters: $\boldsymbol{\theta}\leftarrow ADAM(\hat{\textbf{g}})$
    }
    \caption{Convolutional Recurrent Autoencoder training algorithm \cite{gonzalez2018deep}}
    \label{algo:Ch6train}
\end{algorithm}

\begin{algorithm}
    \SetKwInOut{Input}{Input}
    \SetKwInOut{Result}{Result}
    \Input{Initial condition $\textbf{x}^{0}\in [0,1]^{N_{x}\times N_{y}}$, number of prediction steps $N_{t}$.}
    \Result{Model prediction $\hat{\textbf{X}} = [\hat{\textbf{x}}^{1},\dots,\hat{\textbf{x}}^{N_{t}}]\in [0,1]^{N_{x}\times N_{y}\times N_{t}}$}
    \BlankLine
    Load trained parameters $\boldsymbol{\theta}^{*}$\;
    Endcoder forward pass: $\textbf{h}^{0}\leftarrow f_{enc}(\textbf{x}^{0})$\;
    Initialize RNN subgraph loop: $\hat{\textbf{h}}^{1}\leftarrow f_{LSTM}(\textbf{h}^{0})$\;
    \For{$n\in \{1,\dots,N_{t-1}\}$}
    {%
        $\hat{\textbf{h}}^{n+1} \leftarrow f_{LSTM}(\hat{\textbf{h}})^{n}$
    }
    Decoder forward pass: $\hat{\textbf{X}}\leftarrow f_{dec}(\hat{\textbf{H}})$, where $\hat{\textbf{H}} = [\hat{\textbf{h}}^{1},\dots,\hat{\textbf{h}}^{N_{t}}]$;
    \caption{Convolutional Recurrent Autoencoder prediction algorithm \cite{gonzalez2018deep}}
    \label{algo:Ch6pred}
\end{algorithm}

\section*{Flow past an isolated cylinder}

\begin{figure} 
\centering
\includegraphics[width=0.45\textwidth]{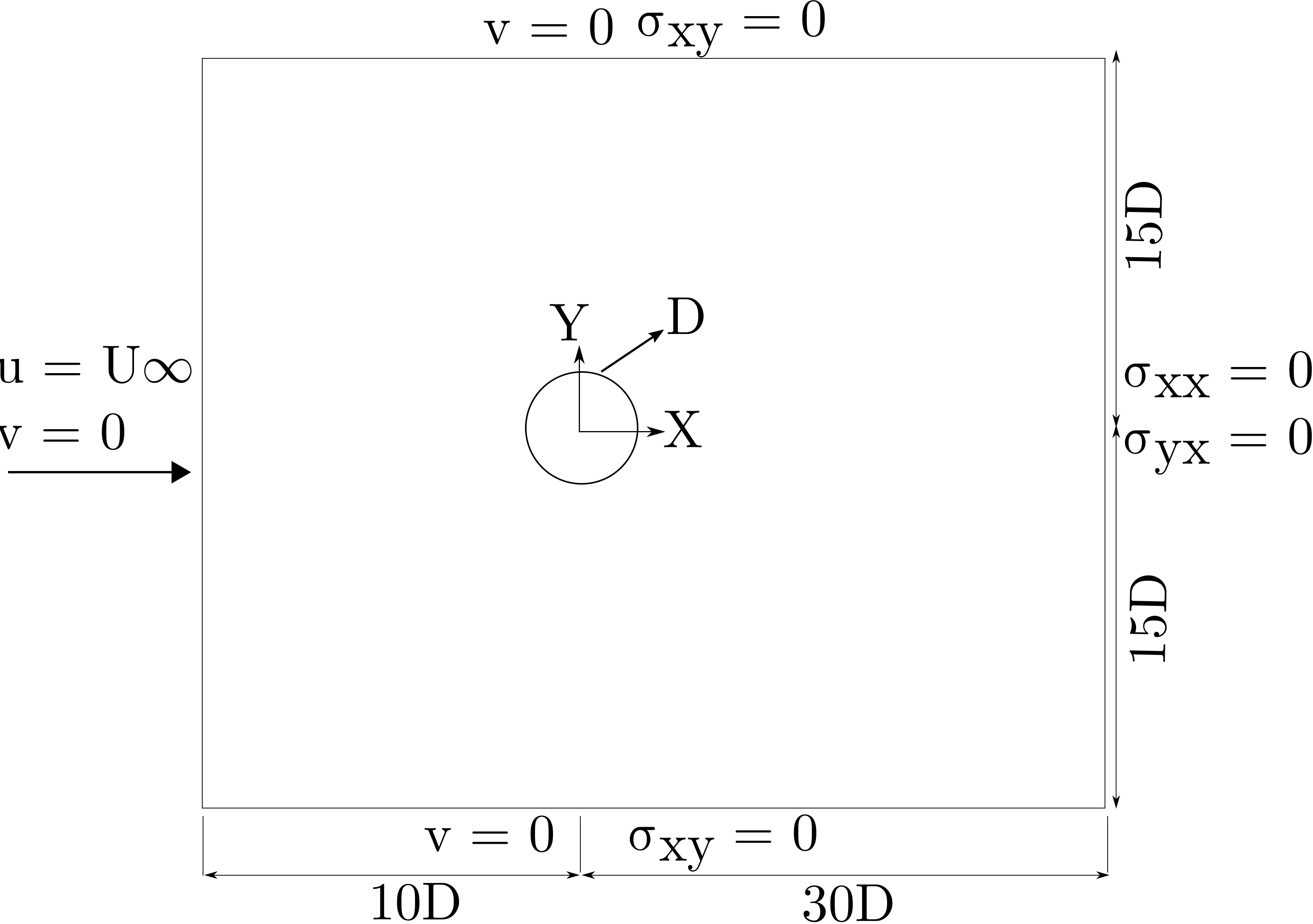}
\caption{Schematic diagram of the problem setup: flow past a plain cylinder}
\label{fig5}
\end{figure}

\begin{figure}
    \centering
    \subfloat[]
    {\includegraphics[width = 0.25 \textwidth]{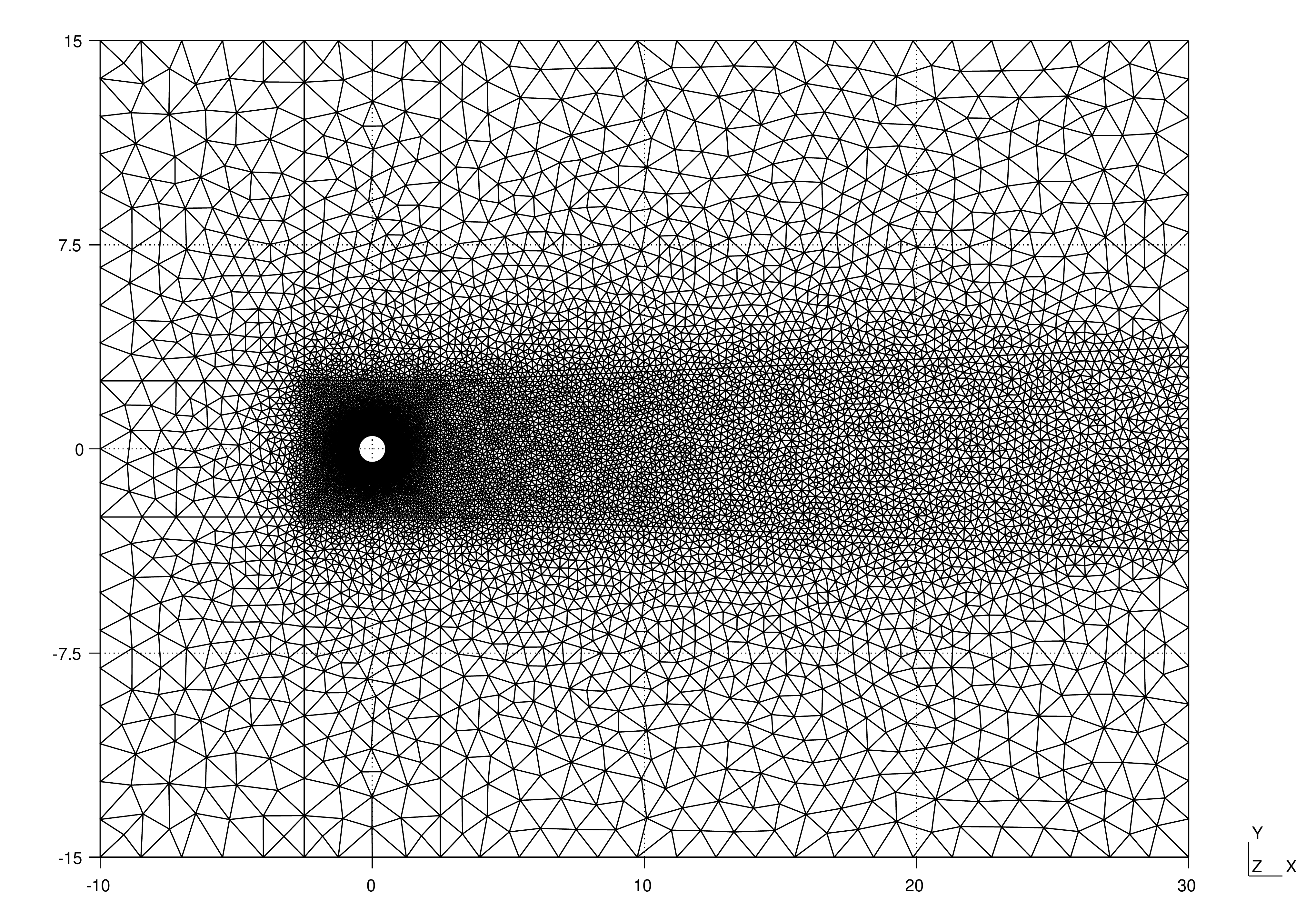}}
    \subfloat[]
    {\includegraphics[width = 0.25\textwidth]{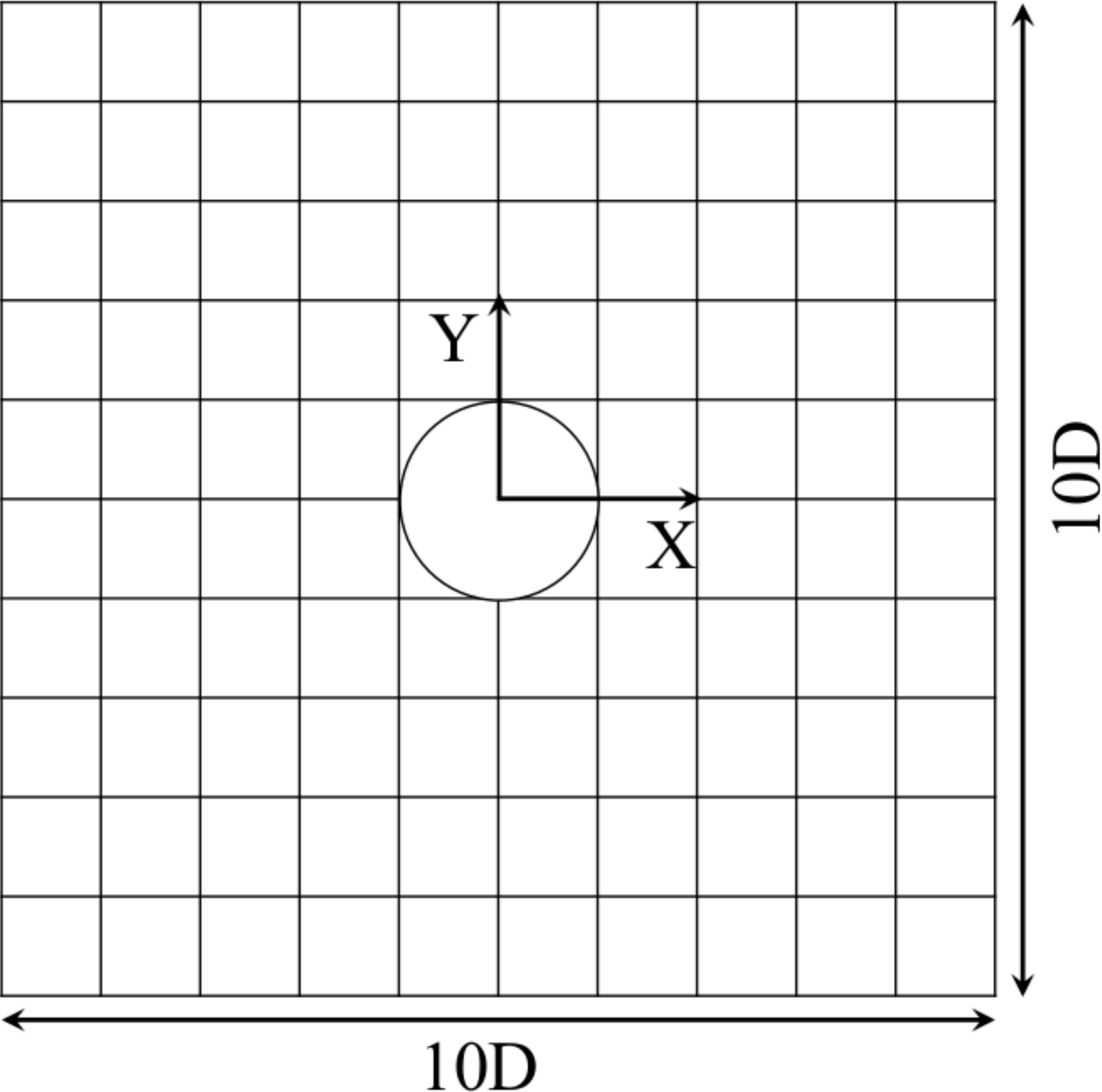}}
    \caption{(a) Unstructured mesh used in full order model simulation, (b) Square uniform grid used as input function in the convolutional recurrent autoencoder model.}
    \label{figcompre}
\end{figure}

\begin{figure}
\centering
\includegraphics[width = 0.45\textwidth]{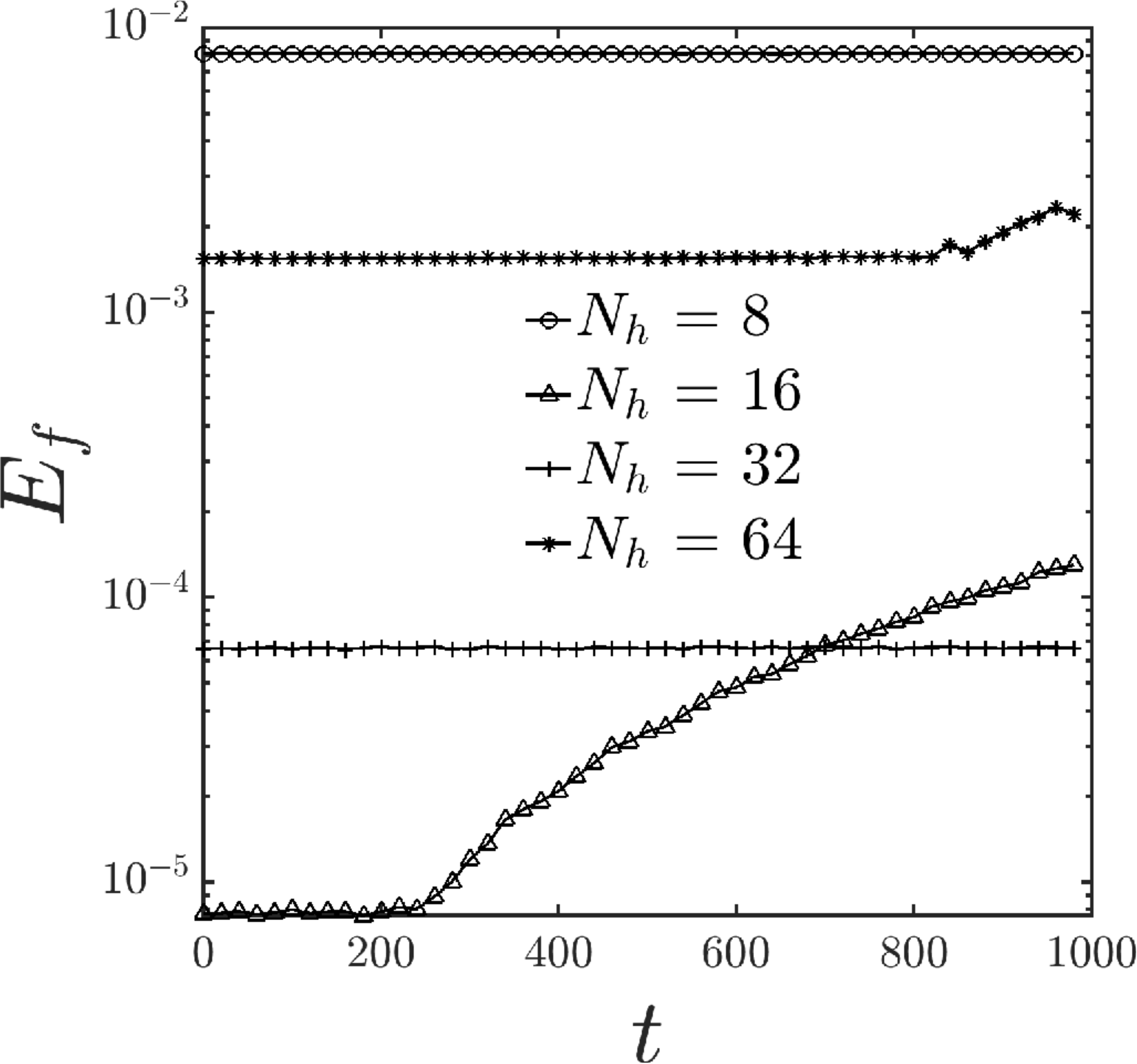}
\caption{Normalized squared error $E_{f}$ for the velocity field in X-direction ($U$): flow past a plain cylinder}
\label{fig6_1}
\end{figure}

\begin{figure}
\centering
\includegraphics[width = 0.15\textwidth]{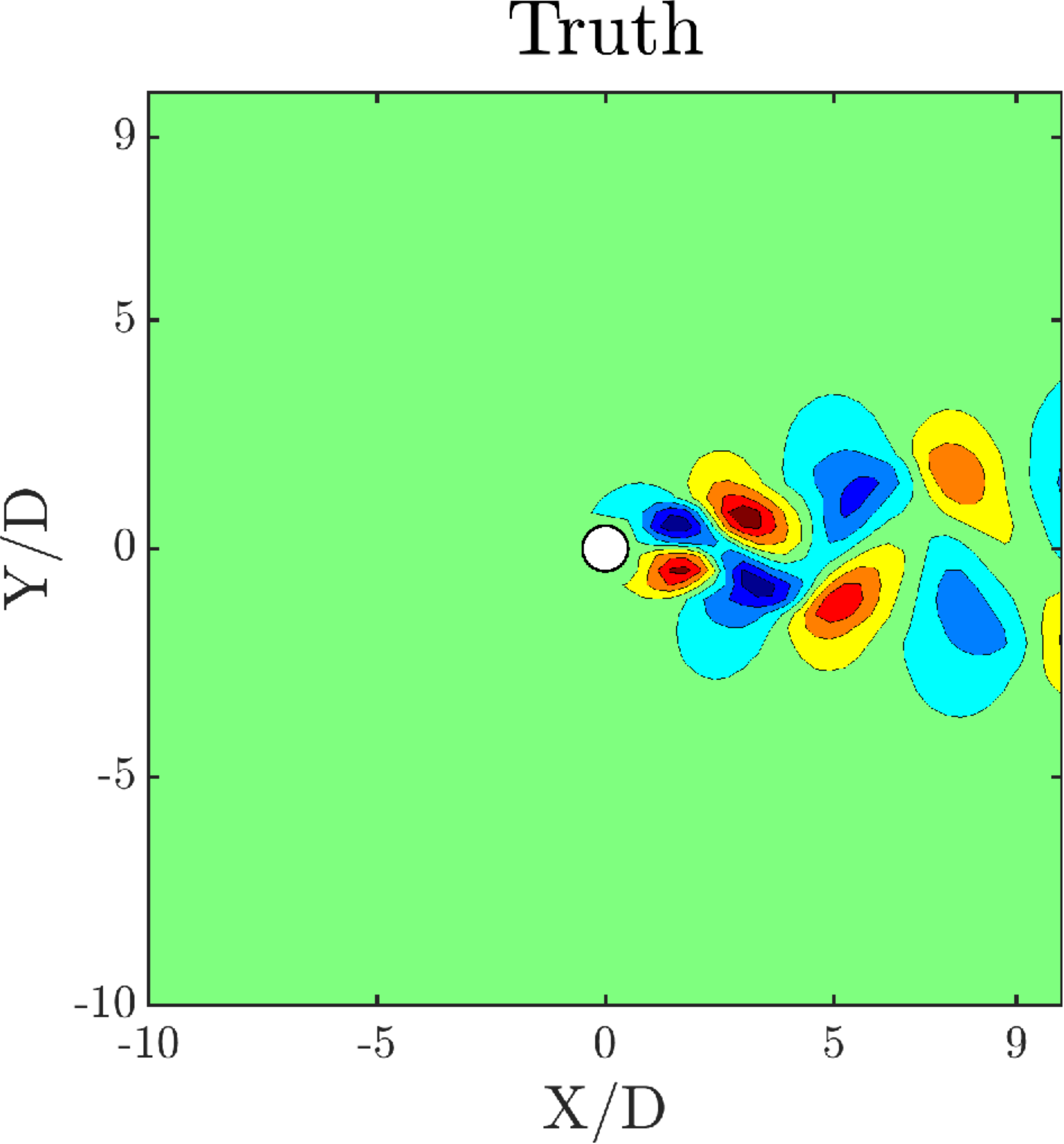}
\includegraphics[width = 0.15\textwidth]{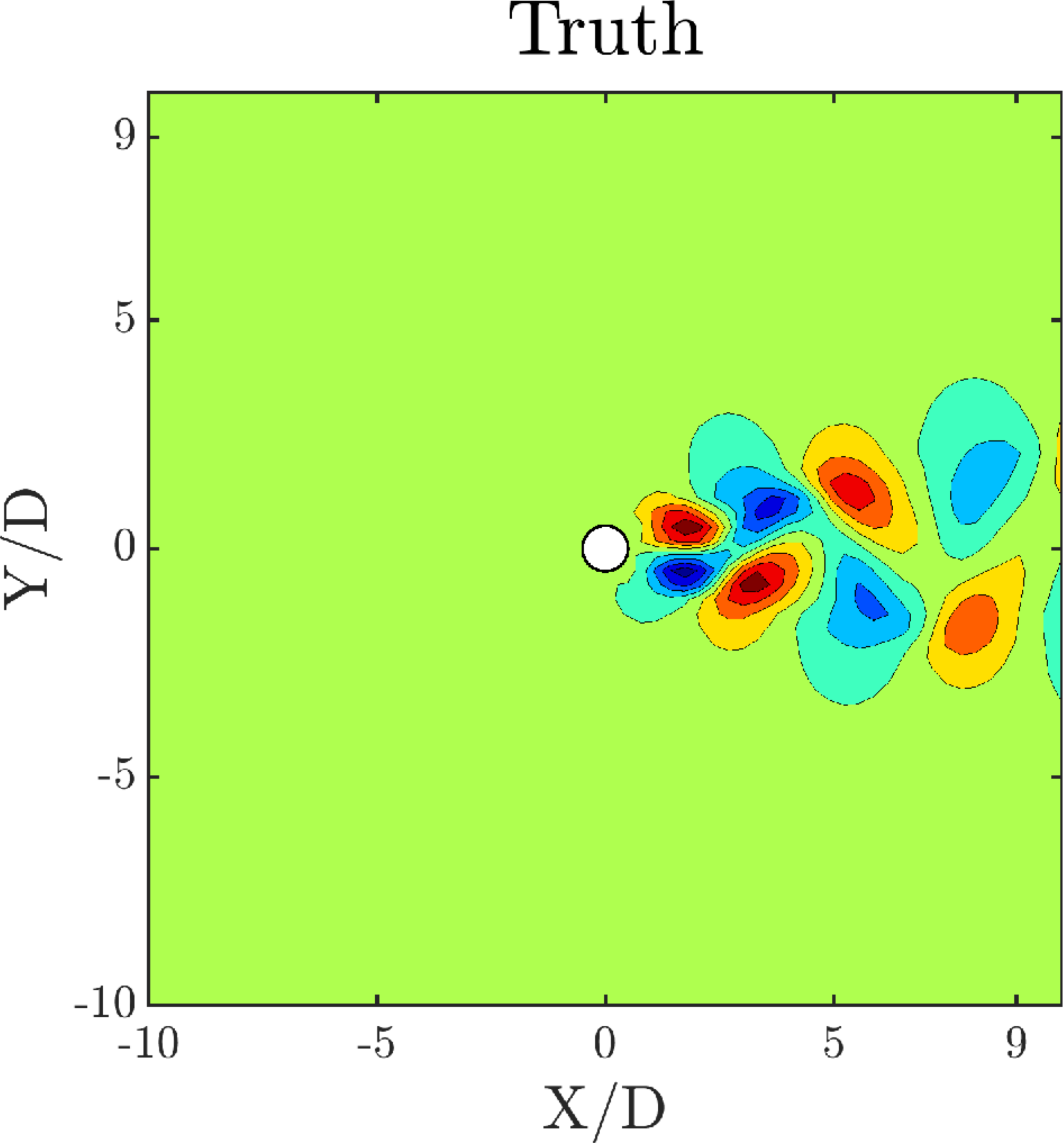}
\includegraphics[width = 0.15\textwidth]{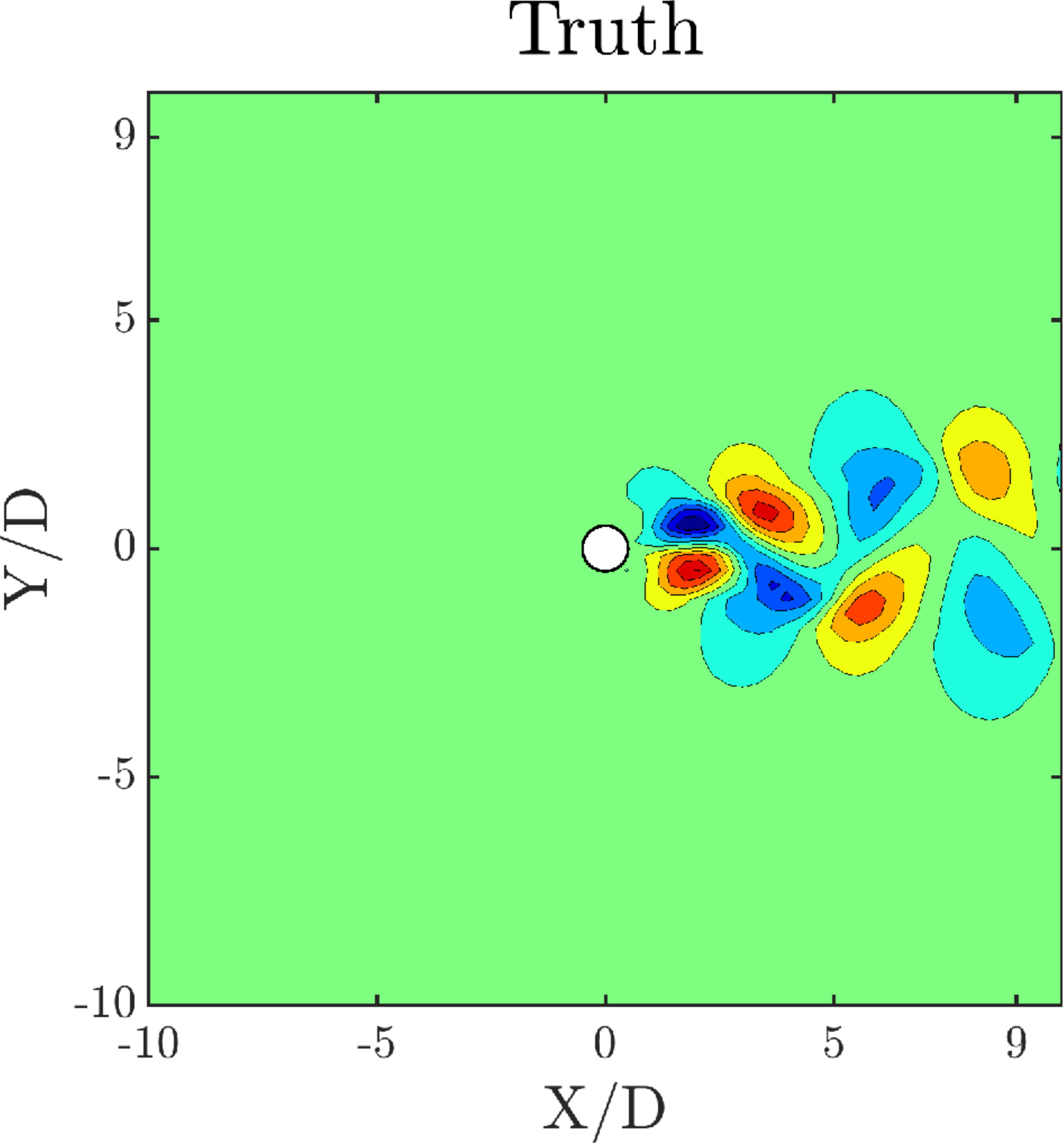}
\\
\vspace{0.05\textwidth}
\includegraphics[width = 0.15\textwidth]{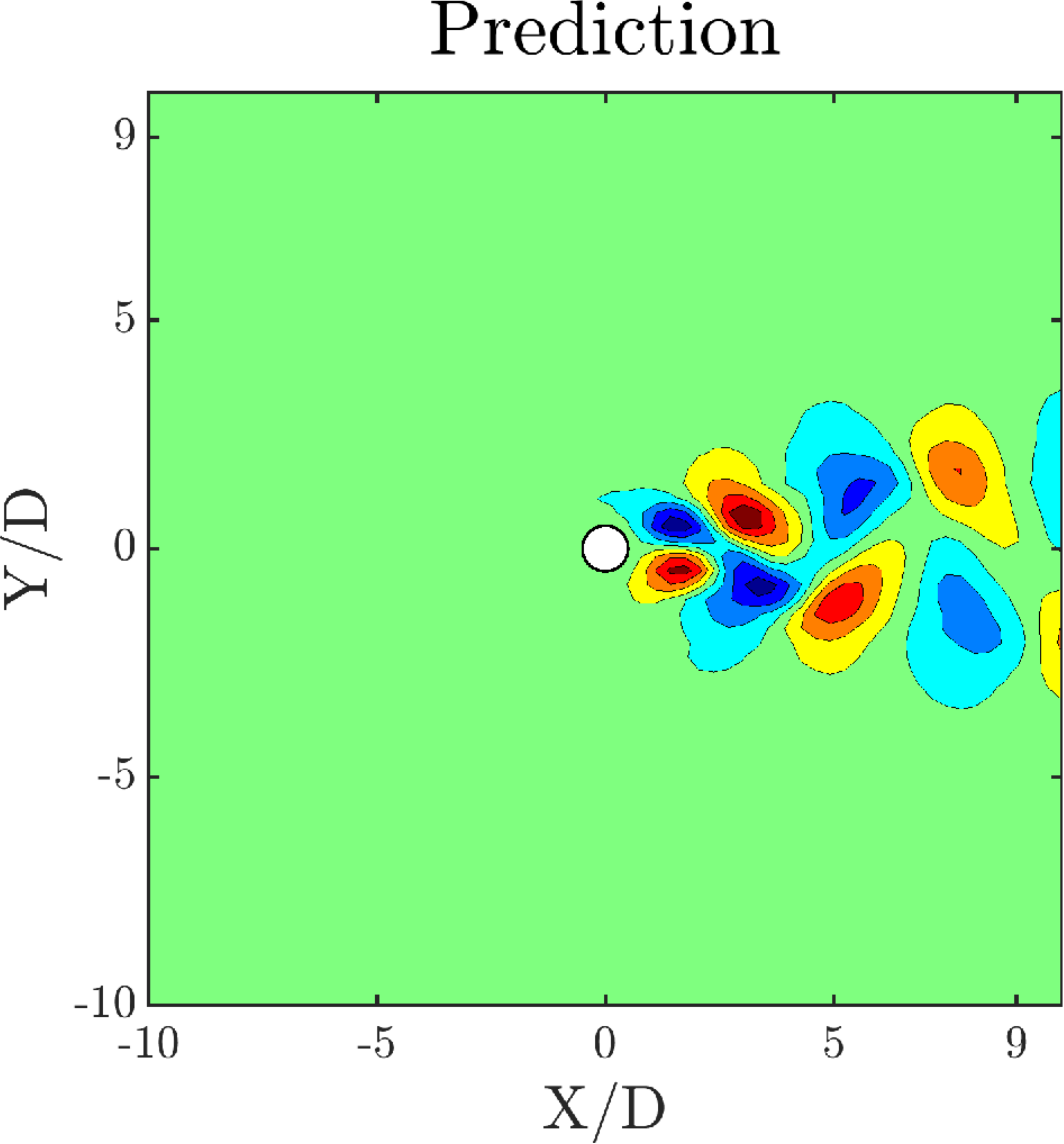}
\includegraphics[width = 0.15\textwidth]{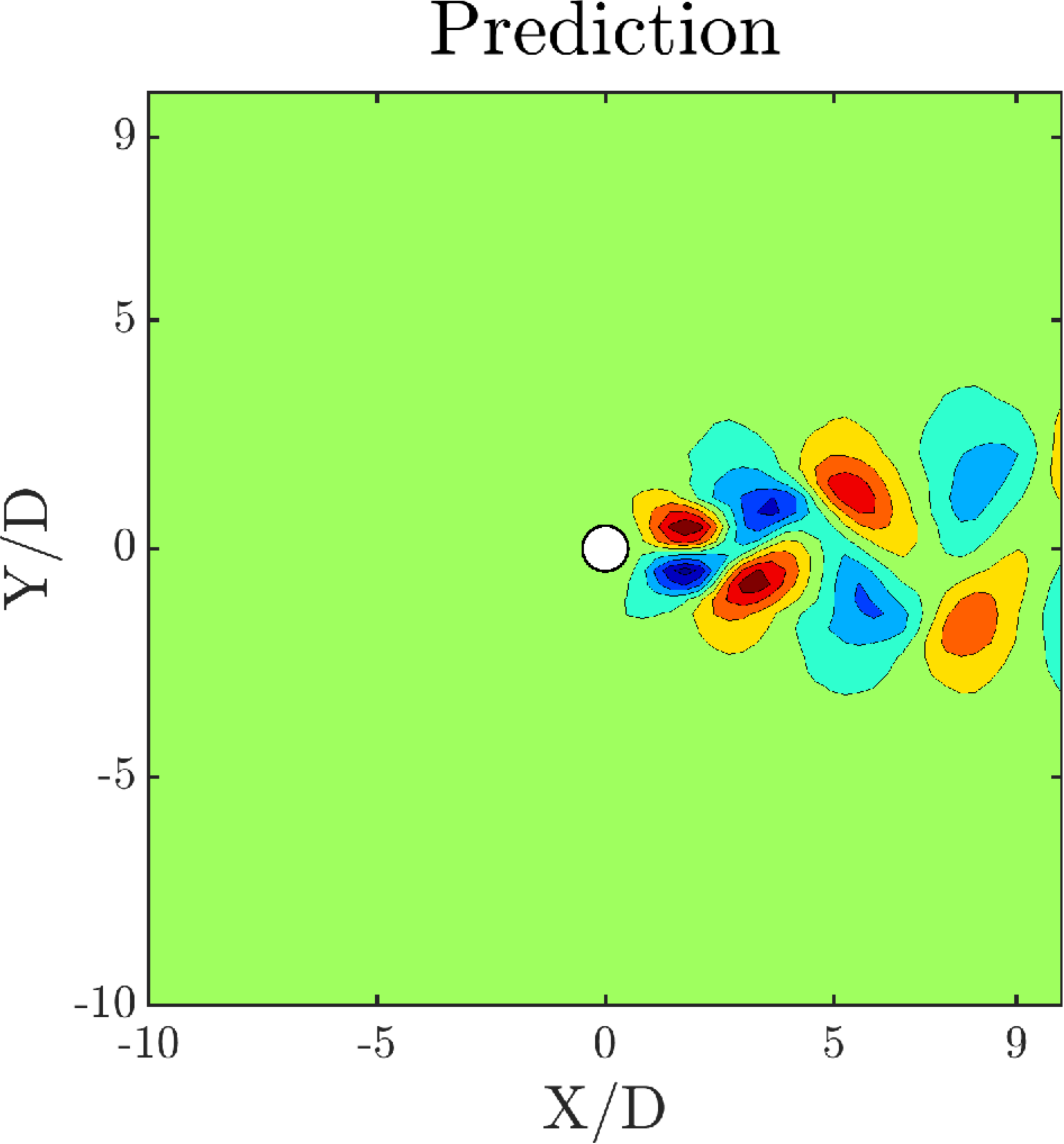}
\includegraphics[width = 0.15\textwidth]{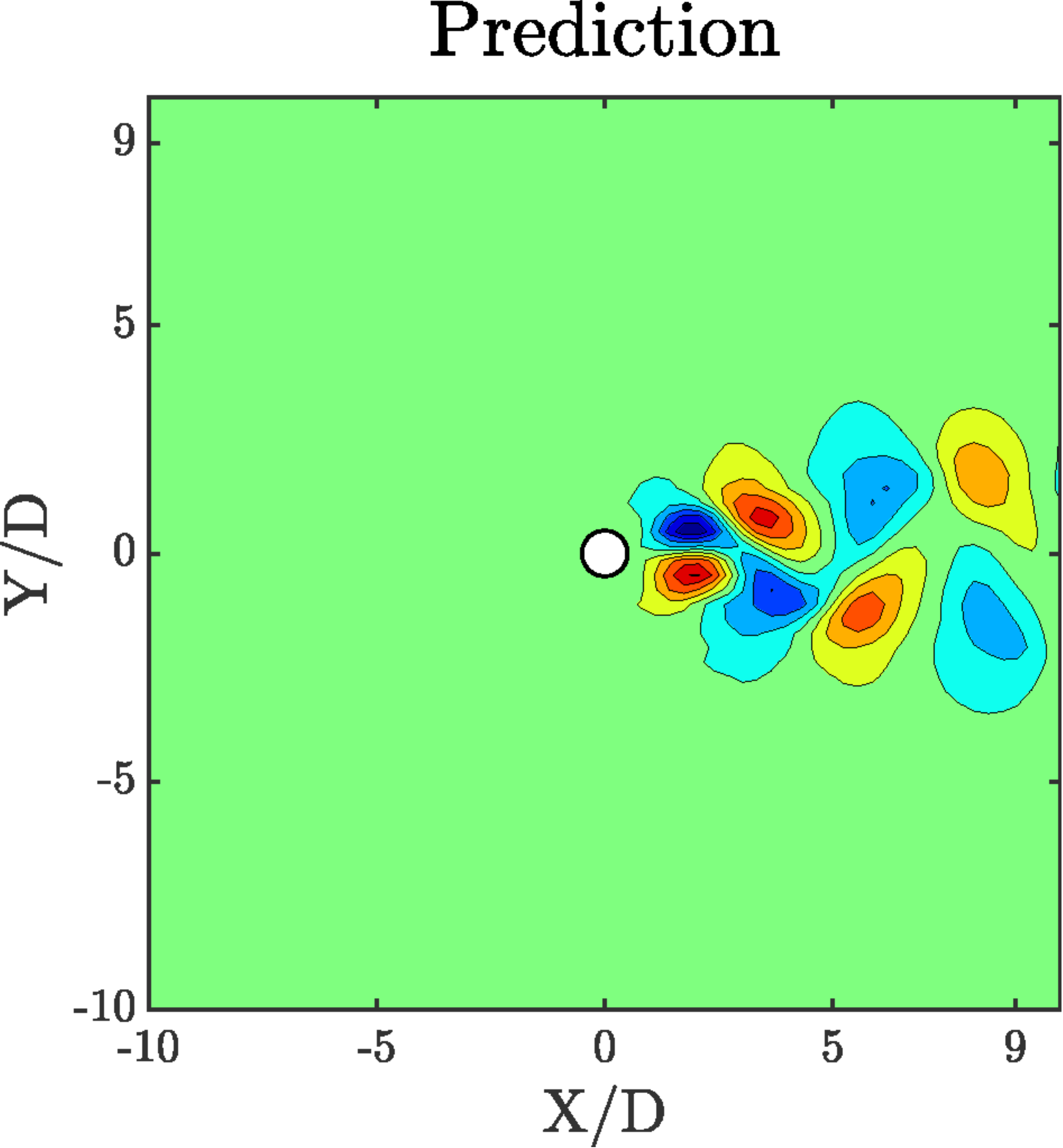}
\\
\vspace{0.05\textwidth}
\subfloat[]
{\includegraphics[width = 0.15\textwidth]{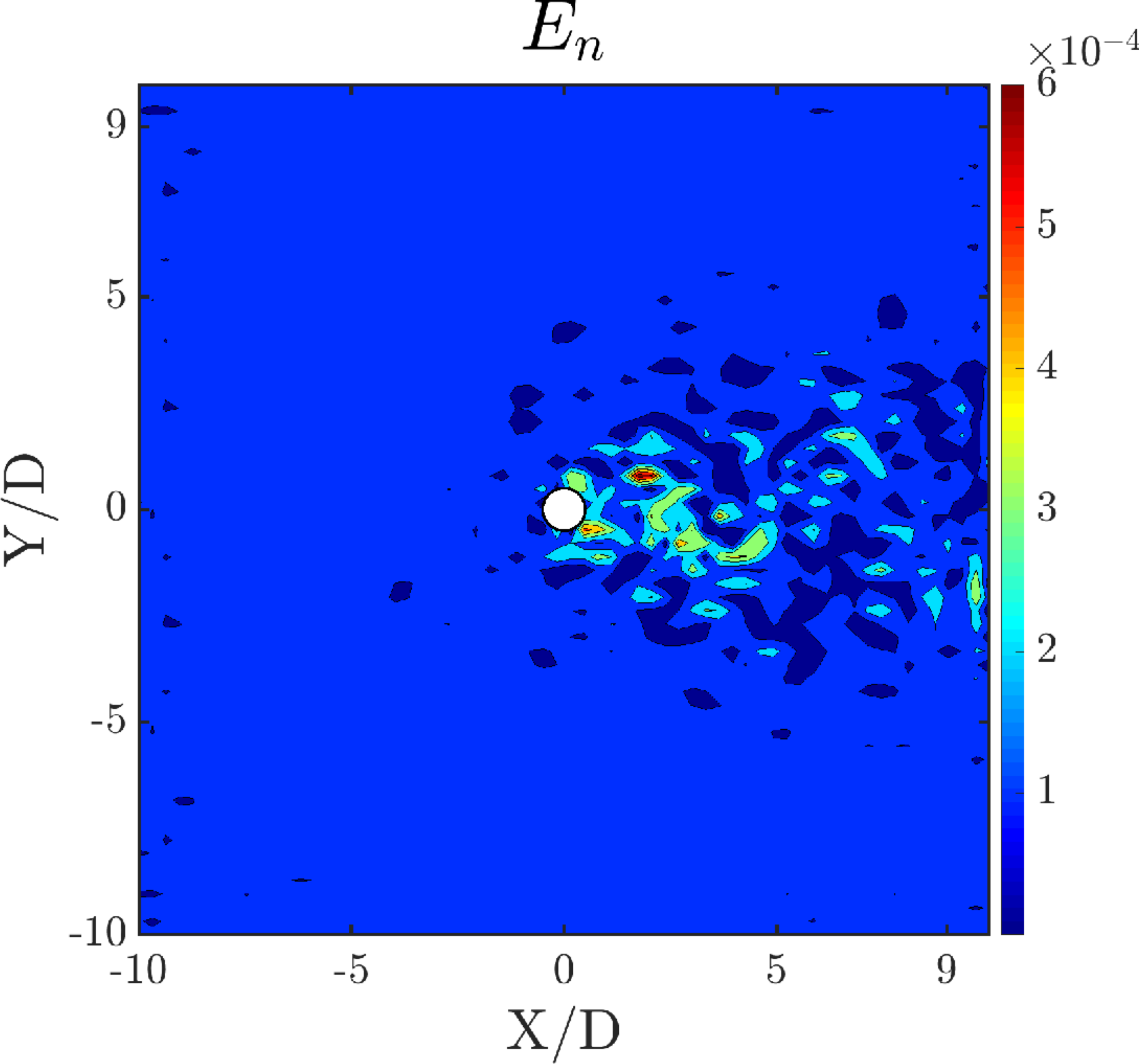}}
\subfloat[]
{\includegraphics[width = 0.15\textwidth]{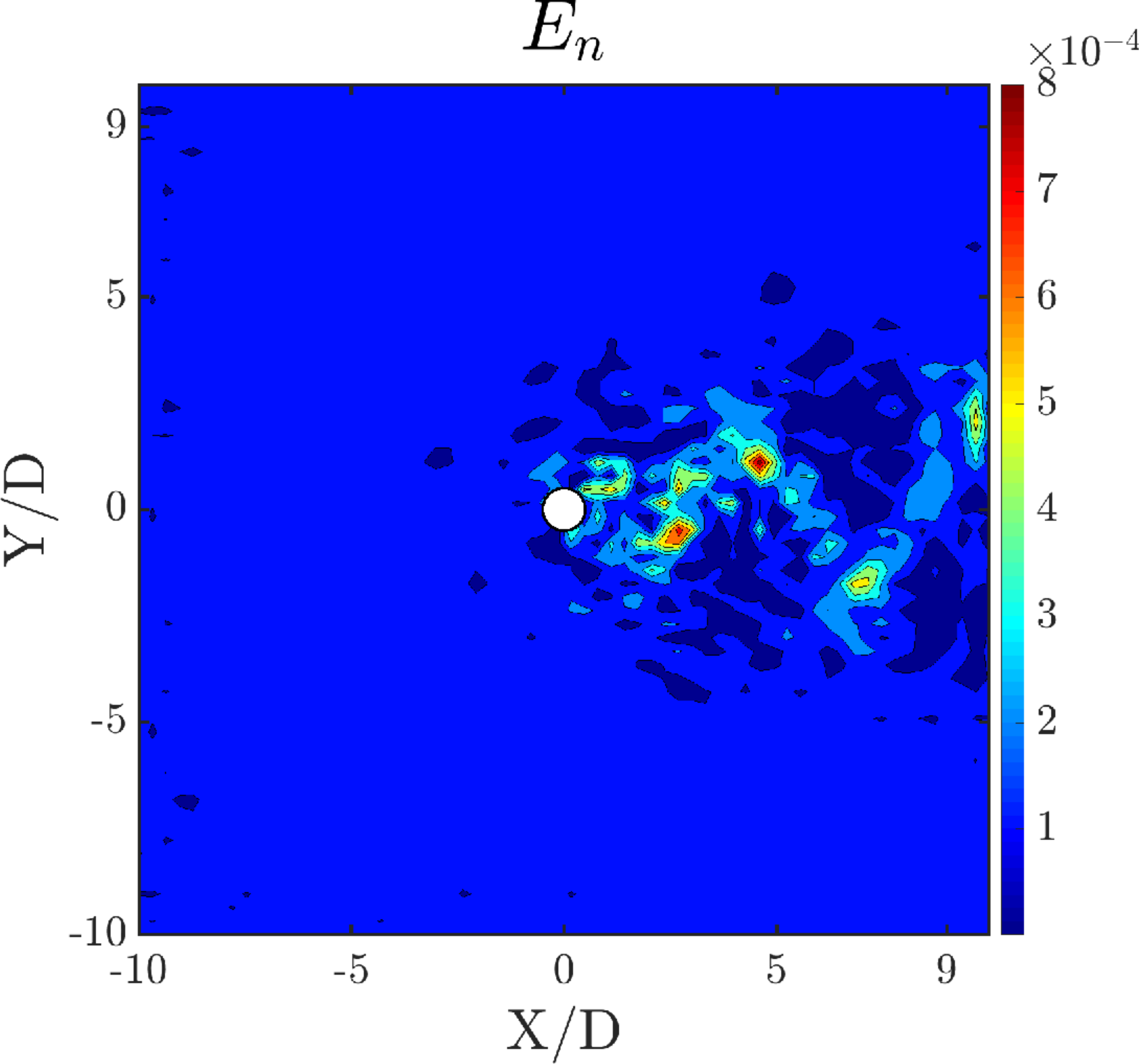}}
\subfloat[]
{\includegraphics[width = 0.15\textwidth]{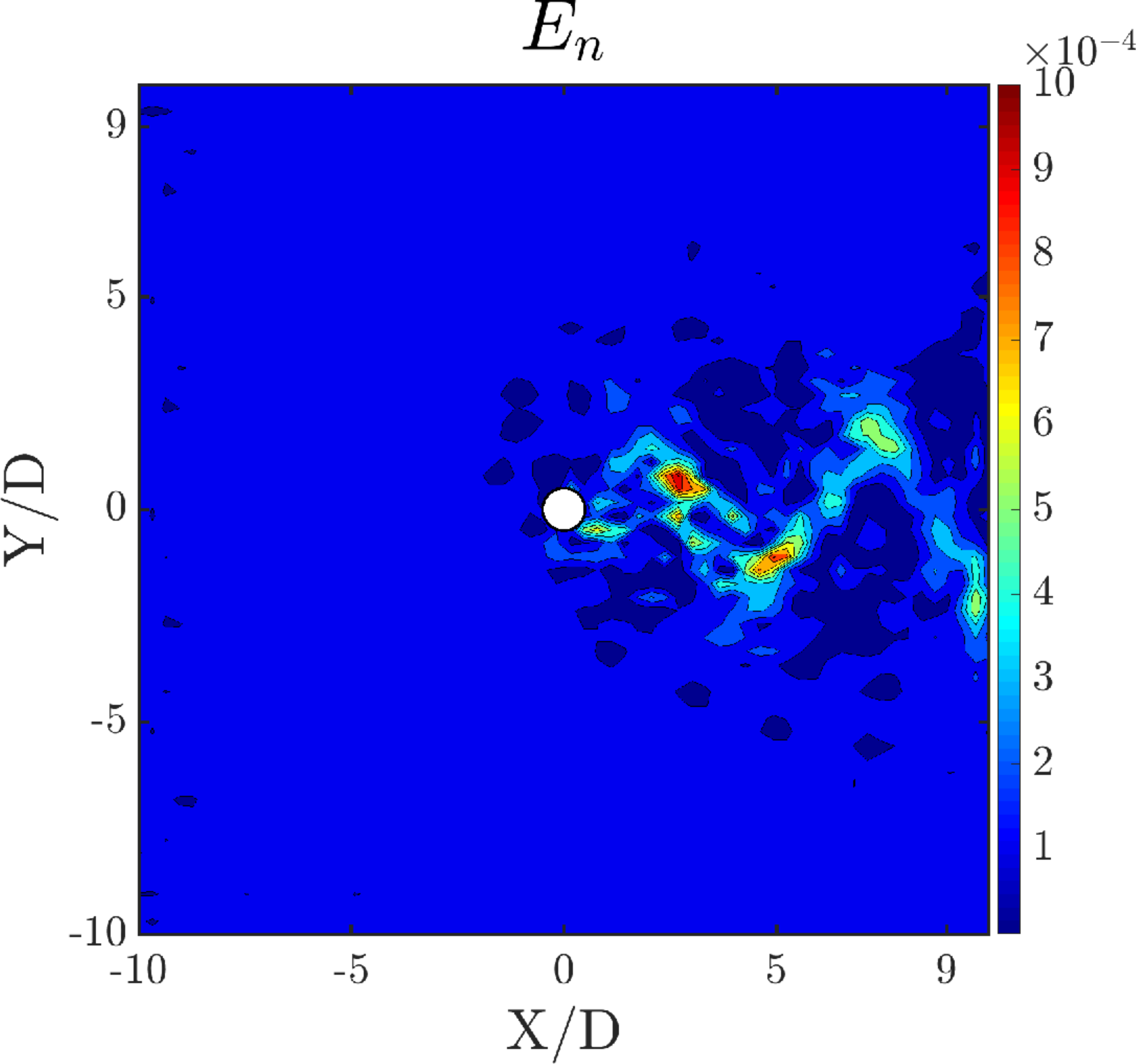}}

\caption{Comparison of truth and predicted fields along with normalized reconstruction error $E_{n}$ at (a) $t = 1050s$, (b) $t = 1125s$, (c) $t= 1200s$ for velocity field in X-direction ($U$): flow past plain cylinder }
\label{fig6_3}
\end{figure}

\begin{figure}
\centering
\includegraphics[width = 0.45\textwidth]{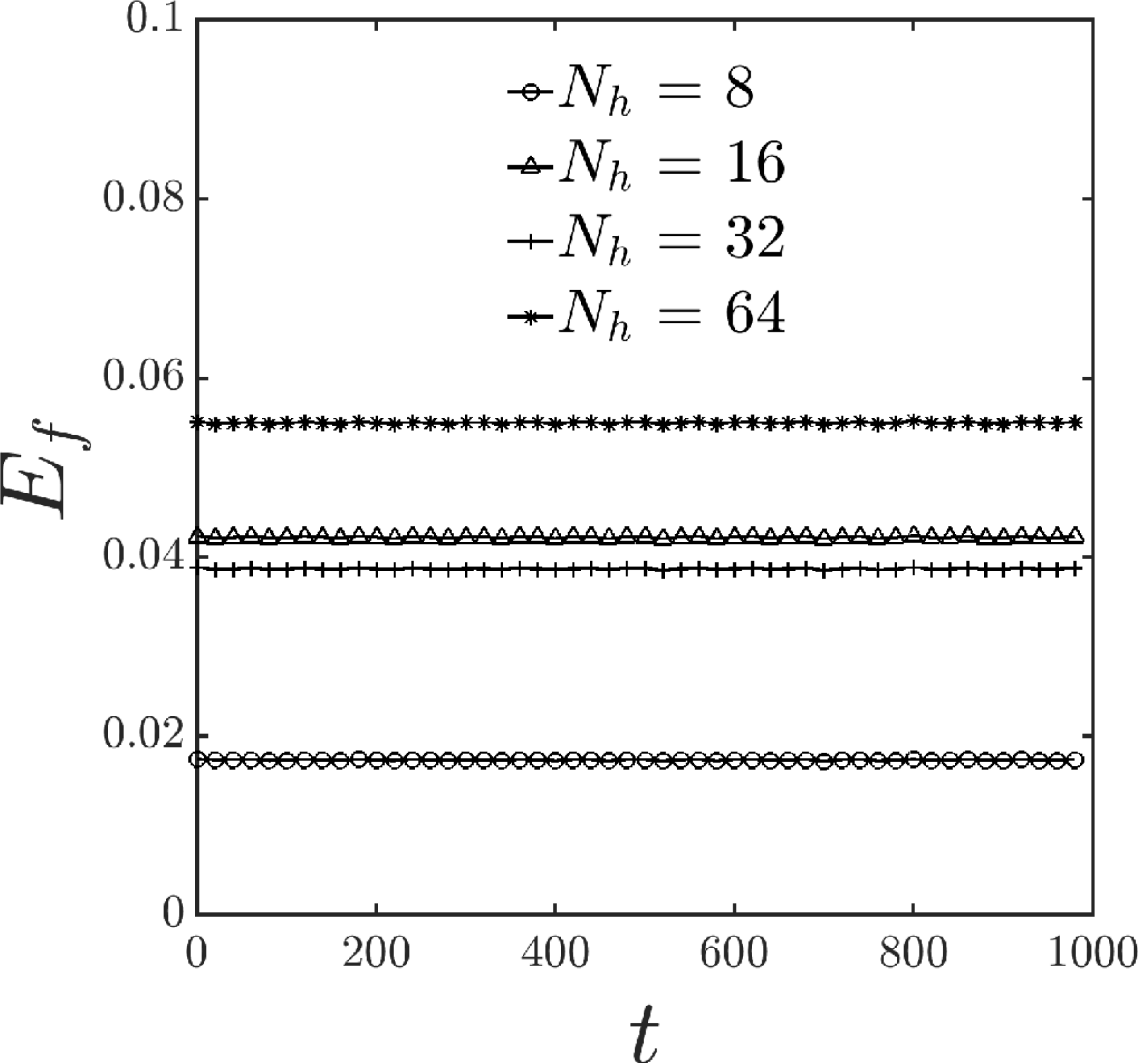}
\caption{Normalized squared error $E_{f}$ for the Pressure field ($P$): flow past a plain cylinder}
\label{fig6_2}
\end{figure}

The end to end nonlinear model reduction tool based on convolutional recurrent autoencoder netoworks is applied first on the canonical problem of the flow past a plain cylinder. The schematic of the problem set up is the same as the one depicted in Fig.~\ref{fig5}. The simulations were carried out via finite element Navier-Stokes solver described above to generate the training and test data.

The Reynolds number of the problem is taken as $Re = 100$. The full order simulation was carried out for $1250s$ with a time step of $0.05s$. A total of $5000$ snapshots of the simulation were collected for every $0.25s$. First we collect the velocity field in X-direction ($U$) for the analysis. The first and foremost step before proceeding with the training of the model is to map the unstructured data coming from full order simulations onto a square uniform grid of size $64\times 64$. It is essential that the input to the convolutional recurrent autoencoder to be a square matrix of uniform grid. We use the size $64\times 64$ throughout all our numerical examples for the sake of simplicity. We map the data from the unstructured grid to the uniform grid via $MATLAB$ function $``griddata".$ Figure \ref{figcompre} depicts both the unstructured grid used in the full order simulations and the uniform grid which is used an input to the convolutional recurrent autoencoder model. The presence of the rigid body is ensured by masking the region covered by the cylinder with a mandatory function which zeroes the predictions made by the model in the points inside the cylinder. The procedure described in the previous sections is followed to generate the feature scaled data for training.

\begin{equation}
    \mathcal{U} = \{\textbf{U}^{'1}_{s},\dots,\textbf{U}^{'N_{s}}_{s}\}\in [0,1]^{N_{x}\times N_{y}\times N_{t}\times N_{s}}
    \label{eqU}
\end{equation}
where each training sample $\textbf{U}^{'i}_{s} = [\textbf{u}^{'1}_{s,i},\dots,\textbf{u}_{s,i}^{'N_{t}}]$ with $N_{x},\;N_{y} = 64$ and $N_{t} = 25$, $N_{s} = 150$. Of the total $5000$ samples, $3000$ are used for training whereas $1000$ are used for validation and the final $1000$ are used for testing.

Four convolutional recurrent autoencoder models were trained using the above dataset, with low-dimensional representations of sizes $N_{h} = 8,\;16,\;32,\;64$. All the four models are trained on single Intel E5-2690v3 (2.60GHz, 12 cores) CPU node for $N_{train} = 500000$ iterations. The training took approximately $48$ hours. The online prediction is carried for 1000 steps with the trained model. Figure \ref{fig6_1} shows the normalized squared error for the velocity field in X-direction $(U)$.
\begin{equation}
   E_f =  \frac{\|\textbf{u}^{'n}_{s}-\hat{\textbf{u}}^{'n}_{s}\|^{2}_{2}}{\|\textbf{u}_{s}^{'n}\|^{2}_{2}}
\end{equation}

It is interesting to note that there is an order of magnitude difference in normalized squared error $E_{f}$ for different sizes of low-dimensional representations. The highest is with $N_{h} = 8$, followed by $N_{h} = 64$, $N_{h} = 16$ and $N_{h} = 32$. The  error is diverging rapidly for the case of $N_{h} = 16$, whereas it is diverging slowly for the case of $N_{h} = 64$. We have chosen $N_{h} = 32$ as the final trained model to depict the flow field predictions in detail.

Figure \ref{fig6_3} depicts the comparison of truth and predicted values for the velocity field in $X$-direction ($U$). The normalized reconstruction error $E_{n}$ is constructed by taking the absolute value of differences between and truth and predictions and normalizing it with $L_{2}$ norm of the truth data and is given by
\begin{equation}
    E_{n} = \frac{|\textbf{u}_{s}^{'n}-\hat{\textbf{u}}_{s}^{'n}|}{\|\textbf{u}_{s}^{'n}\|_{2}}
\end{equation}
The predictions from the convolutional recurrent autoencoder model are quite satisfactory. The alternating vortices are clearly captured in the predictions as shown in figure \ref{fig6_3}. The predictions are also quite smooth without any oscillations. The reconstruction error $E_{n}$ is quite low in the order of $10^{-4}$ for the predictions.

The same study is repeated with the pressure field. All the other details i.e., the size of training, validation and testing samples are the same as the case with the velocity field. Four convolutional recurrent autoencoder models are evaluated for the pressure field predictions by changing the sizes of the low-dimensional representations. Figure \ref{fig6_2} depicts the normalized squared error $E_{f}$ for the pressure field predictions from convolutional recurrent autoencoder model. The normalized squared error $E_{f}$ for this case is calculated by replacing the variable $\textbf{u}$ with $\textbf{p}$. The errors from all the four models are the in the same order of magnitude. There is no evidence for divergence in the error but the magnitude of errors are higher when compared to velocity field predictions. Without any loss of generality, we have chosen $N_{h} = 32$ for the depiction of full flow field predictions.

\begin{figure}
\centering
\includegraphics[width = 0.15\textwidth]{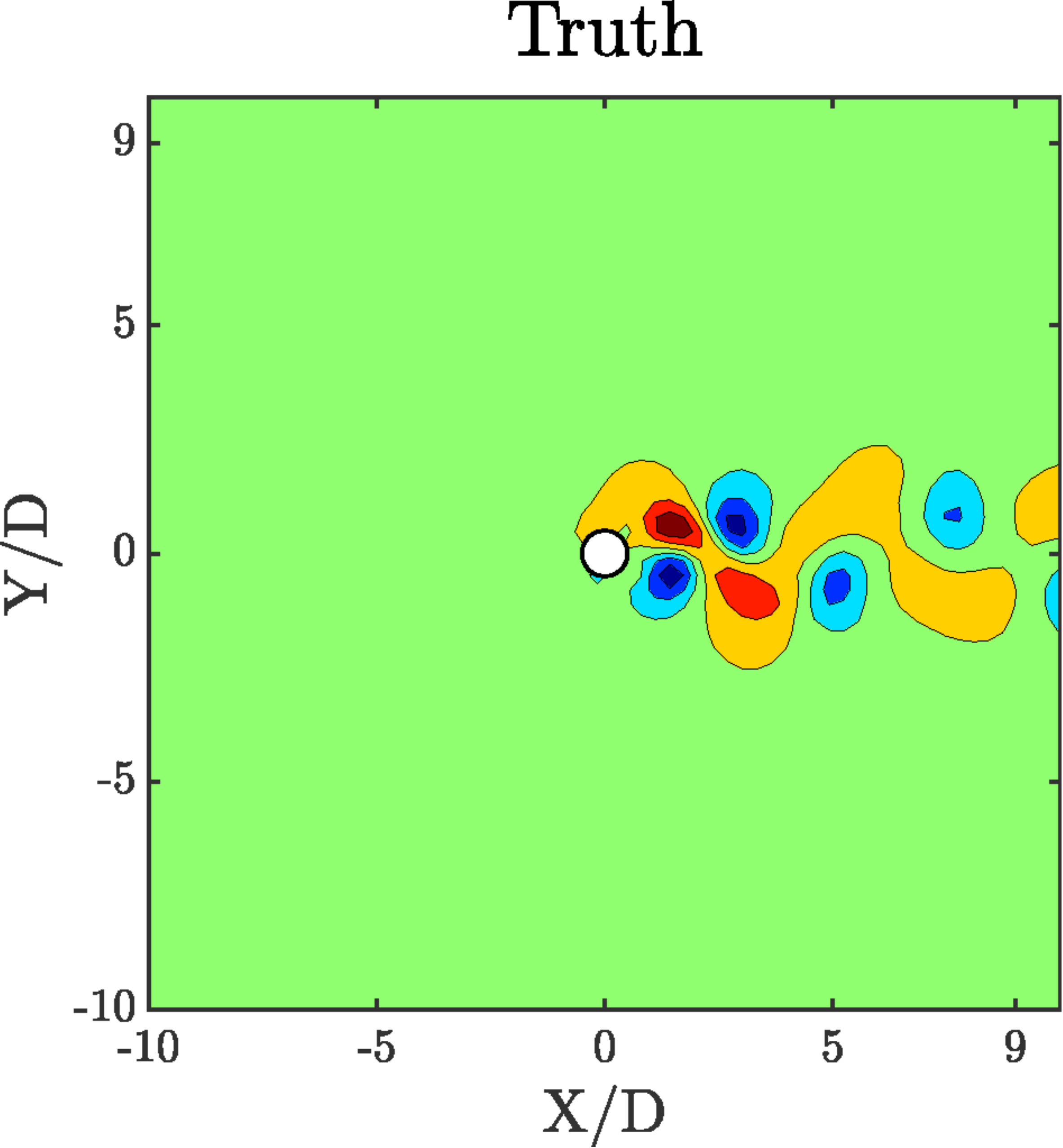}
\includegraphics[width = 0.15\textwidth]{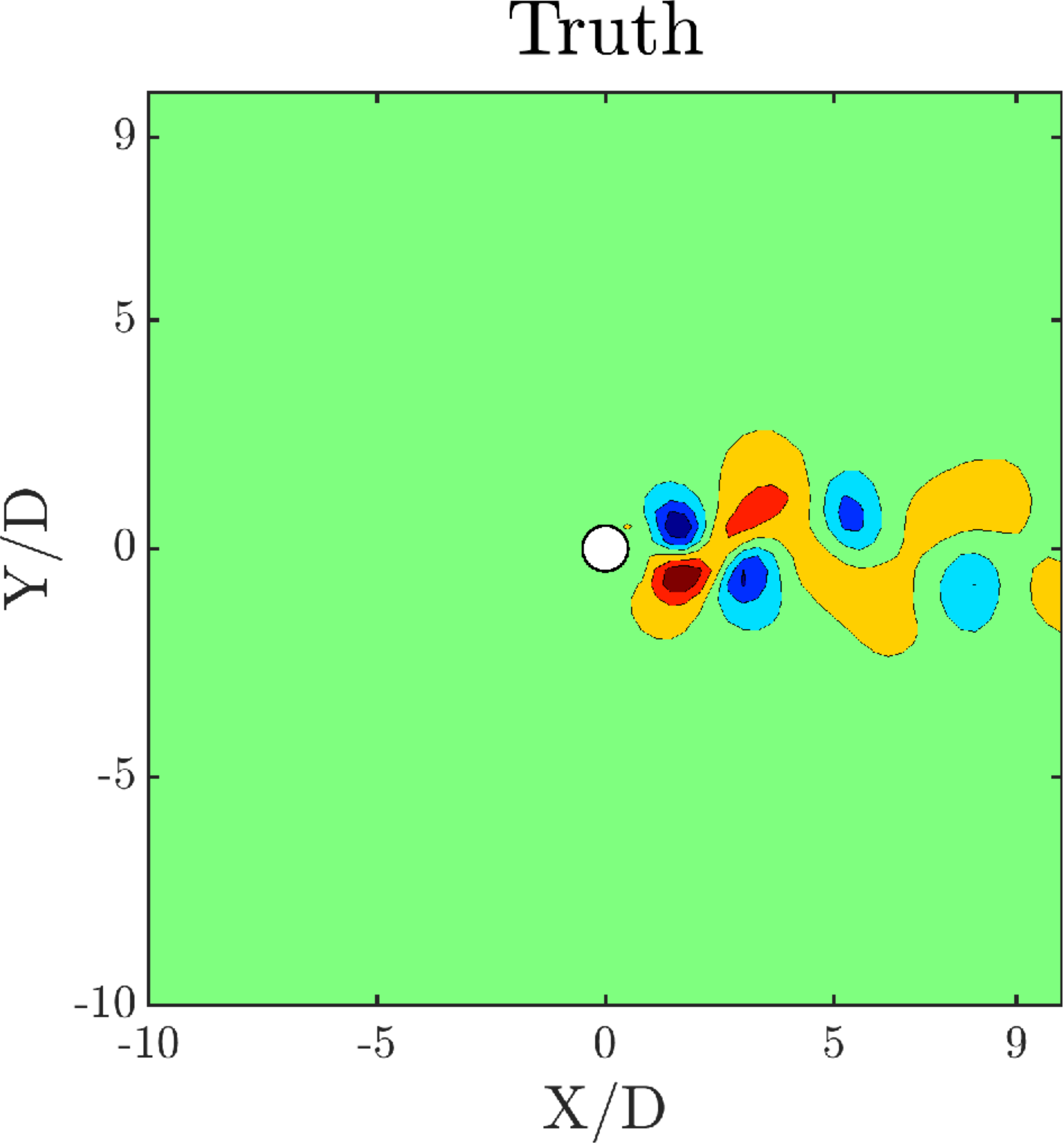}
\includegraphics[width = 0.15\textwidth]{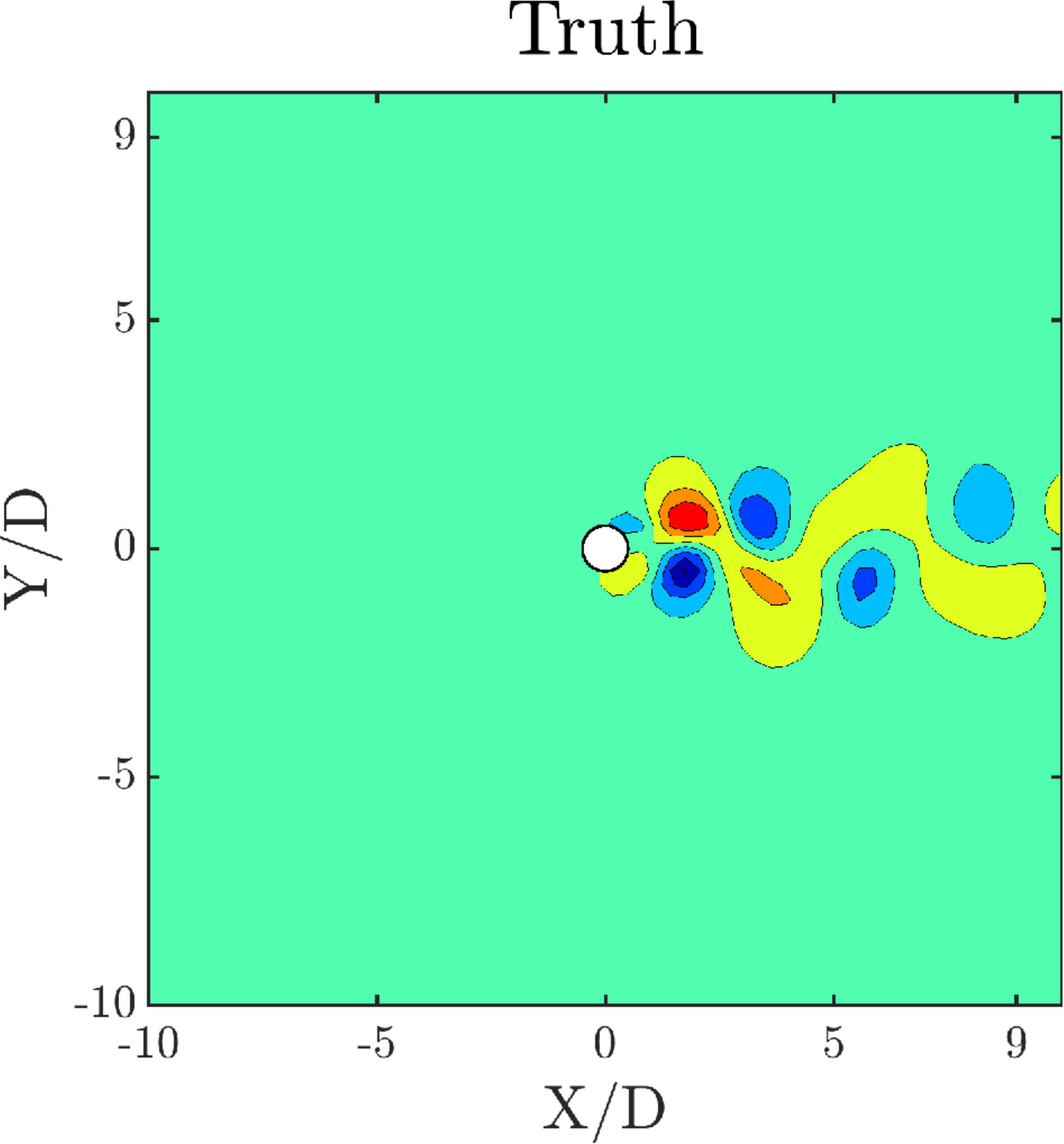}
\\
\vspace{0.05\textwidth}
\includegraphics[width = 0.15\textwidth]{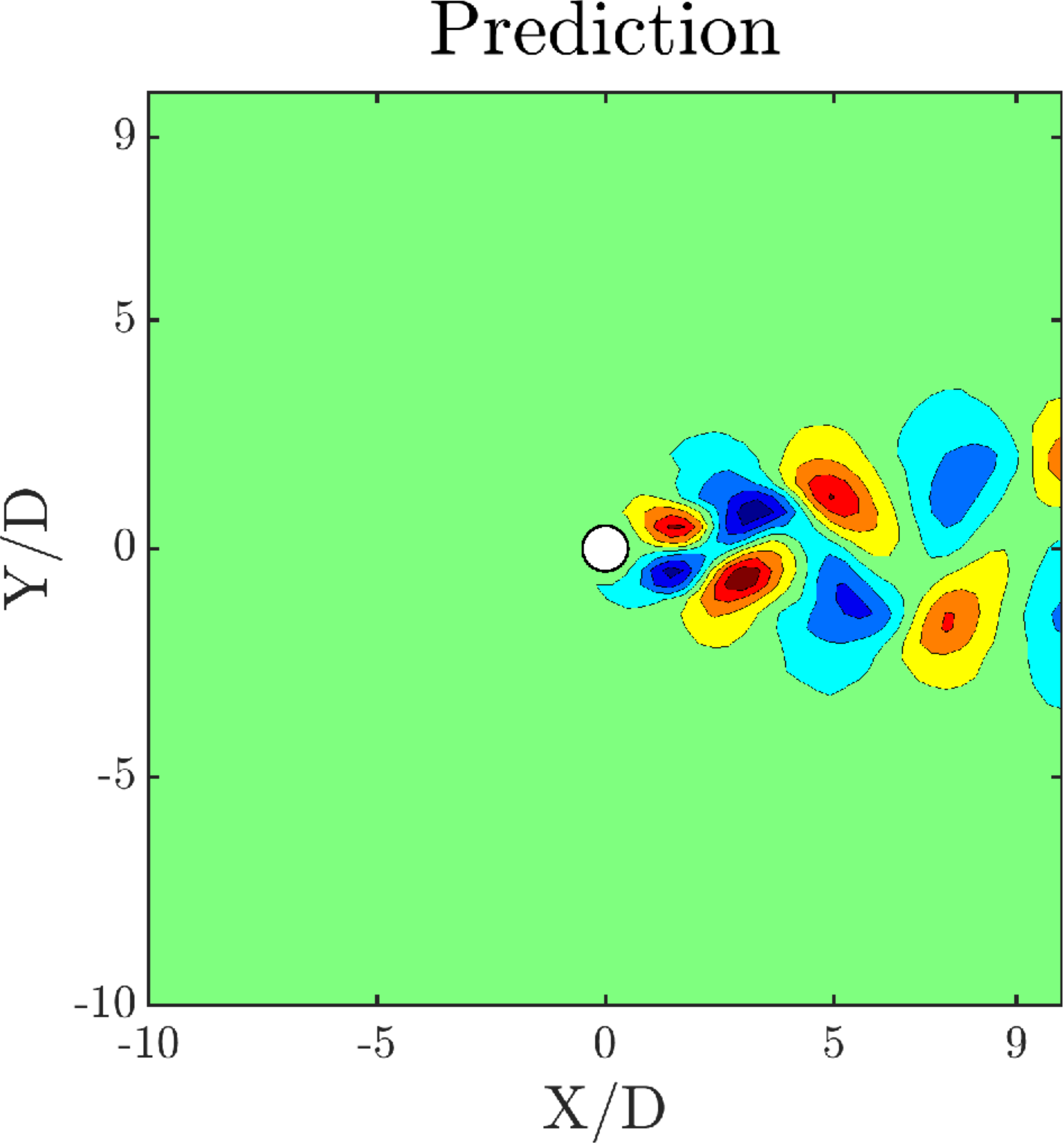}
\includegraphics[width = 0.15\textwidth]{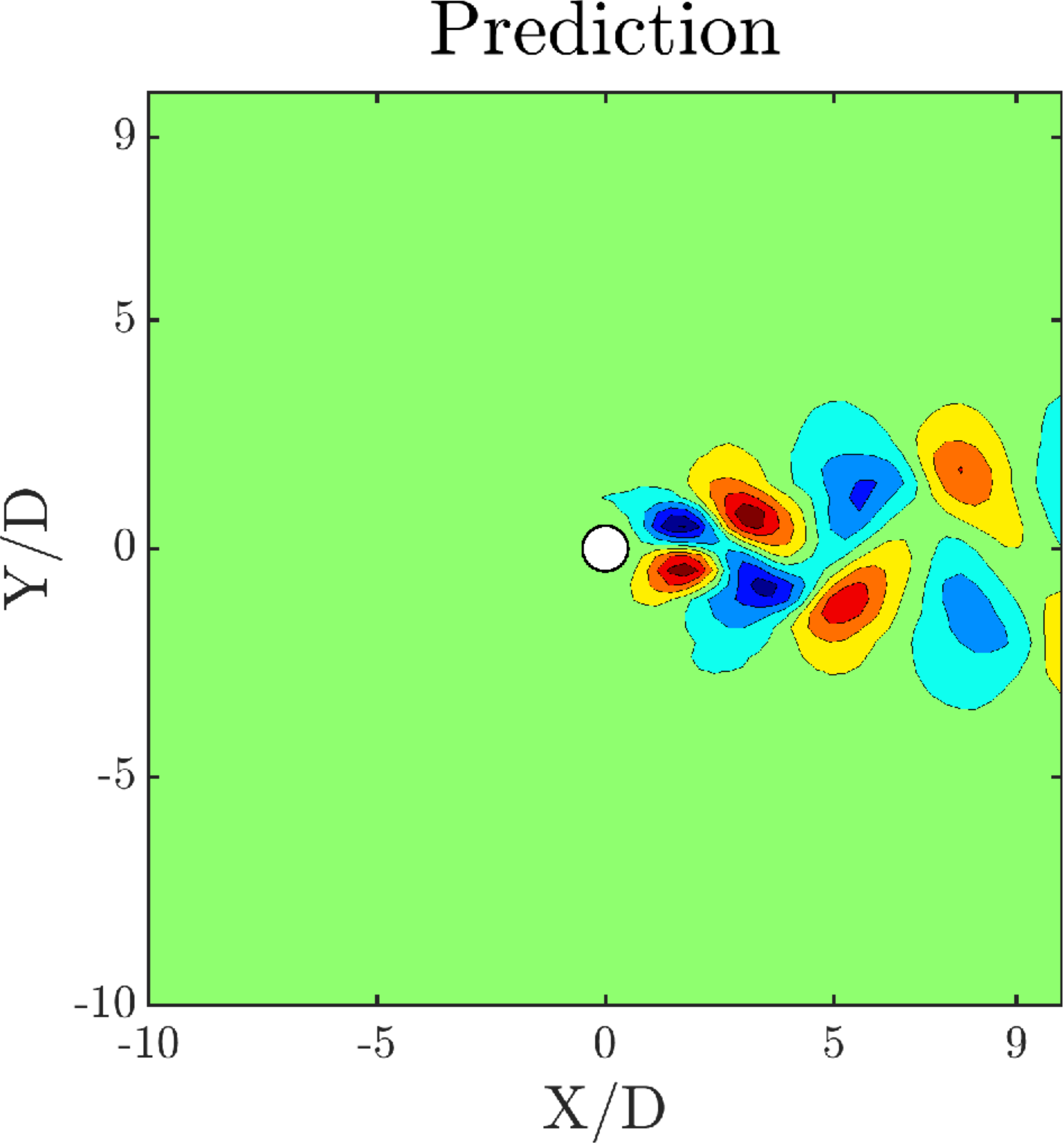}
\includegraphics[width = 0.15\textwidth]{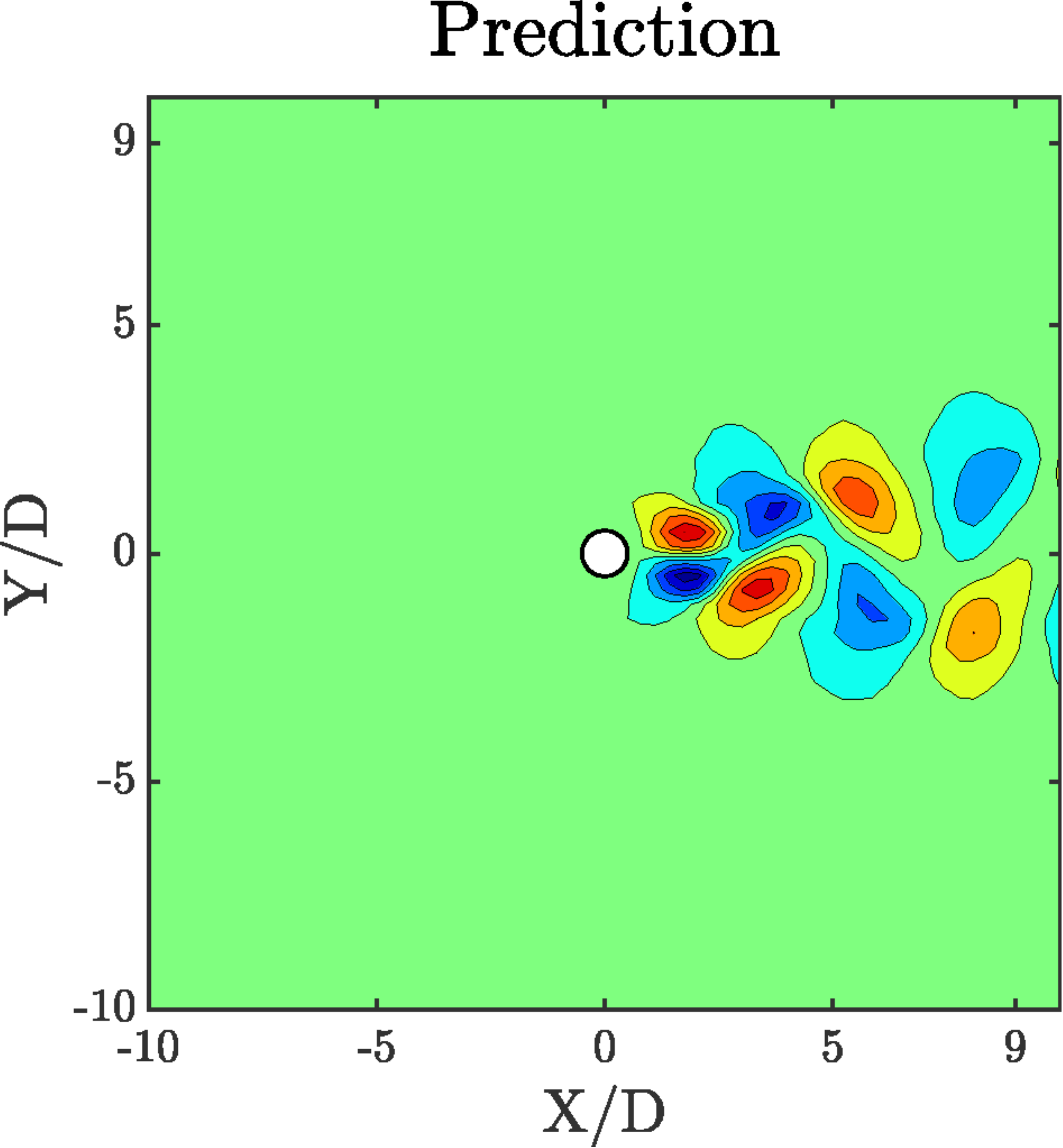}
\\
\vspace{0.05\textwidth}
\subfloat[]
{\includegraphics[width = 0.15\textwidth]{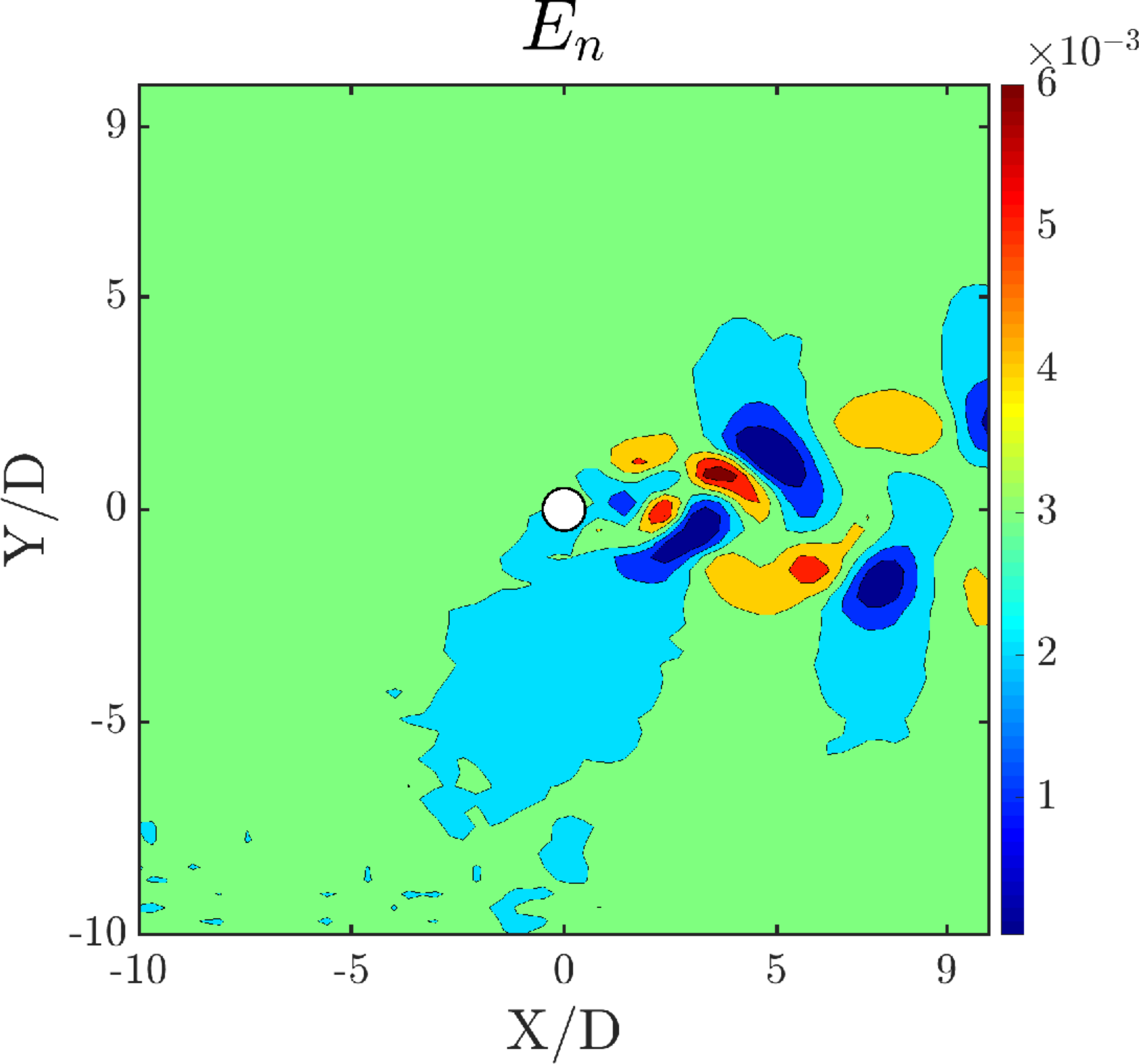}}
\subfloat[]
{\includegraphics[width = 0.15\textwidth]{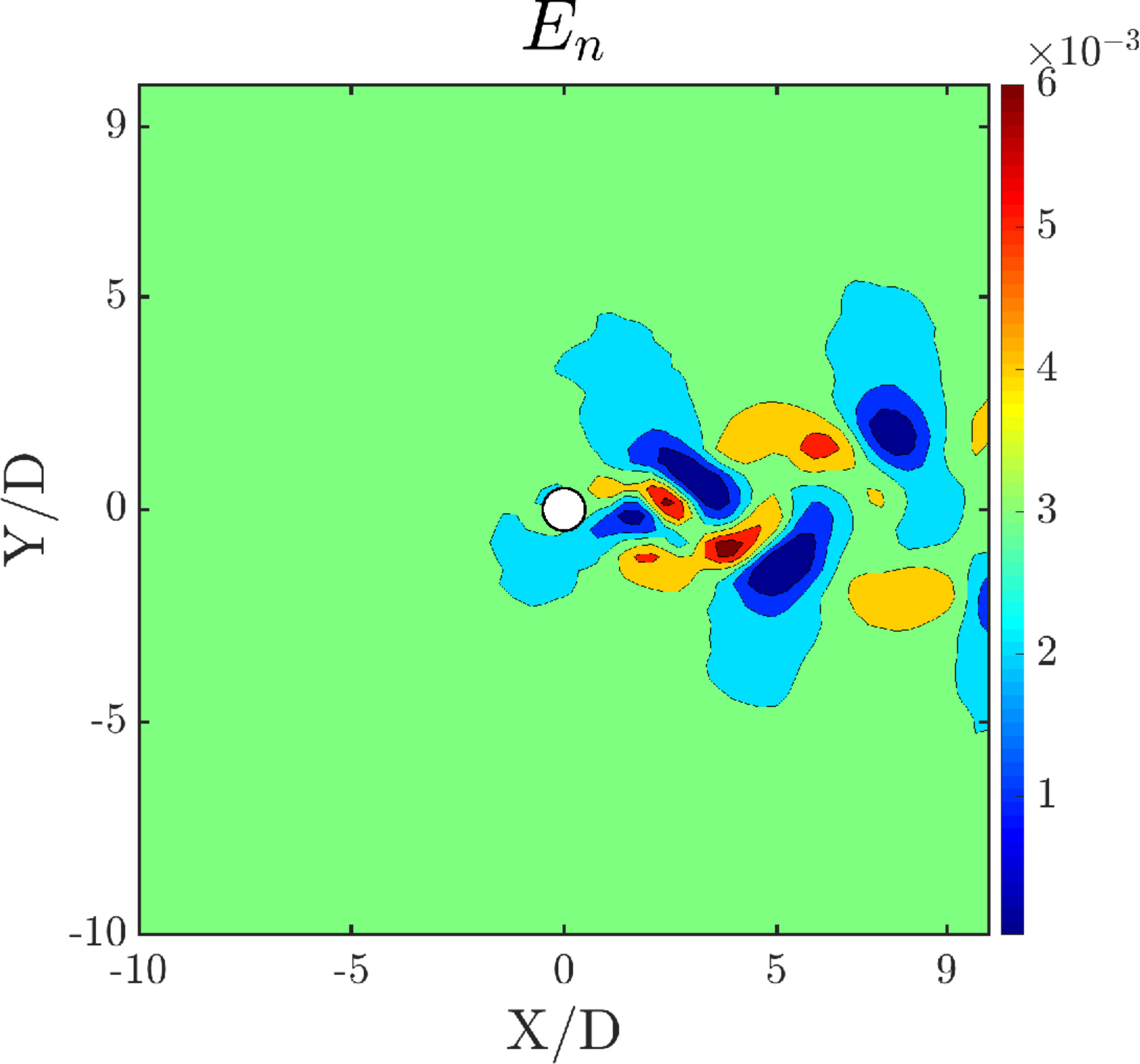}}
\subfloat[]
{\includegraphics[width = 0.15\textwidth]{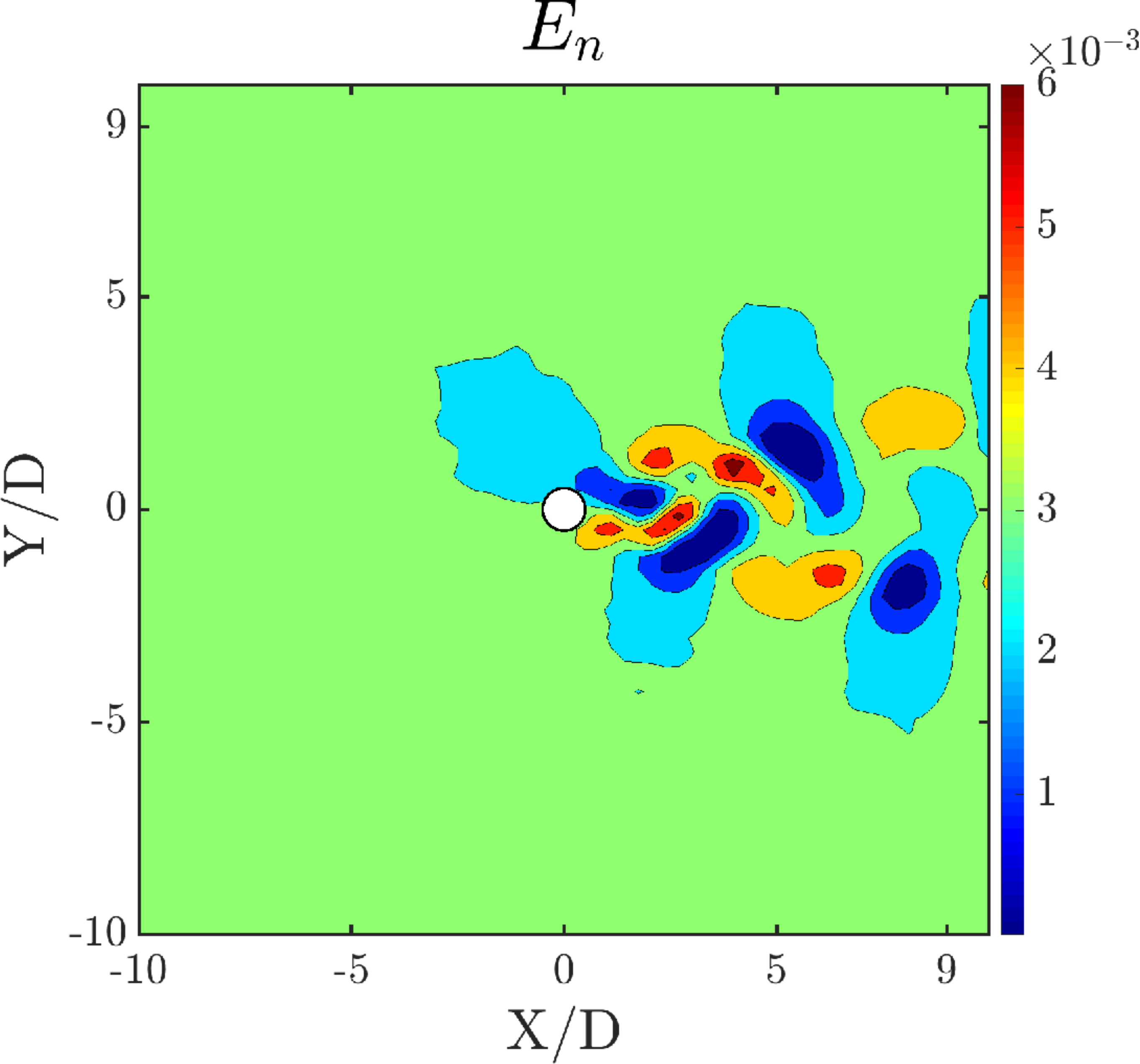}}

\caption{Comparison of truth and predicted fields along with normalized reconstruction error $E_{n}$ at (a) $t = 1050s$, (b) $t = 1125s$, (c) $t= 1200s$ for pressure field ($P$): flow past plain cylinder}
\label{fig6_4}
\end{figure}

Figure \ref{fig6_4} depicts the comparison of truth and predicted values for the pressure field $P$ along with the normalized reconstruction error $E_{n}$. It is very clear that the predictions are devoid of any spurious oscillations. The normalized reconstruction error $E_{n}$ is in the order of $10^{-3}$ and is higher when compared to velocity field predictions.

There appears to be some common points in the predictions of both velocity field $U$ and the pressure field $P$ for the problem of flow past an isolated single cylinder. The predictions are quite smooth and close to the truth data. In both the cases, the near wake region behind the cylinder appears to have majority of the reconstruction error. The nonlinearities associated with the near wake can be the reason for the errors. 

It should be noted that we are using closed recurrent neural network in the convolutional recurrent autoencoder model i.e., the prediction from the previous step is used as input to the next prediction. The main advantage of using such a model is that, there is no requirement of true data to be fed into the network in a time delayed fashion. One can predict for a finite time horizon with a single true sample which is used for the initialization of the network. We have also used closed loop recurrent neural network in the POD-RNN for the problem of flow past plain cylinder and the results were satisfactory. In that sense, one can argue that the convolutional recurrent autoencoder model does not provide any significant improvements over the POD-RNN model for this problem. However, the same inference is expected since the problem at hand is the simplest of fluid flow problems and any standard prediction tool should be able to perform at equal capabilities. 

\section*{Side-by-Side Cylinders Arrangement}

The same closed loop recurrent neural network in the POD-RNN model failed in the case of flow past a side-by-side cylinder arrangement. We have implemented the encoder-decoder networks which takes a certain length of true data to predict a finite time horizon \cite{reddy2019reduced}. The flow past a side-by-side cylinder arrangement is known to be a more complex problem with a bi-stable solution, flip-flopping regime and gap flow \cite{liu2016interaction}. In the next section, the same convolutional recurrent autoencoder model with closed loop recurrent neural network is employed on the problem of flow past side-by-side cylinders.

\begin{figure}
    \centering
    \includegraphics[width = 0.45\textwidth]{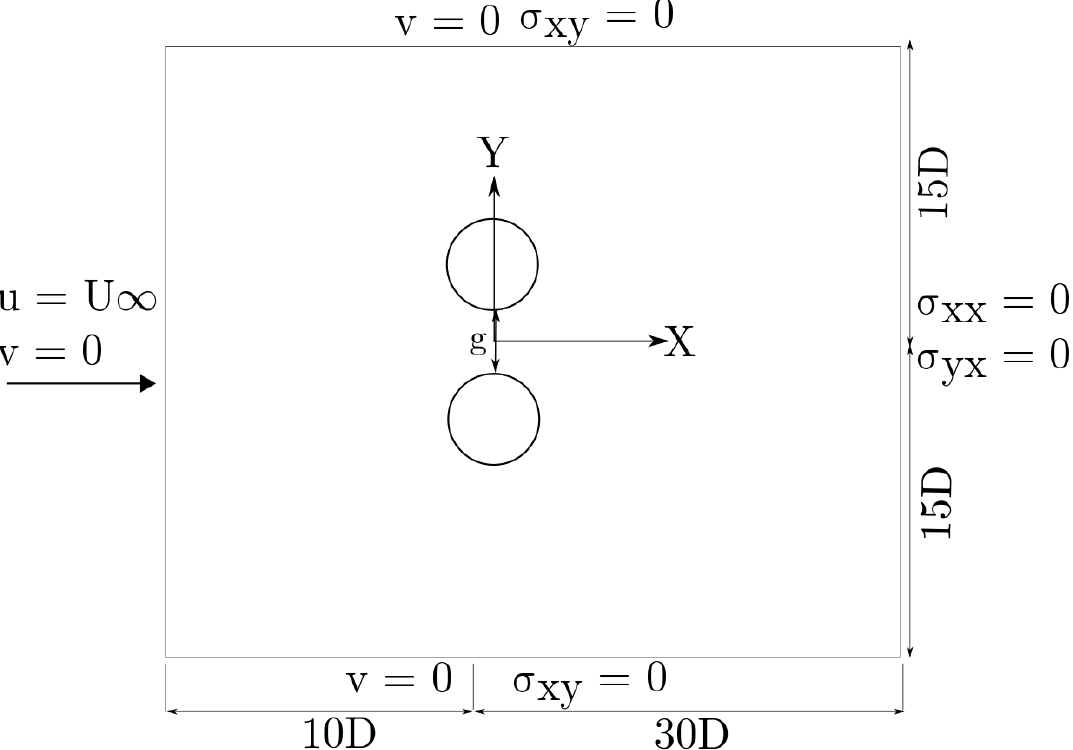}
    \caption{Schematic diagram of the problem setup: the side-by-side cylinder arrangement}
    \label{figsbs}
\end{figure}

\begin{figure}
    \centering
    \includegraphics[width = 0.45\textwidth]{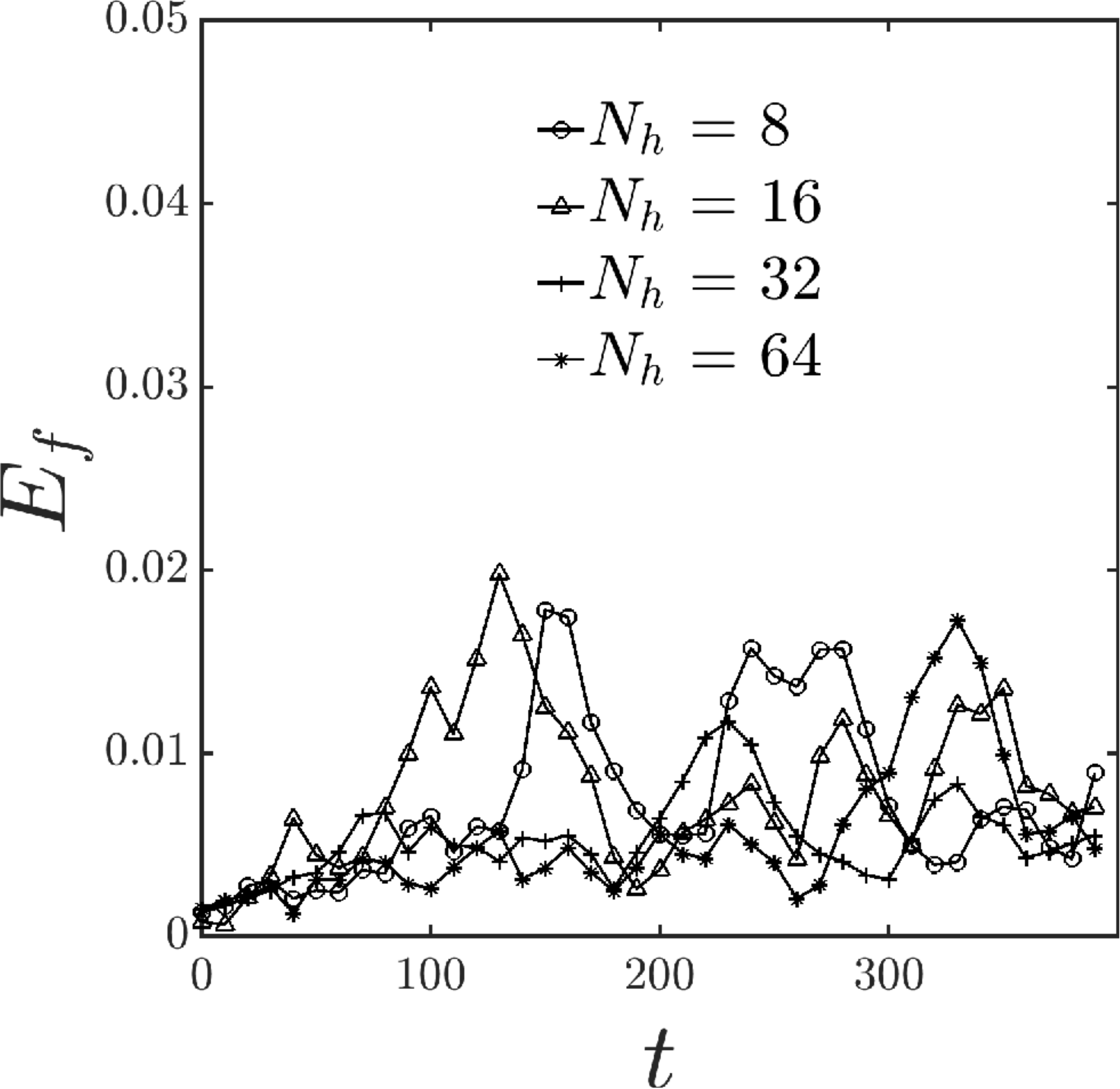}
    \caption{Normalized squared error $E_{f}$ for the velocity field in X-direction ($U$): flow past side-by-side cylinders}
    \label{figsbserror_u}
\end{figure}

The canonical problem of flow past side-by-side cylinders can act as a representative case for many offshore multibody systems. The dynamics of the problem is still a popular topic of research in the fluid mechanics community. Therefore this problem has been chosen to test the capabilities of our data-driven framework. The schematic of the problem set up is given in Fig.~\ref{figsbs} where all the information about the domain boundaries and the boundary conditions are clearly outlined. 

The Reynolds number of the problem is considered as $Re = 100$.
Although it is in the regime of low Reynolds number, the dynamics associated with this problem are far more complex when compared to the flow past a single cylinder. The full order simulation was carried out for $600s$ with a time step of $0.02s$. A total of $3000$ snapshots of the simulation were collected for every $0.2s$. Similar to the previous problem, we map the unstructured data coming from full order simulations onto a square uniform grid of size $64\times 64$. 
Both the velocity field $U$ and the pressure field $P$ are considered for the training and prediction. The training data is generated with $N_{s} = 100$ samples, each with a size of $N_{t} = 20$, which accounts for first $2000$ samples. The next $600$ samples are used for the validation and the final $400$ samples are used for testing the model.

\begin{figure}
    \centering
    \includegraphics[width = 0.45\textwidth]{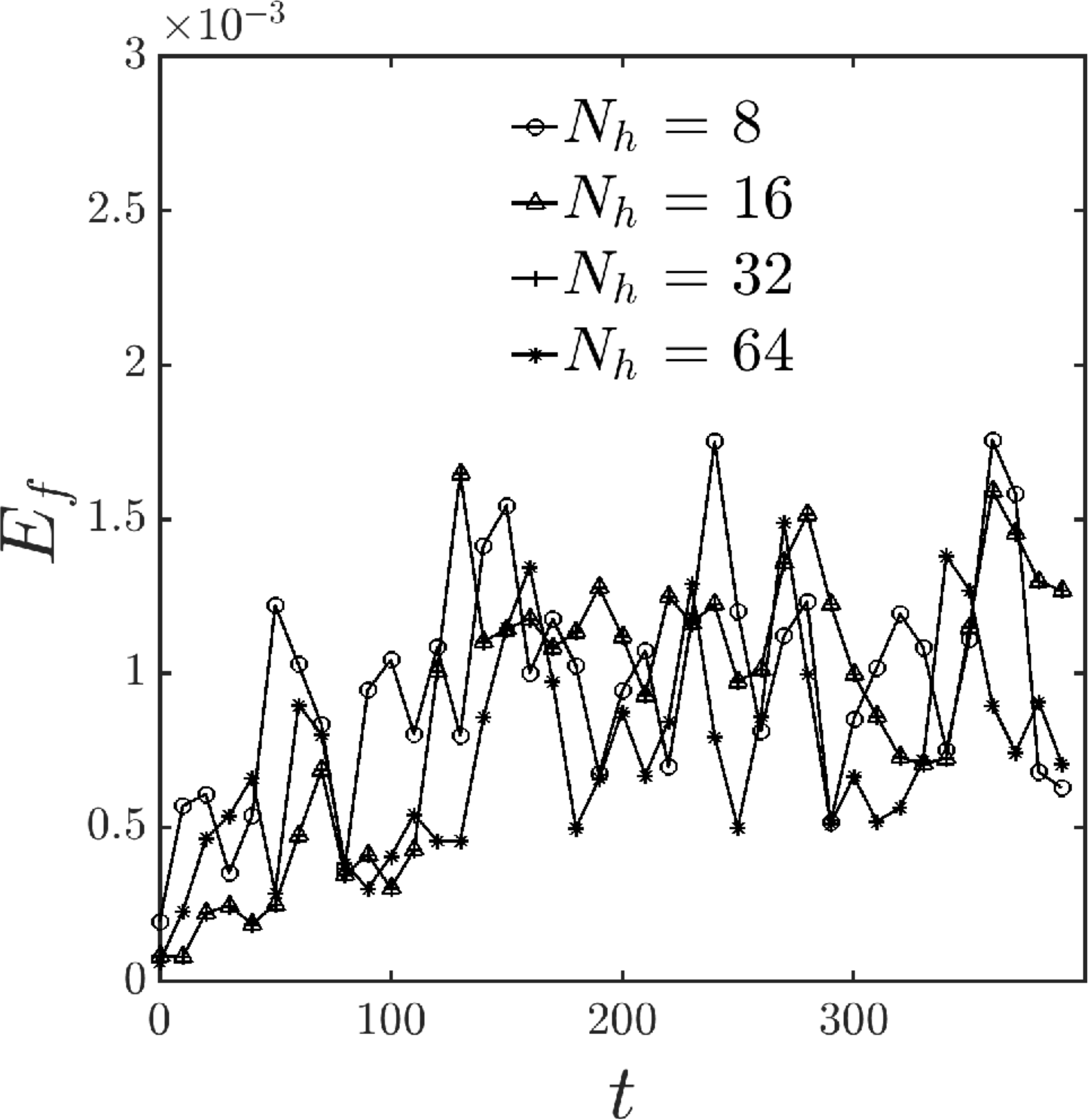}
    \caption{Normalized squared error $E_{f}$ for pressure field ($P$): flow past side-by-side cylinders}
    \label{figsbserror_p}
\end{figure}

Similar to the previous section, four models with different sizes of low dimensional representation were trained. All the four models are trained on single Intel E5-2690v3 (2.60GHz, 12 cores) CPU node for $N_{train} = 1000000$ iterations. The training took approximately $48$ hours. The online prediction is carried for 400 steps with the trained model. Figure \ref{figsbserror_u} shows the normalized squared error $E_{f}$ for the velocity field in $X$-direction $(U)$ and Fig.~\ref{figsbserror_p} shows the same error for pressure field $P$
 
In contrast to the case with an isolated plain cylinder, the velocity error here for all the four models appears to be higher than the pressure error. Also for the predictions of both $U$ and $P$, the errors are in the same order of magnitude for all the four low dimensional representations.
We have chosen $N_{h} = 64$ as the final trained model to depict the flow field predictions in detail since it has a better performance when compared to other models.

\begin{figure}
\centering
\includegraphics[width = 0.15\textwidth]{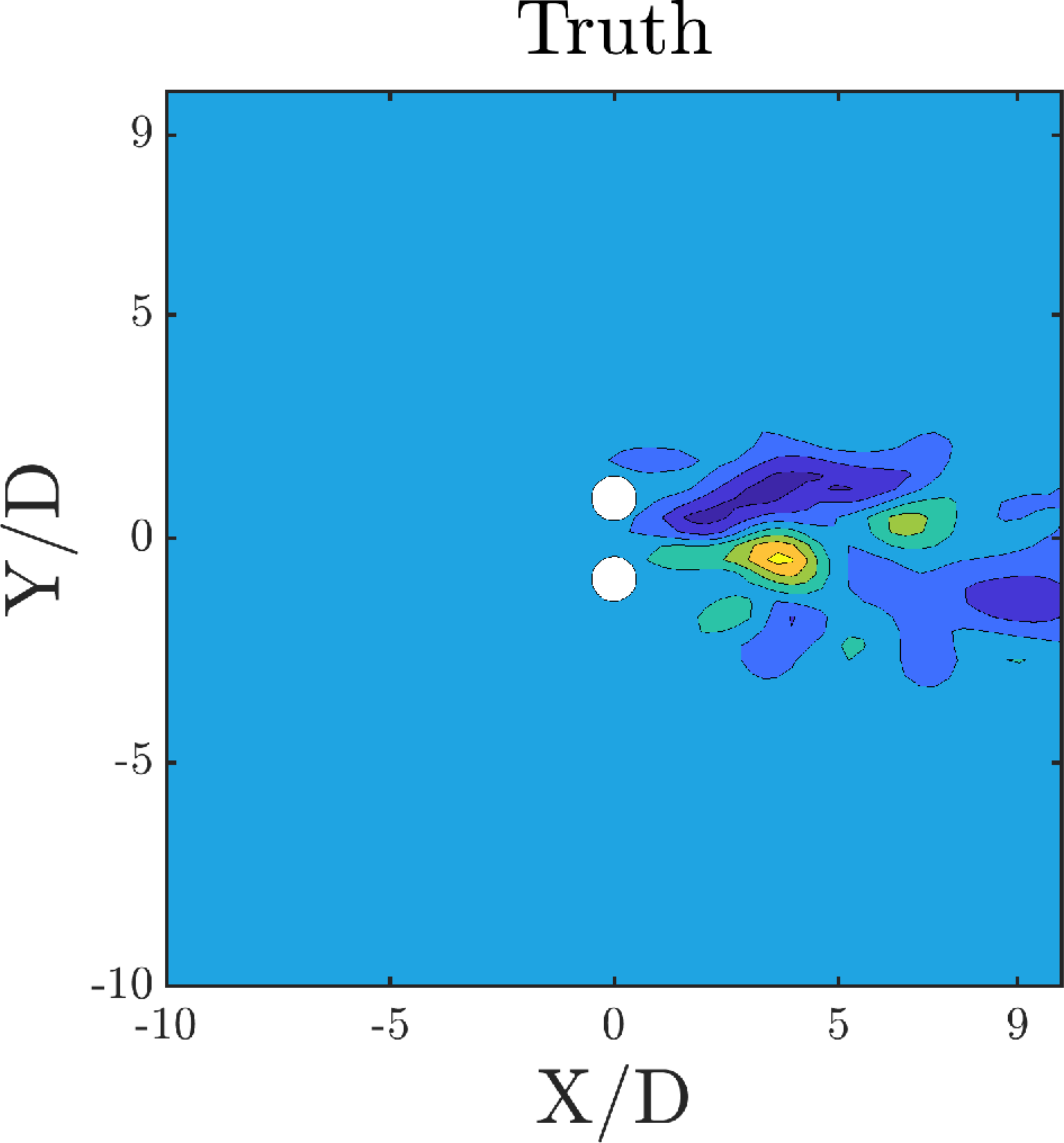}
\includegraphics[width = 0.15\textwidth]{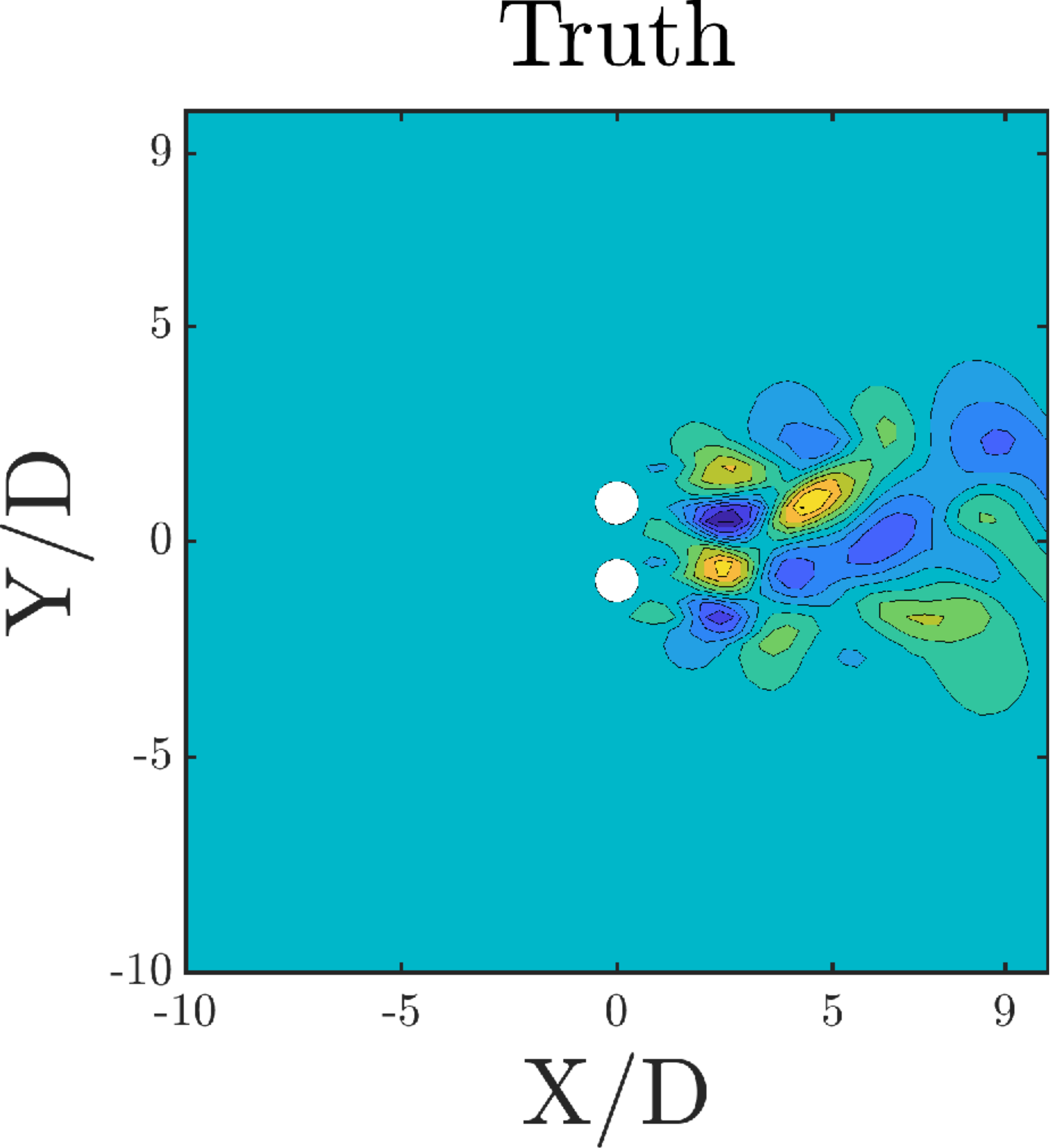}
\includegraphics[width = 0.15\textwidth]{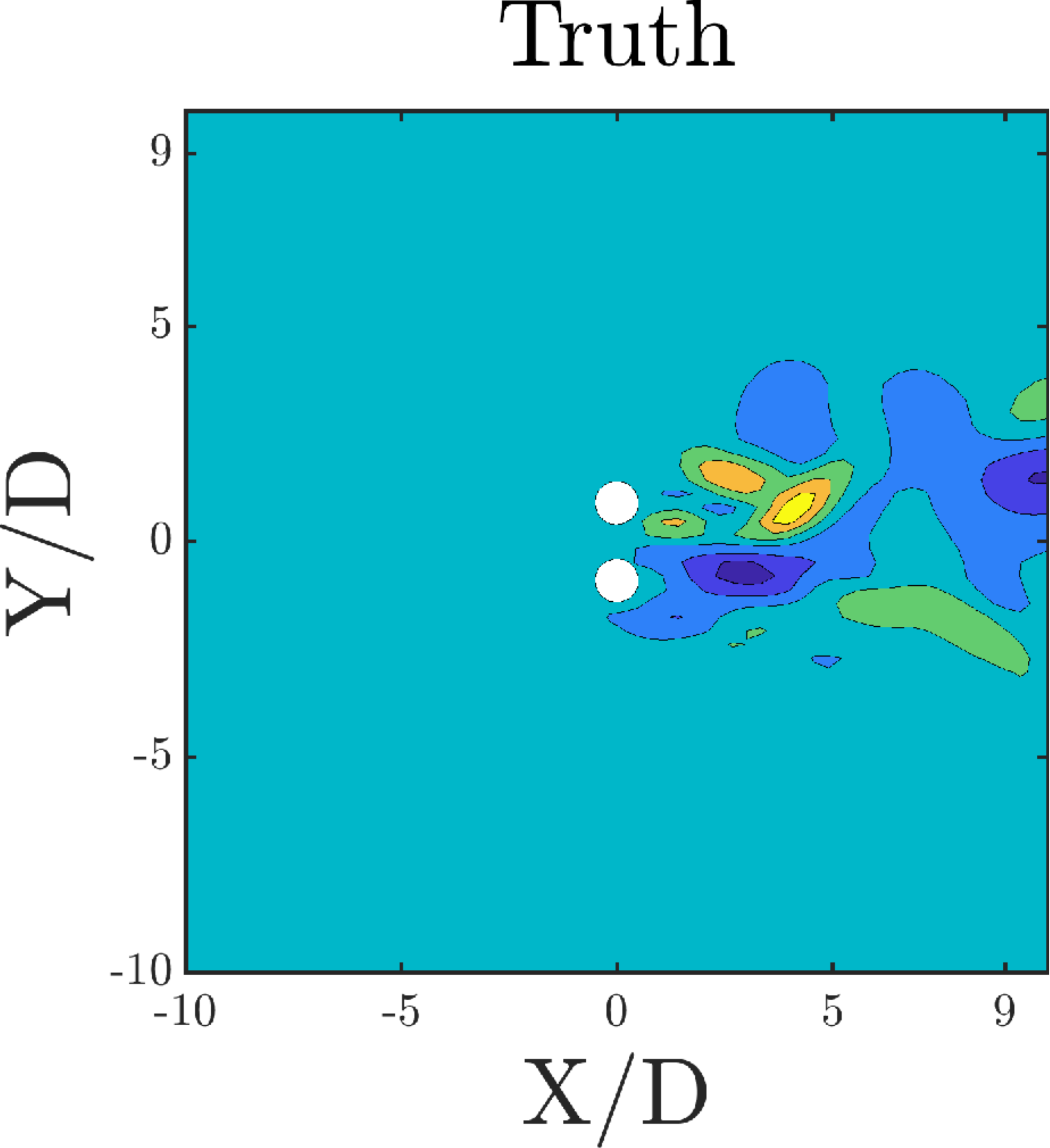}
\\
\vspace{0.05\textwidth}
\includegraphics[width = 0.15\textwidth]{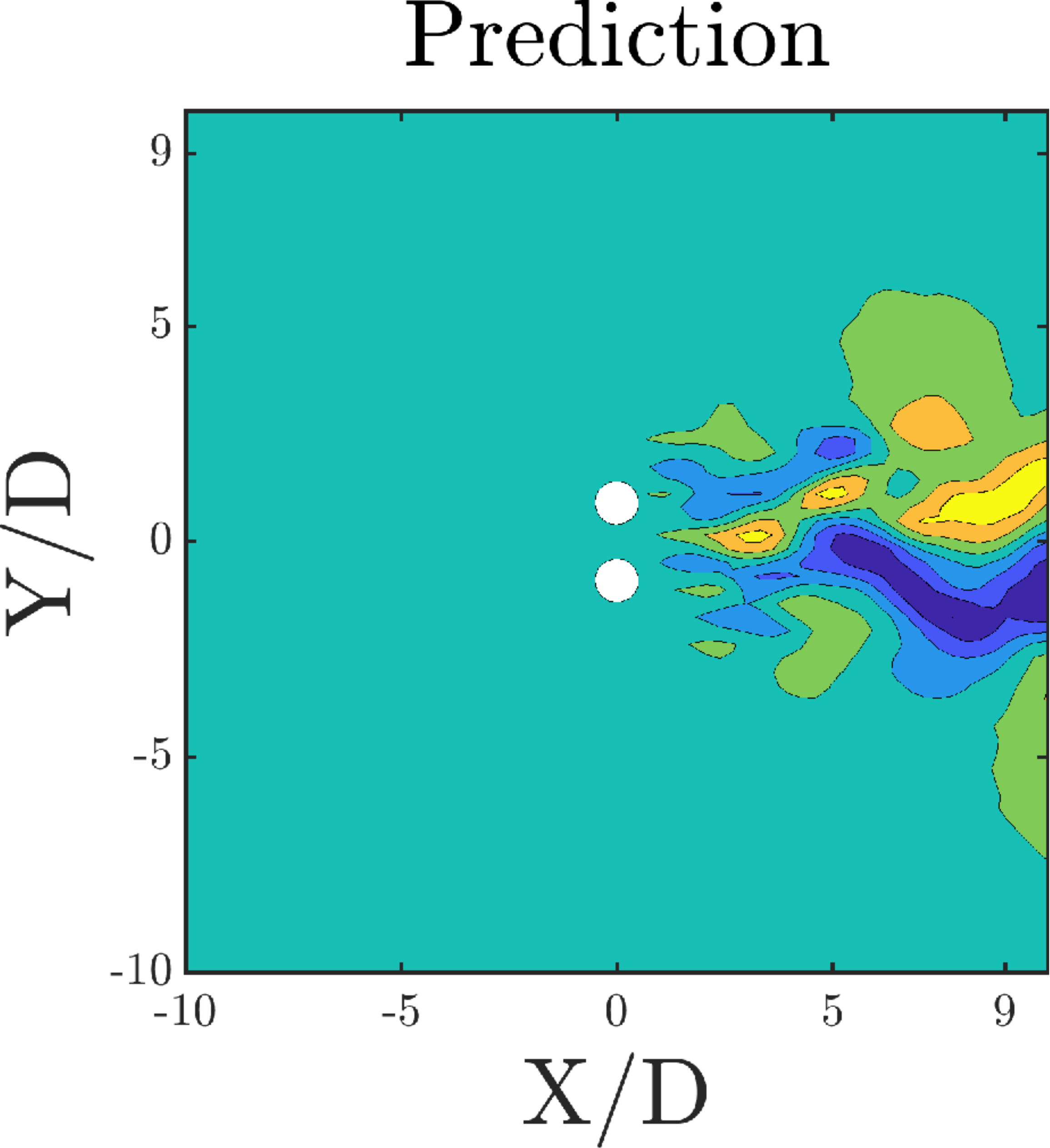}
\includegraphics[width = 0.15\textwidth]{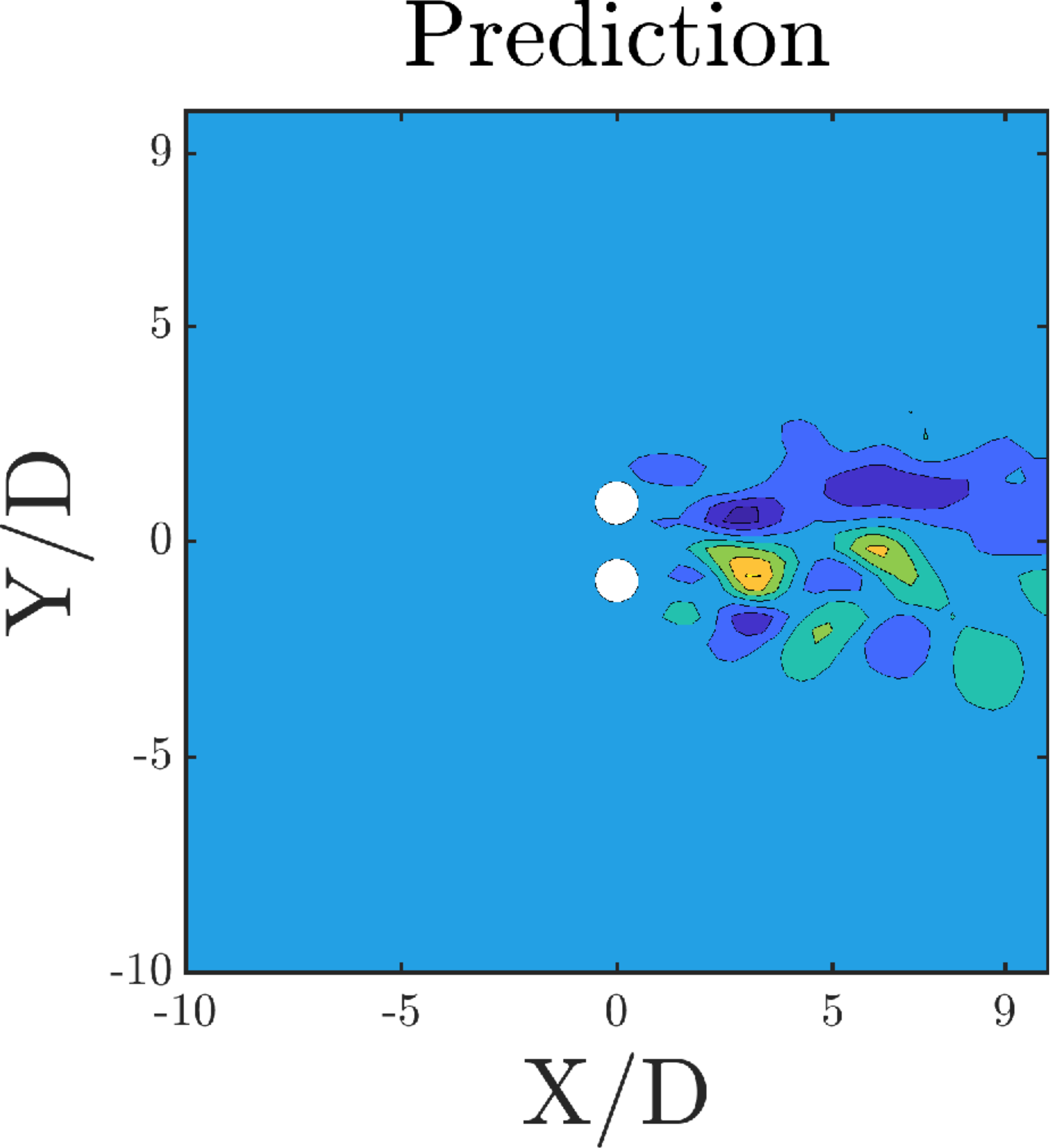}
\includegraphics[width = 0.15\textwidth]{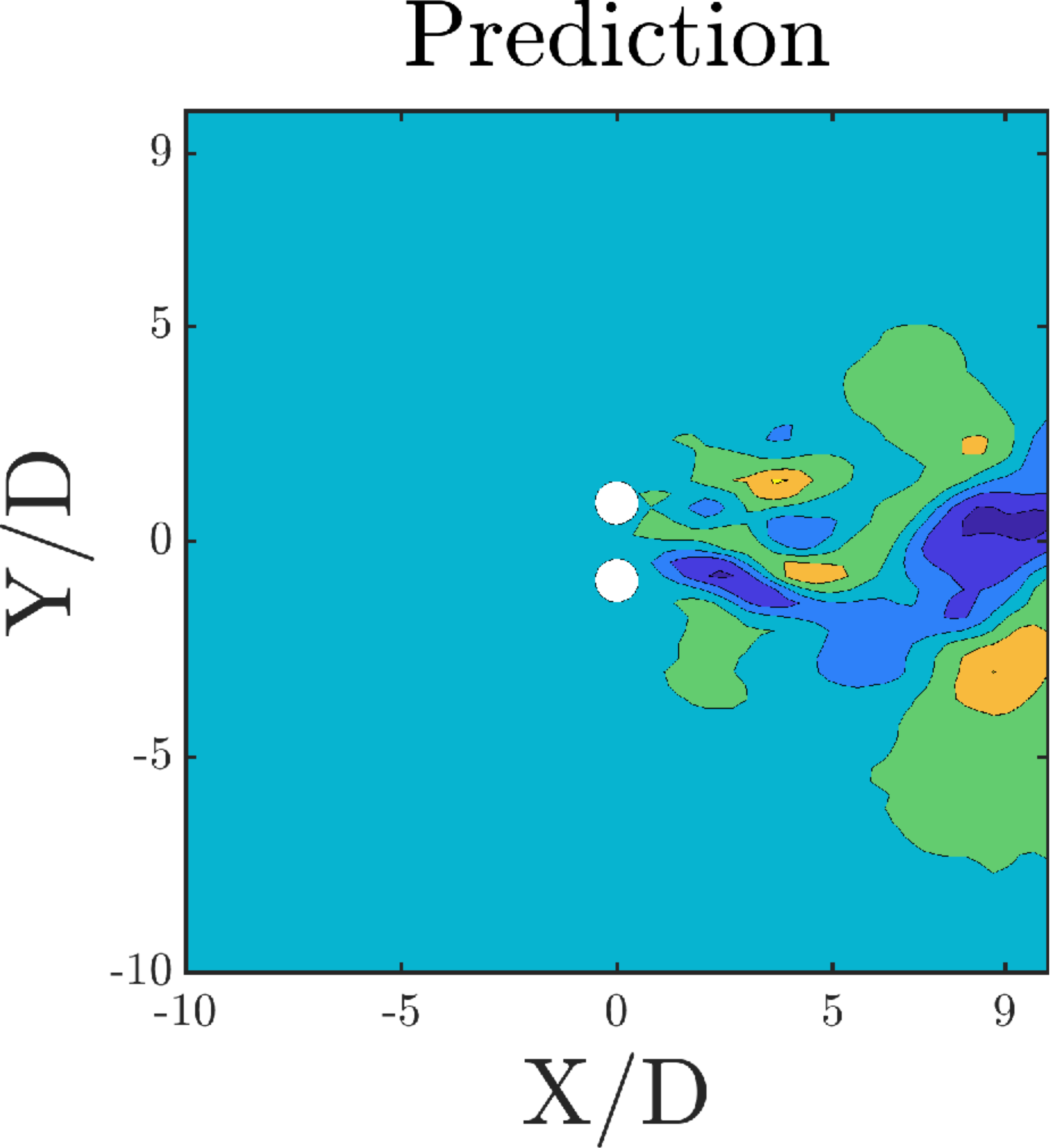}
\\
\vspace{0.05\textwidth}
\subfloat[]
{\includegraphics[width = 0.15\textwidth]{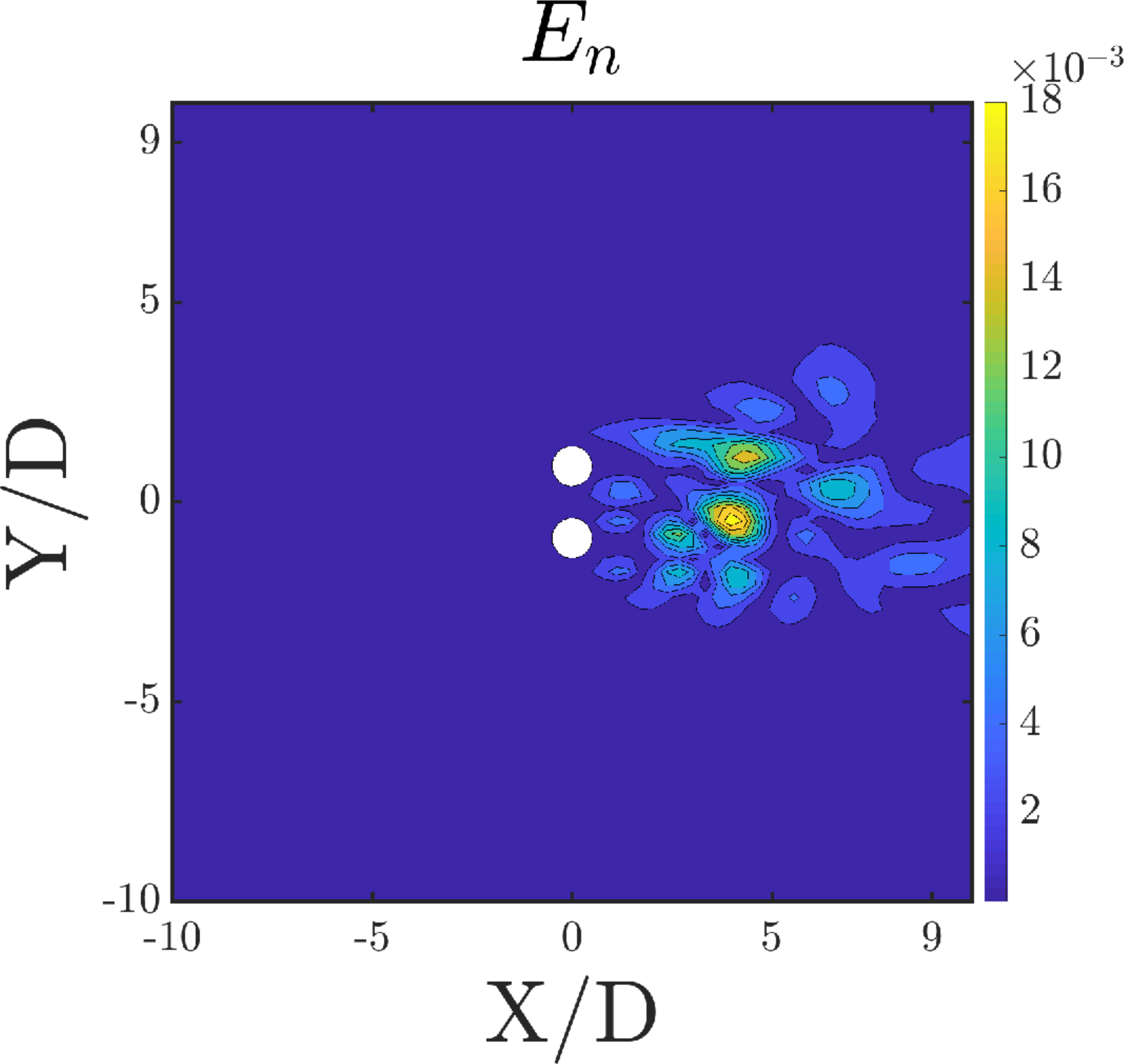}}
\subfloat[]
{\includegraphics[width = 0.15\textwidth]{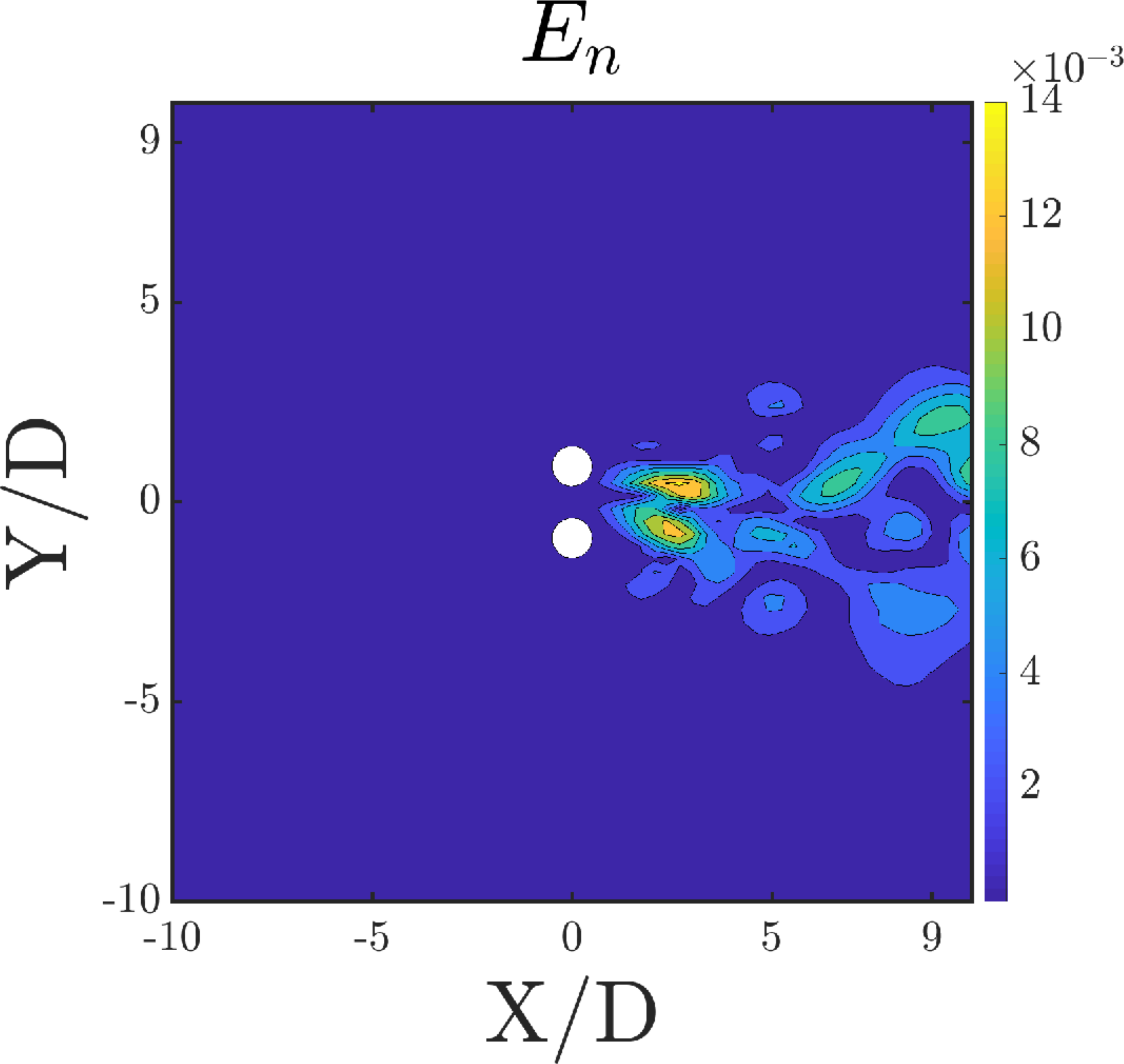}}
\subfloat[]
{\includegraphics[width = 0.15\textwidth]{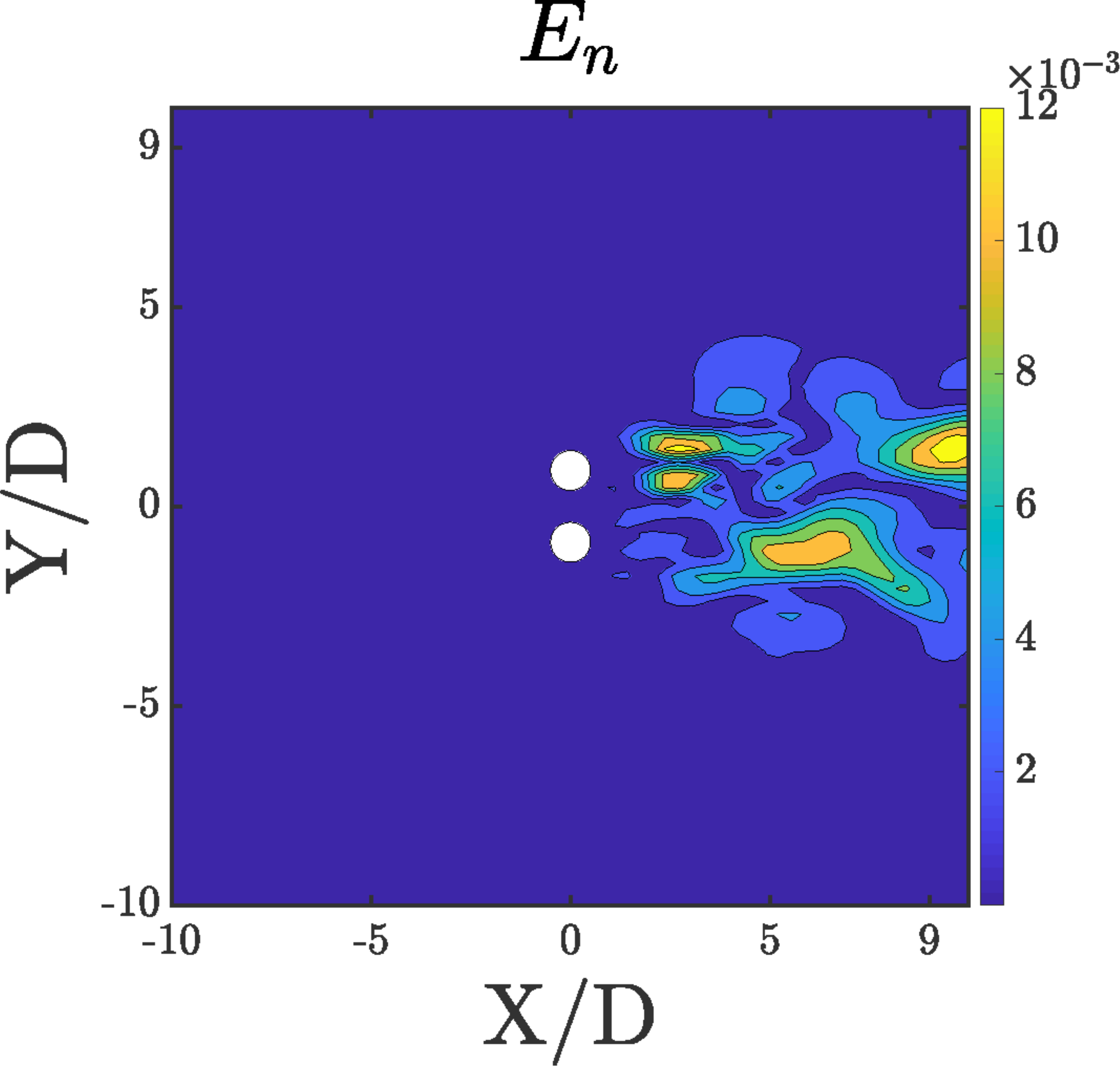}}

\caption{Comparison of truth and predicted fields along with normalized reconstruction error $E_{n}$ at (a) $t = 540s$, (b) $t = 560s$, (c) $t= 580s$ for velocity field in X-direction ($U$): flow past side-by-side cylinders}
\label{fig6_5}
\end{figure}

\begin{figure}
\centering
\includegraphics[width = 0.15\textwidth]{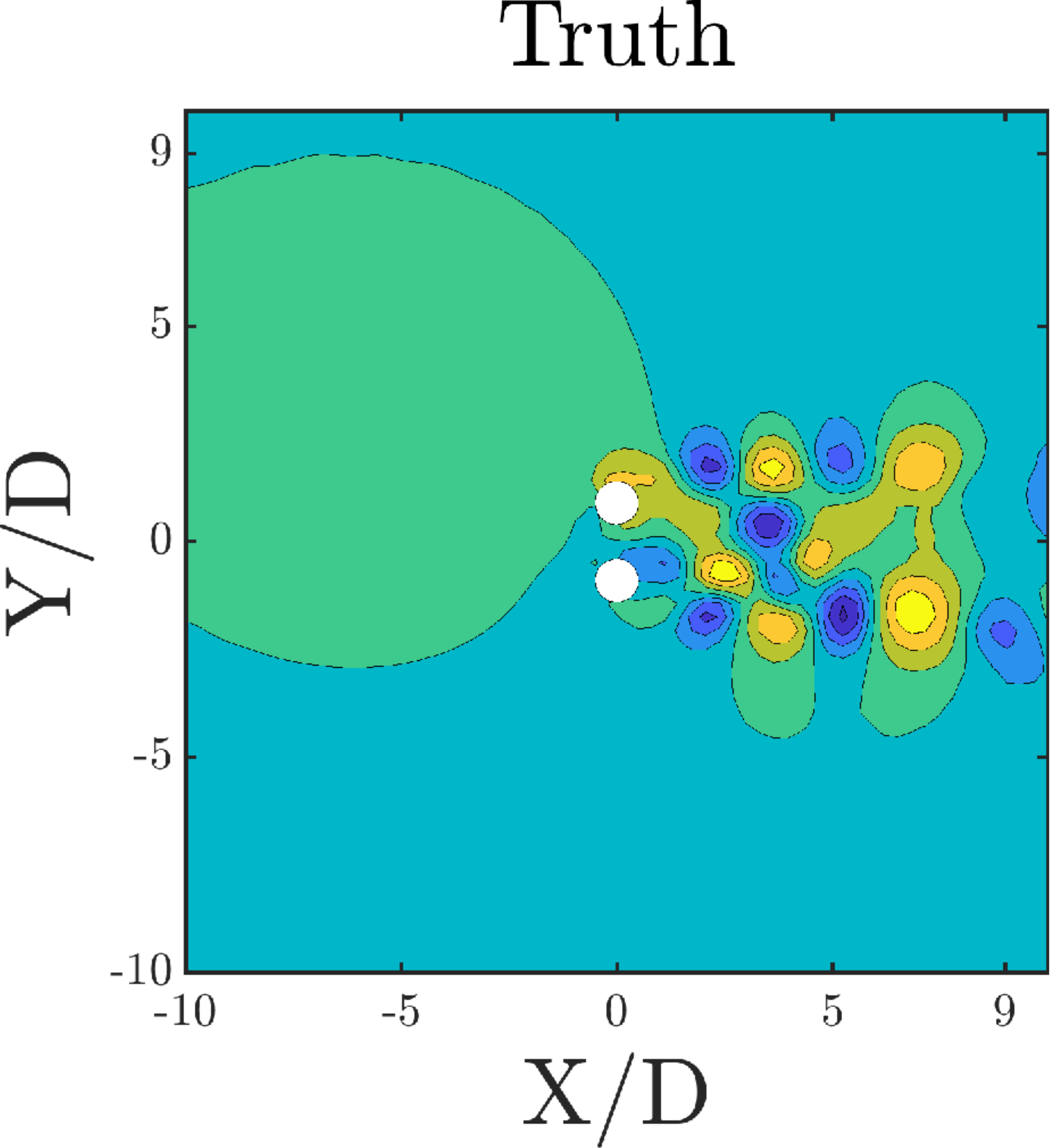}
\includegraphics[width = 0.15\textwidth]{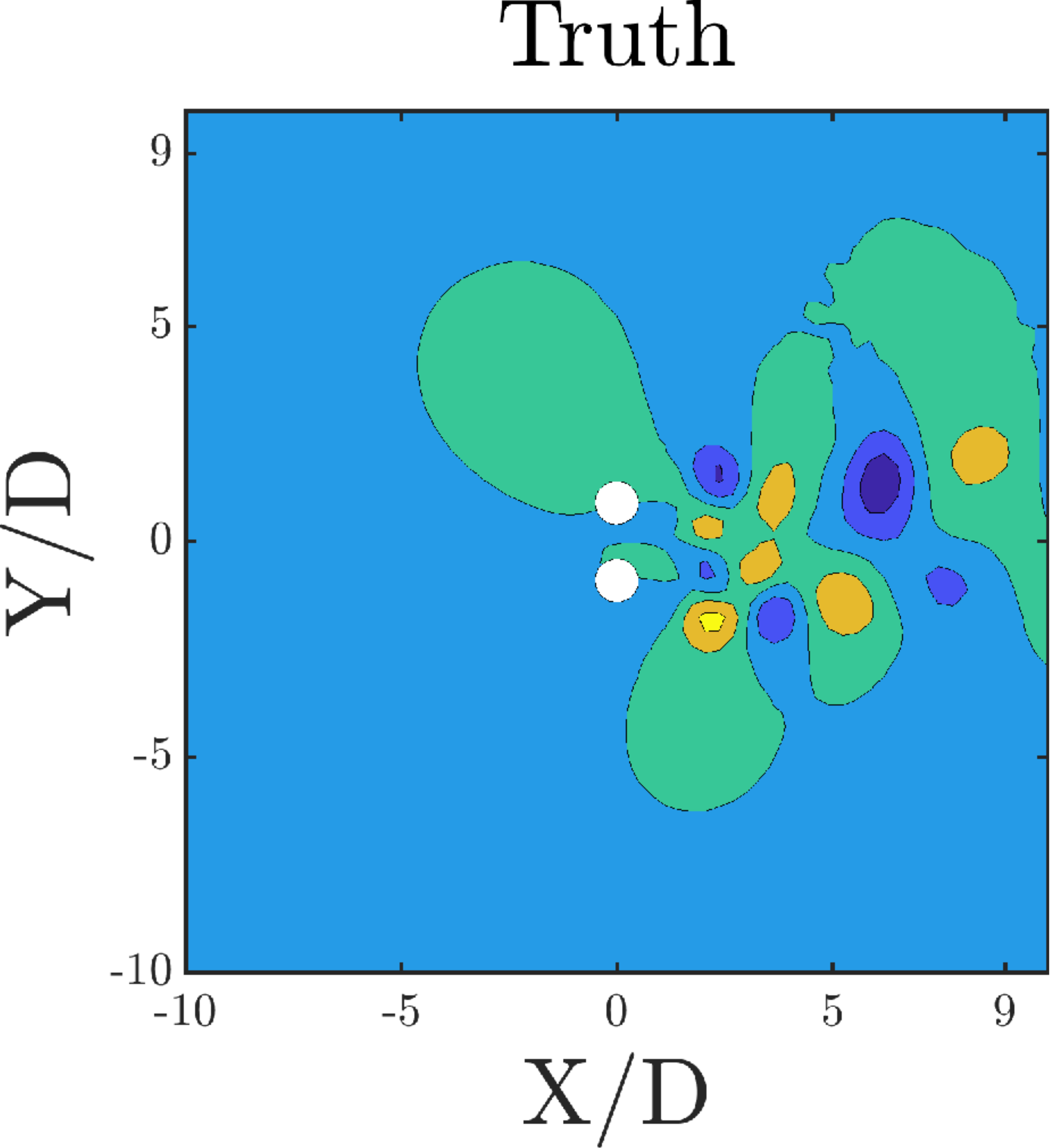}
\includegraphics[width = 0.15\textwidth]{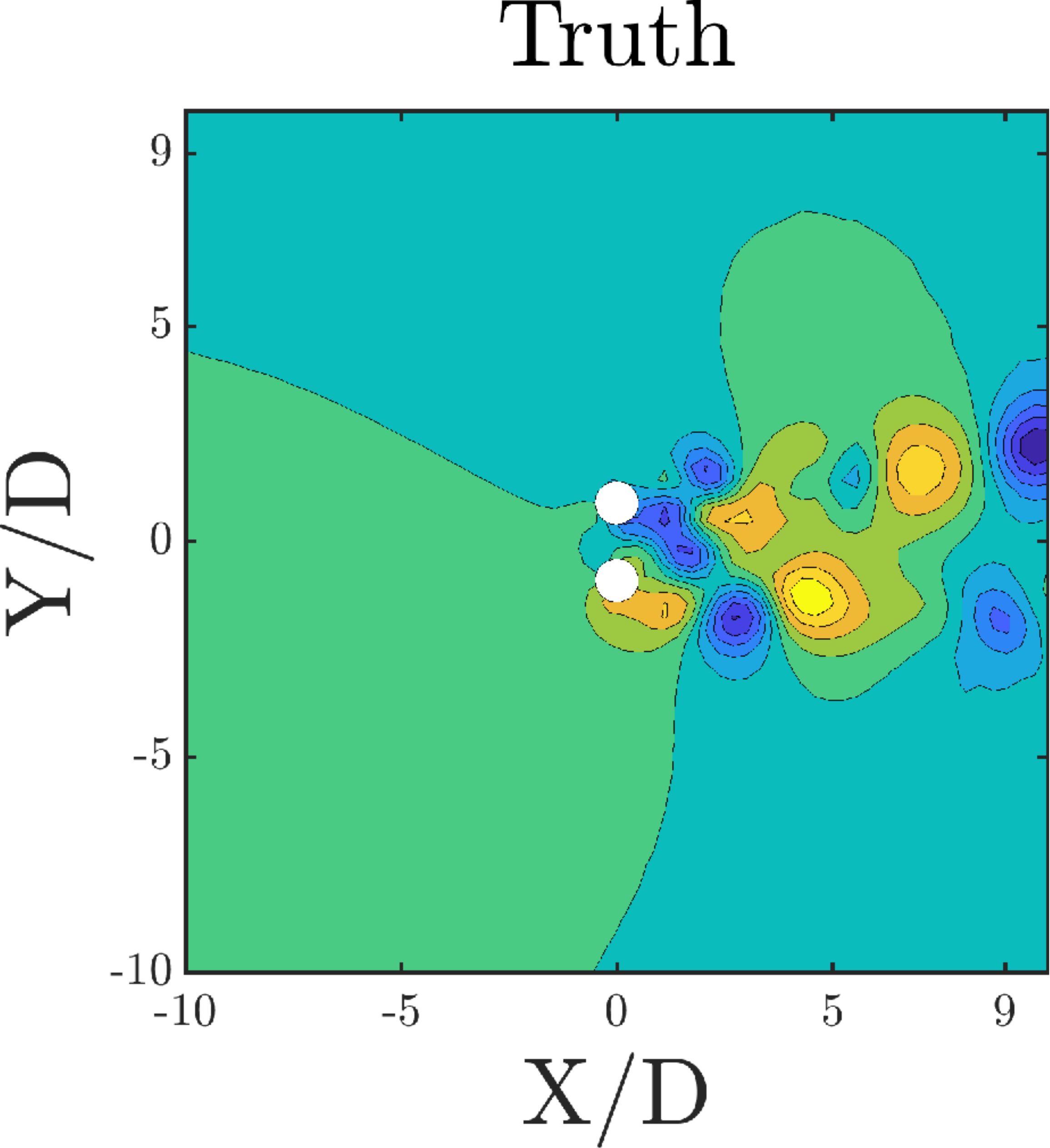}
\\
\vspace{0.05\textwidth}
\includegraphics[width = 0.15\textwidth]{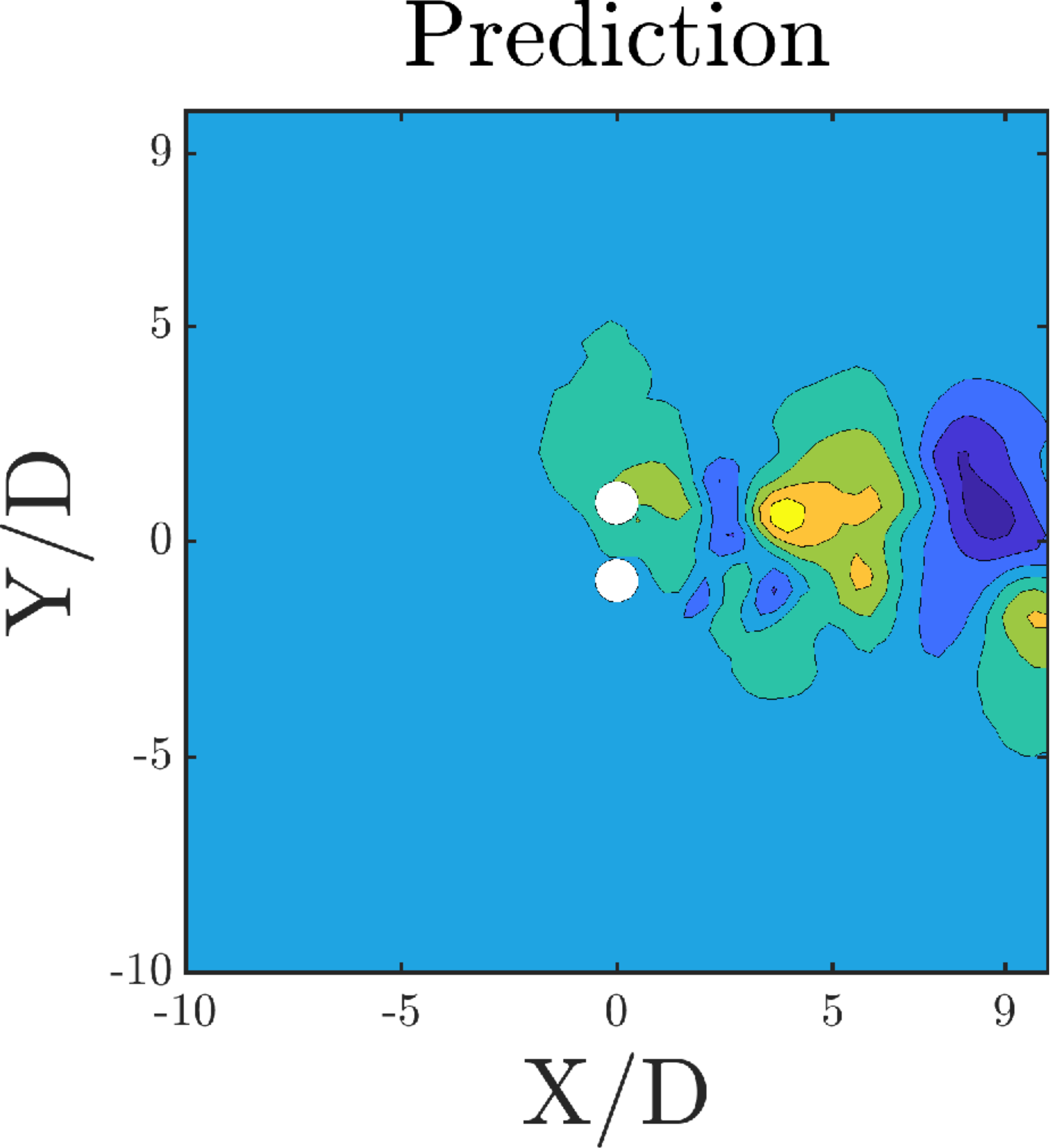}
\includegraphics[width = 0.15\textwidth]{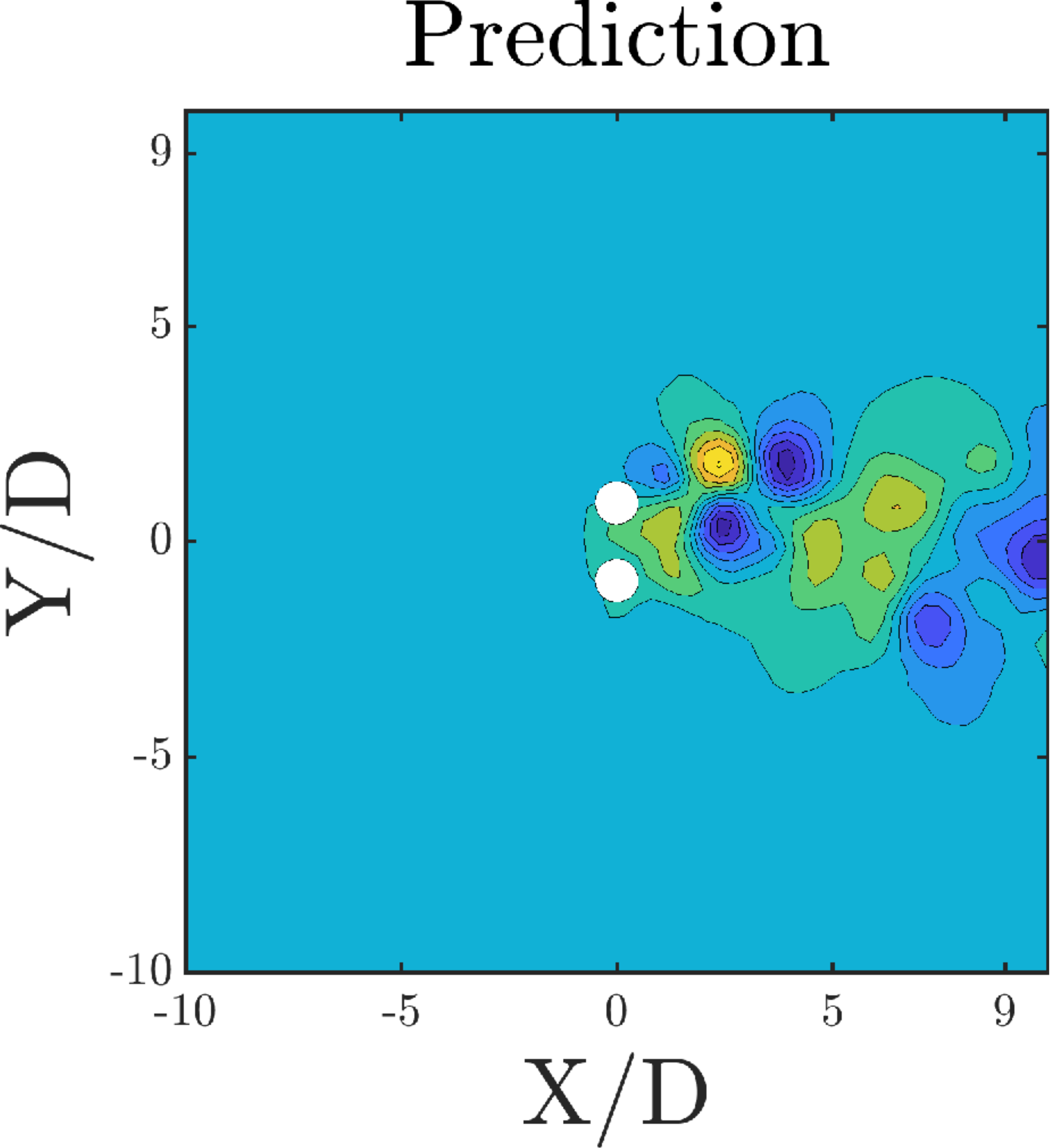}
\includegraphics[width = 0.15\textwidth]{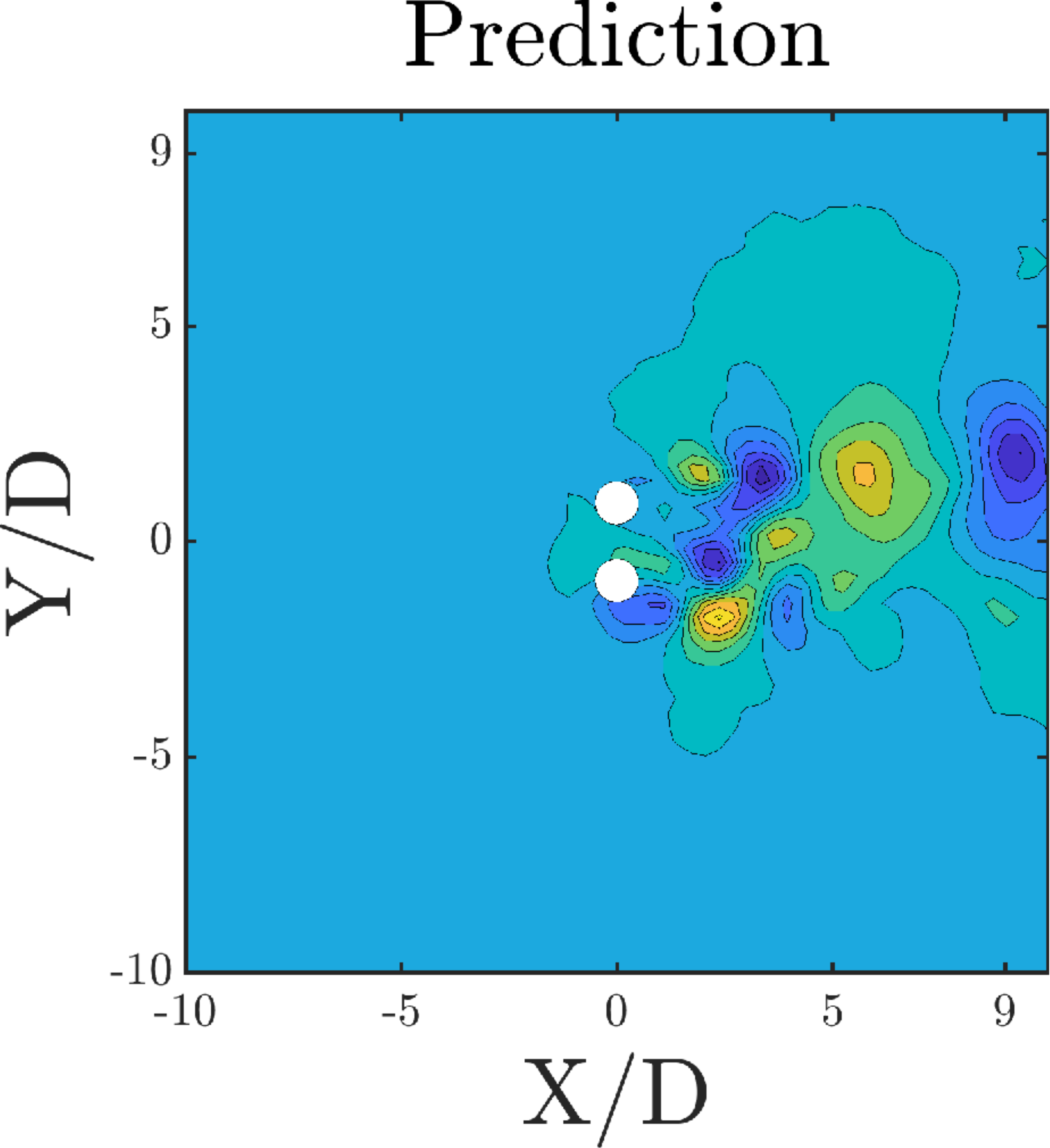}
\\
\vspace{0.05\textwidth}
\subfloat[]
{\includegraphics[width = 0.15\textwidth]{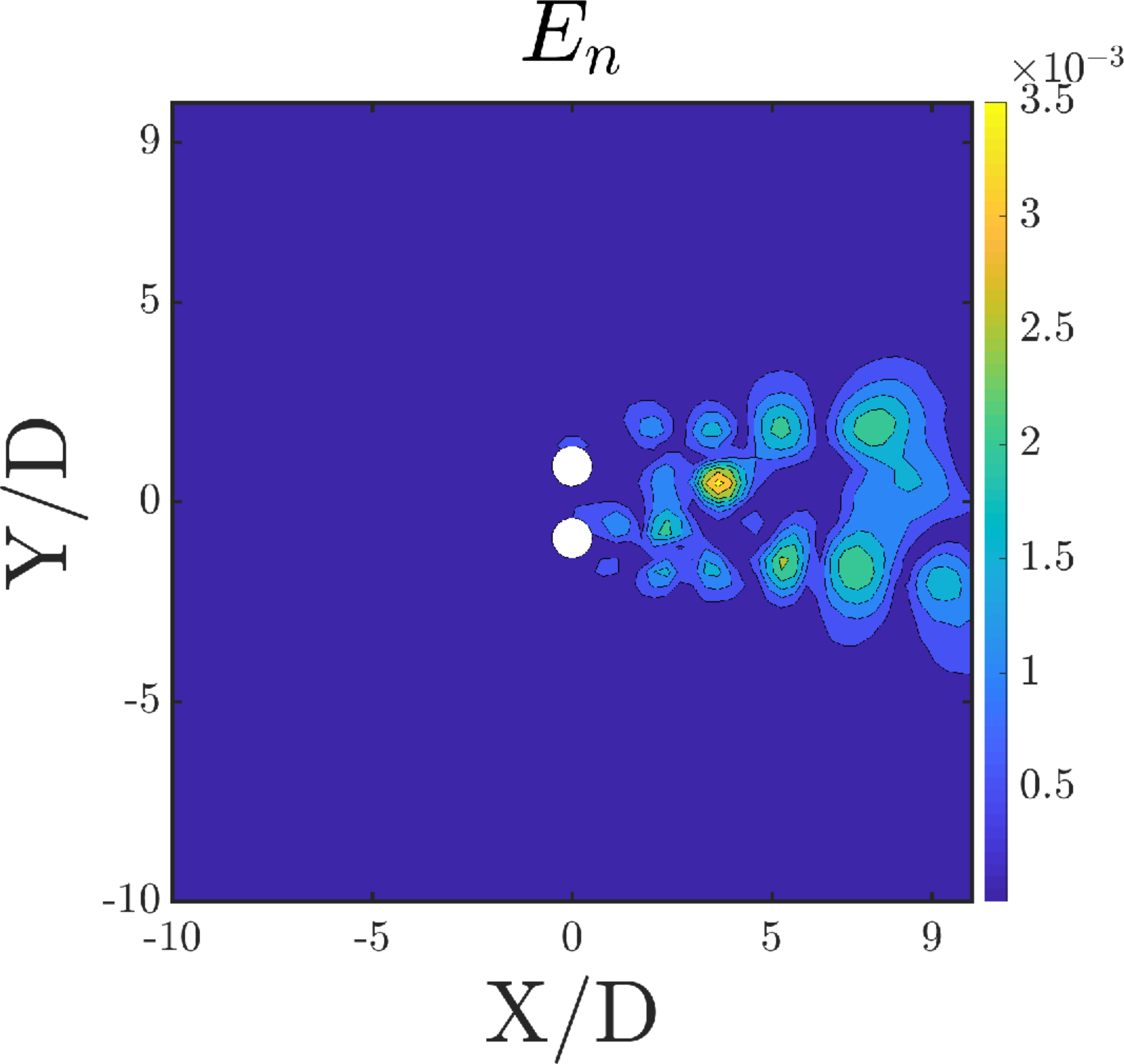}}
\subfloat[]
{\includegraphics[width = 0.15\textwidth]{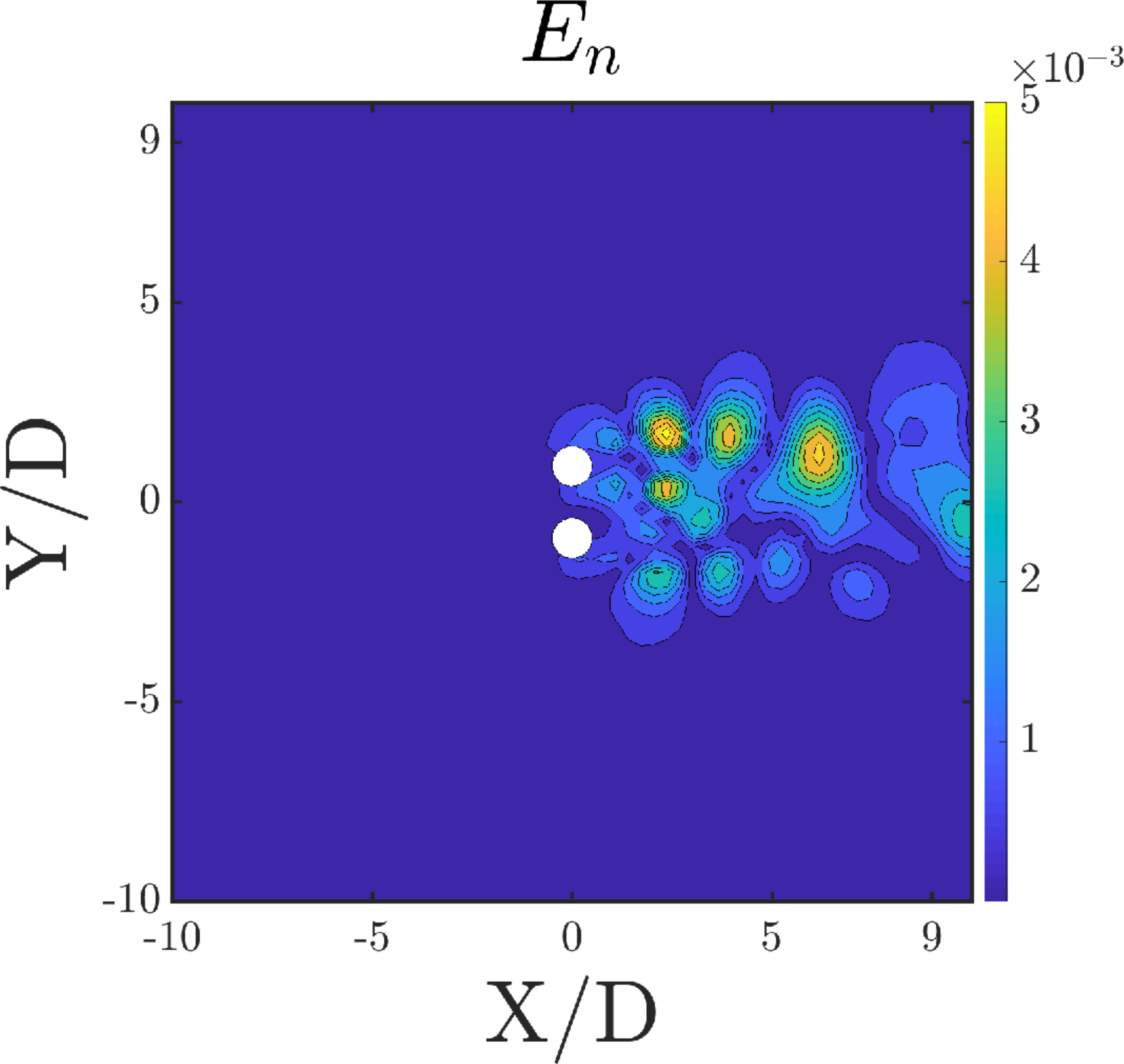}}
\subfloat[]
{\includegraphics[width = 0.15\textwidth]{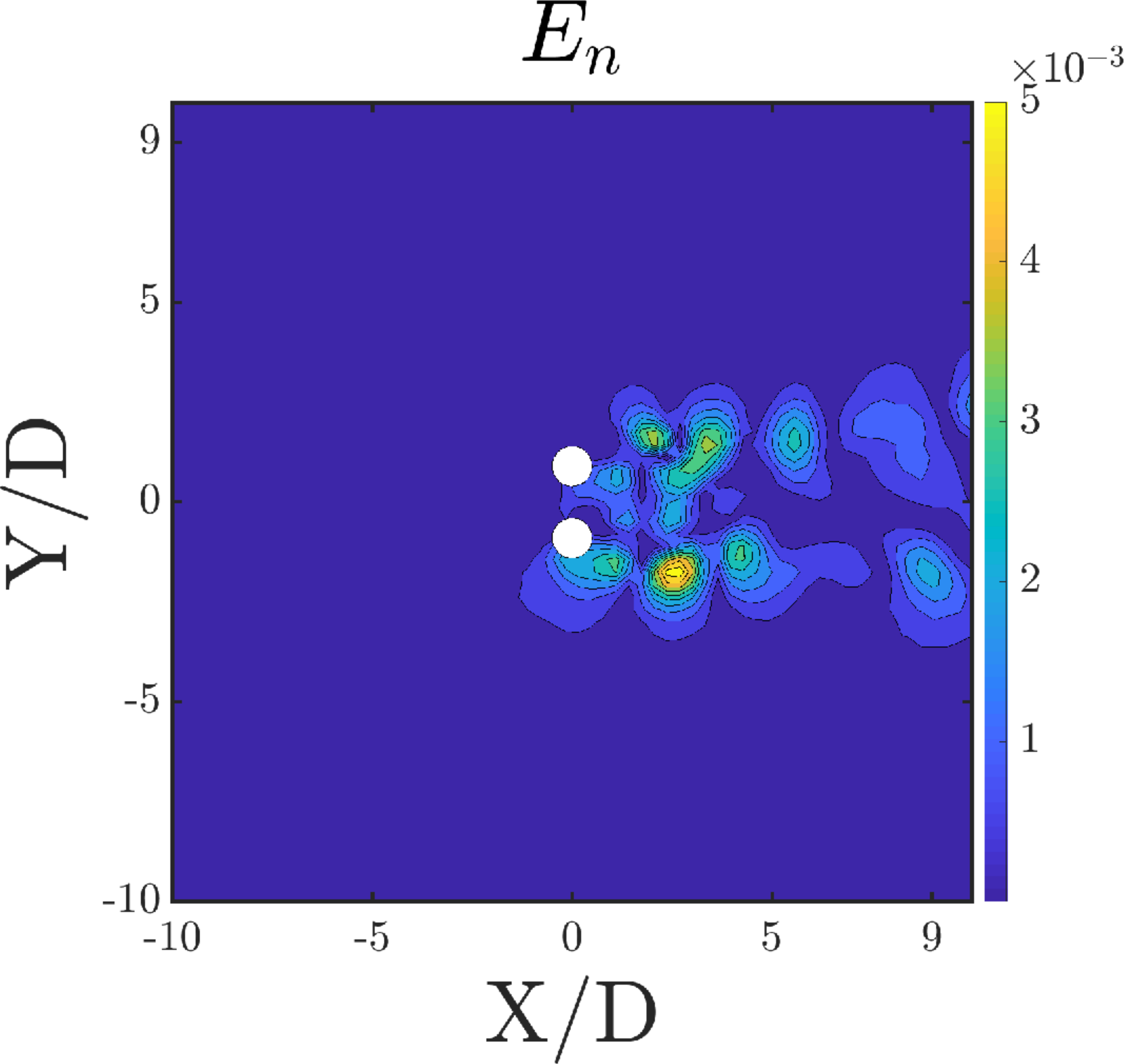}}

\caption{Comparison of truth and predicted fields along with normalized reconstruction error $E_{n}$ at (a) $t = 540s$, (b) $t = 560s$, (c) $t= 580s$ for pressure field ($P$): flow past side-by-side cylinders}
\label{fig6_6}
\end{figure}

Figure \ref{fig6_5} shows the comparison of truth and predicted values for the velocity field in X-direction ($U$) along with the normalized reconstruction error $E_{n}$ and Fig.\ref{fig6_6} illustrates the same comparison for the pressure field.
Overall predictions from the convolutional recurrent autoencoder model are quite satisfactory. It is remarkable to note that the convolutional recurrent autoencoder model is able to capture the flip-flopping phenomenon in both  velocity $U$ and pressure $P$ predictions. The predictions are also quite smooth without any oscillations. The reconstruction error $E_{n}$ is quite low in the order of $10^{-3}$ for both the predictions. Similar to the isolated cylinder, the reconstruction errors are concentrated in the near wake region where there is strong presence of nonlinearity.

It should be noted that the POD-RNN model with the closed loop recurrent neural network completely failed to predict the flow past an arrangement of side-by-side cylinders. Whereas, the convolutional recurrent autoencoder model was able to predict the flow fields up to 400 time steps satisfactorily without any divergence in the error. It is also able to capture the randomness in the flow field with respect to dominant flow features.  In the case of POD-RNN model, the spatial and temporal parts of the problem are dealt independently which might work for simple problems of the flow past a plain isolated cylinder. Such linear decomposition procedures completely fail at highly non-linear problems such as side-by-side cylinder. The improvement compared to the POD-RNN model can be attributed to the low dimensional features obtained by convolutional neural networks of convolutional recurrent autoencoder model. They are very powerful in identifying dominant local flow features \cite{thanrindu2018}. Also, the complete end to end architecture of the network which enables us to integrate the encoding, evolution and decoding in a complete non-linear fashion is also the primary reason for this model to work for this problem. This is a very promising result and motivates us to take forward this concept of convolutional recurrent autoencoder models for more complicated problems of fluid mechanics and fluid-structure interaction.

\subsection*{CONCLUSION}
An overall framework for the data-driven reduced order model is presented in this work. The current work can be considered as a extension of our previous work on POD-RNN \cite{reddy2019reduced}. The shortcomings of the POD-RNN model are discussed and the idea of completely bypassing the POD for obtaining the low dimensional features is presented via convolutional recurrent autoencoder networks. The proposed convolutional recurrent autoencoder networks serves as an end to end nonlinear model reduction tool for unsteady flow predictions. To demonstrate the effectiveness, the proposed model was applied on the problem of the flow past an isolated cylinder and more complicated problem of the flow past side-by-side cylinders. Pressure and velocity fields were predicted ahead in time for a finite time horizon and the model performed satisfactorily in agreement with the error metrics. We are currently working to extend this framework to more complicated cases of free-surface wave prediction, turbulent flows and fluid-structure interaction of realistic geometries.

\section*{ACKNOWLEDGEMENT}
This work done at the Keppel-NUS Corporate Laboratory was supported by the National Research Foundation, Keppel
Offshore and Marine Technology Centre (KOMtech) and
National University of Singapore (NUS). The conclusions put
forward to reflect the views of the authors alone and not necessarily
those of the institutions within the Corporate Laboratory. The computational work for this article was (fully/partially) performed on resources of the National Supercomputing Centre, Singapore (https://www.nscc.sg).









\bibliographystyle{asmems4}


%
% ...

\bibliography{main}


\end{document}